\documentclass[aps,prx,article,twocolumn,preprintnumbers,amsmath,amssymb,superscriptaddress]{revtex4-2}

\date{\today}
\usepackage{epsfig}
\usepackage{cases}
\usepackage{subfigure}
\usepackage{amsmath}
\usepackage{graphicx}
\usepackage{dcolumn}
\usepackage{bm}
\usepackage[section]{placeins}
\usepackage{float}
\usepackage{mathtools}
\usepackage{dblfloatfix}
\usepackage{color}
\usepackage{comment}
\usepackage{bbm}
\usepackage[normalem]{ulem}
\usepackage[section]{placeins}
\usepackage{xr-hyper}

\makeatletter
\newcommand*{\addFileDependency}[1]{
  \typeout{(#1)}
  \@addtofilelist{#1}
  \IfFileExists{#1}{}{\typeout{No file #1.}}
}
\makeatother

\newcommand*{\myexternaldocument}[1]{
    \externaldocument{#1}
    \addFileDependency{#1.tex}
    \addFileDependency{#1.aux}
}

\myexternaldocument{supplement_v2_arXiv}

\usepackage[colorlinks,linkcolor=blue,hyperindex,CJKbookmarks]{hyperref}

\begin{document}
\newcommand{\Ham}{\mathcal{H}}
\newcommand{\kbf}{\mathbf{k}}
\newcommand{\qbf}{\mathbf{q}}
\newcommand{\Qbf}      {\textbf{Q}}
\newcommand{\lbf}      {\textbf{l}}
\newcommand{\ibf}      {\textbf{i}}
\newcommand{\jbf}      {\textbf{j}}
\newcommand{\rbf}      {\textbf{r}}
\newcommand{\Rbf}      {\textbf{R}}
\newcommand{\Schrdg} {{Schr\"{o}dinger}}
\newcommand{\aband} {{(\alpha)}}
\newcommand{\bband} {{(\beta)}}
\newcommand{\eps} {{\bm{\varepsilon}}}
\newcommand{\probA}      {{\mathsf{A}}}
\newcommand{\HamPump}      {{\Ham_{\rm pump}}}
\newcommand{\HamPr}      {{\Ham_{\rm probe}}}
\newcommand{\timeMax} {{t_{\rm m}}}
\newcommand{\qin} {{\qbf_{\rm i}}}
\newcommand{\qout} {{\qbf_{\rm s}}}
\newcommand{\epsin} {{\eps_{\rm i}}}
\newcommand{\epsout} {{\eps_{\rm s}}}
\newcommand{\win} {{\omega_{\rm in}}}
\newcommand{\wout} {{\omega_{\rm s}}}

\setlength{\parindent}{2ex}

\title{Ultrafast renormalization of the onsite Coulomb repulsion in a cuprate superconductor}

\author{Denitsa R. Baykusheva}
 \email[All correspondence should be addressed to D.B. (\href{mailto:dbaykusheva@g.harvard.edu}{dbaykusheva@g.harvard.edu}), Y.W. (\href{mailto:yaowang@g.clemson.edu}{yaowang@g.clemson.edu}) and M.M. (\href{mailto:mmitrano@g.harvard.edu}{mmitrano@g.harvard.edu}) 
]{}
\affiliation{Department of Physics, Harvard University, Cambridge, Massachusetts 02138, USA}
\author{Hoyoung Jang}
\affiliation{PAL-XFEL, Pohang Accelerator Laboratory, Pohang, Gyeongbuk 37673, Republic of Korea}
\author{Ali A. Husain}
\affiliation{Department of Physics and Seitz Materials Research Laboratory, University of Illinois, Urbana, IL 61801, USA}
\affiliation{Quantum Matter Institute and Department of Physics and Astronomy, University of British Columbia, Vancouver, BC V6T 1Z4, Canada}
\author{Sangjun Lee}
\affiliation{Department of Physics and Seitz Materials Research Laboratory, University of Illinois, Urbana, IL 61801, USA}
\author{Sophia F. R. TenHuisen}
\affiliation{Department of Physics, Harvard University, Cambridge, Massachusetts 02138, USA}
\affiliation{John A. Paulson School of Engineering and Applied Sciences, Harvard University, Cambridge, Massachusetts 02138, USA}
\author{Preston Zhou}
\affiliation{Department of Physics, Harvard University, Cambridge, Massachusetts 02138, USA}
\author{Sunwook Park}
\affiliation{Department of Physics, Pohang University of Science and Technology, Pohang 790-784, South Korea}
\affiliation{Center for Artificial Low Dimensional Electronic Systems, Institute for Basic Science (IBS), 77 Cheongam-Ro, Pohang 790-784, South Korea}
\author{Hoon Kim}
\affiliation{Department of Physics, Pohang University of Science and Technology, Pohang 790-784, South Korea}
\affiliation{Center for Artificial Low Dimensional Electronic Systems, Institute for Basic Science (IBS), 77 Cheongam-Ro, Pohang 790-784, South Korea}
\author{Jinkwang Kim}
\affiliation{Department of Physics, Pohang University of Science and Technology, Pohang 790-784, South Korea}
\affiliation{Center for Artificial Low Dimensional Electronic Systems, Institute for Basic Science (IBS), 77 Cheongam-Ro, Pohang 790-784, South Korea}
\author{Hyeong-Do Kim}
\affiliation{PAL-XFEL, Pohang Accelerator Laboratory, Pohang, Gyeongbuk 37673, Republic of Korea}
\author{Minseok Kim}
\affiliation{PAL-XFEL, Pohang Accelerator Laboratory, Pohang, Gyeongbuk 37673, Republic of Korea}
\author{Sang-Youn Park}
\affiliation{PAL-XFEL, Pohang Accelerator Laboratory, Pohang, Gyeongbuk 37673, Republic of Korea}
\author{Peter Abbamonte}
\affiliation{Department of Physics and Seitz Materials Research Laboratory, University of Illinois, Urbana, IL 61801, USA}
\author{B. J. Kim}
\affiliation{Department of Physics, Pohang University of Science and Technology, Pohang 790-784, South Korea}
\affiliation{Center for Artificial Low Dimensional Electronic Systems, Institute for Basic Science (IBS), 77 Cheongam-Ro, Pohang 790-784, South Korea}
\author{G. D. Gu}
\affiliation{Condensed Matter Physics and Materials Science, Brookhaven National Laboratory (BNL), Upton, NY 11973 USA}

\author{Yao Wang}
 \email[All correspondence should be addressed to D.B. (\href{mailto:dbaykusheva@g.harvard.edu}{dbaykusheva@g.harvard.edu}), Y.W. (\href{mailto:yaowang@g.clemson.edu}{yaowang@g.clemson.edu}) and M.M. (\href{mailto:mmitrano@g.harvard.edu}{mmitrano@g.harvard.edu}) 
]{}
 \affiliation{Department of Physics and Astronomy, Clemson University, Clemson, South Carolina 29631, USA}
\author{Matteo Mitrano}
 \email[All correspondence should be addressed to D.B. (\href{mailto:dbaykusheva@g.harvard.edu}{dbaykusheva@g.harvard.edu}), Y.W. (\href{mailto:yaowang@g.clemson.edu}{yaowang@g.clemson.edu}) and M.M. (\href{mailto:mmitrano@g.harvard.edu}{mmitrano@g.harvard.edu}) 
]{}
\affiliation{Department of Physics, Harvard University, Cambridge, Massachusetts 02138, USA}
\date{\today}

\begin{abstract}

Ultrafast lasers are an increasingly important tool to control and stabilize emergent phases in quantum materials. Among a variety of possible excitation protocols, a particularly intriguing route is the direct light-engineering of microscopic electronic parameters, such as the electron hopping and the local Coulomb repulsion (Hubbard $U$). In this work, we use time-resolved x-ray absorption spectroscopy to demonstrate the light-induced renormalization of the Hubbard $U$ in a cuprate superconductor, La$_{1.905}$Ba$_{0.095}$CuO$_4$. We show that intense femtosecond laser pulses induce a substantial redshift of the upper Hubbard band, while leaving the Zhang-Rice singlet energy unaffected. By comparing the experimental data to time-dependent spectra of single- and three-band Hubbard models, we assign this effect to a $\sim140$ meV reduction of the onsite Coulomb repulsion on the copper sites. Our demonstration of a dynamical Hubbard $U$ renormalization in a copper oxide paves the way to a novel strategy for the manipulation of superconductivity, magnetism, as well as to the realization of other long-range-ordered phases in light-driven quantum materials.

\end{abstract}
\maketitle

\section{INTRODUCTION}

The electronic dynamics of strongly correlated materials is governed by a subtle competition between itinerancy due to hopping and localization driven by Coulomb repulsion \cite{Hubbard1963electron,Imada1997}. The balance between these two tendencies is responsible for a wide variety of emergent quantum phases and its manipulation through external perturbations is a central focus of modern condensed matter physics \cite{Basov2017towards,Tokura2017}. Ultrafast laser pulses offer an intriguing nonequilibrium control route, particularly when fields are strong enough ($\sim0.1$-$1$~V/\AA\hspace{2pt}) to induce energy changes at the scale of the effective electronic interactions. Electronic hopping can be controlled by transiently bending the band structure (dynamical Franz-Keldysh effect)~\cite{Novelli2013,Schultze2013,Schultze2014,Lucchini2016,Schlaepfer2018,Granas2020}, by displacing atoms via nonlinear phonon excitation \cite{Forst2011, Subedi2014,Mankowsky2014}, or by coherent light-matter dressing of the band structure (Floquet engineering) \cite{Wang2013,Mahmood2016selective,McIver2020}. However, manipulating the effective Coulomb repulsion with ultrafast lasers is less straightforward. Optical spectroscopy in organic solids hints at the possibility of modifying the effective onsite interaction (Hubbard $U$) by coherently driving local molecular vibrations~\cite{Kaiser2014,Singla2015,Buzzi2020} or enhancing intradimer hopping \cite{Kawakami2009}. Yet this route relies on the presence of molecular orbitals coupled to local structural degrees of freedom and cannot be readily extended to other classes of strongly correlated materials, like transition metal oxides. In these systems, the Hubbard $U$ is a strictly atomic property and its modification requires alternative microscopic mechanisms, such as dynamical electronic screening \cite{Tancogne-Dejean2018,Topp2018,Golez2019,Tancogne-Dejean2020,Beaulieu2020} or Floquet-type dressing of the Coulomb repulsion \cite{Valmispild2020dynamically}. Achieving dynamical control of the Hubbard $U$ in transition metal oxides would be particularly consequential for steering their multiple quantum phases, notably magnetism, multiferroicity, charge/spin order, and superconductivity \cite{Zhang2014dynamics,Ahn2021designing}.
\par In the specific case of the high-$T_c$ cuprate superconductors, the local Coulomb repulsion has a pervasive effect on normal state properties, as well as magnetic and superconducting phases \cite{Lee2006doping,Keimer2015}. The Hubbard $U$ increases the quasiparticle effective mass \cite{Georges1996dynamical} and broadens the Fermi surface \cite{Rossi2020renormalized}. Furthermore, it sets the scale of the magnetic superexchange and, in a spin-fluctuation-mediated picture, it directly determines the superconducting pairing \cite{Scalapino2012common} and critical temperature \cite{Nilsson2019dynamically}. Whether ultrafast lasers are able to modify the Hubbard $U$ in cuprate superconductors is an open experimental and theoretical problem.
\par Here, we demonstrate a pump-induced renormalization of the onsite Coulomb repulsion in La$_{2-x}$Ba$_{x}$CuO$_4$ (LBCO), a single-layer cuprate with coexisting high-temperature superconductivity (up to $32$ K), and charge and spin order \cite{Abbamonte2005,Tranquada2004,Hucker2011stripe}. We directly probe the electronic density of states of LBCO by measuring its time-dependent x-ray absorption spectrum and show that intense near-infrared pump pulses induce a significant transient shift of the absorption peaks. By mapping our data onto single- and three-band Hubbard models, we assign the shift to a renormalization of the local onsite Coulomb repulsion on the copper orbitals. Finally, we discuss implications of these results for the understanding of light-enhanced superconductivity, the generation of high harmonics in correlated materials, and the realization of coherent light-driven states.
\begin{widetext}
    \begin{minipage}{\linewidth}
        \begin{figure}[H]
            \centering
            \includegraphics[width=\linewidth]{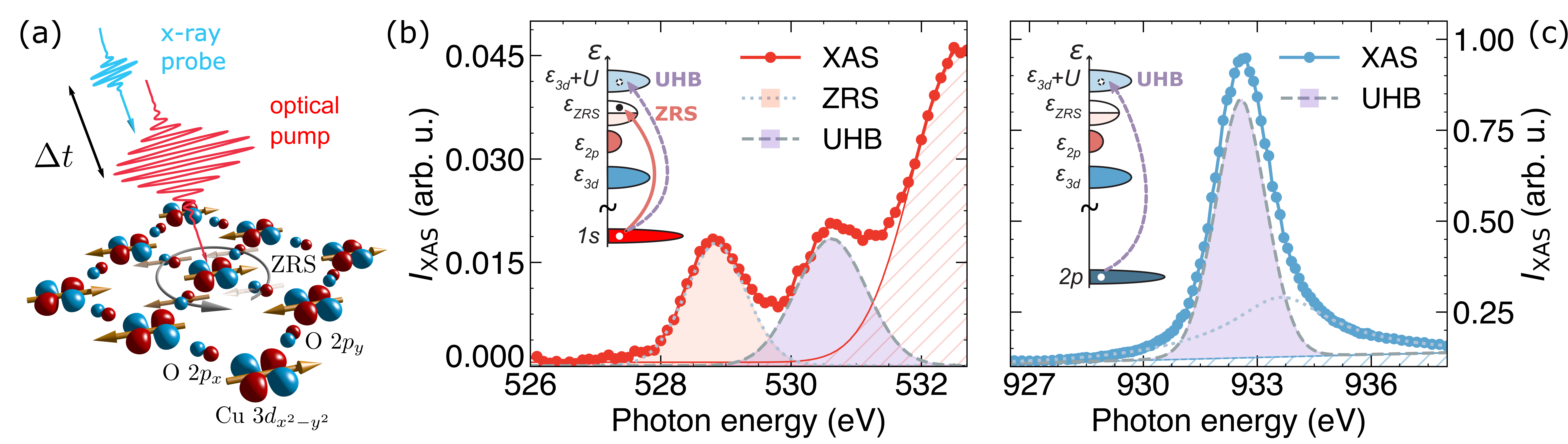}
            \caption{\label{fig:fig1} Probing effective electronic interactions with x-ray absorption spectroscopy. (a) A LBCO ($x=9.5~\%$) sample is driven by intense pump pulses at 1.55 eV, while its low-energy electronic structure is probed by exciting core electrons into unoccupied valence states with soft x-ray pulses. Equilibrium XAS spectrum at (b) O $K$ and (c) Cu $L$ edges, with transitions into Zhang-Rice singlets (ZRS) and upper Hubbard band (UHB) denoted by red and purple areas, respectively. Hatched areas indicate the main absorption edge, while the high-energy shoulder in (c) is a ligand hole sideband which we neglect in the rest of this work. A sketch of the density of states and the relevant x-ray transitions is reported within each panel.}
        \end{figure}
        \vspace{2mm}
    \end{minipage}
\end{widetext}

\section{Experimental Methods}
Time-resolved x-ray absorption spectroscopy (trXAS) is a valuable element-specific probe of the local electronic structure in light-driven materials \cite{Cavalleri2005band,Stamm2007femtosecond,Moulet2017, Carneiro2017, Zurch2017a, Attar2020, Britz2021,Mardegan2021ultrafast}, especially in insulating and/or poorly cleavable samples. This experiment makes use of the resonant soft x-ray scattering (RSXS) endstation of the Pohang Accelerator Laboratory x-ray free electron laser (PAL-XFEL)~\cite{Jang2020}. We acquire trXAS spectra in fluorescence-yield mode at both the O $K$ and Cu $L_3$ edges with an energy resolution of 0.046~eV and 0.34~eV, respectively. The x-ray beam, focused to $120\times 230~\mathrm{\mu m}^2$ and horizontally polarized, impinges at near-normal incidence and is detected by an avalanche photodiode (APD) at $2\theta=150^{\circ}$. Shot-to-shot  x-ray intensity fluctuations are recorded with a gas-monitor detector and used to normalize the XAS signal. We pump the sample with 50-fs pulses centered at 1.55~eV, polarized in the $ab$-plane, and focused to $600\times 600~\mathrm{\mu m}^2$ to obtain a 10~mJ/cm$^2$ fluence ($\sim$12~MV/cm peak electric field). Since the optical penetration depth (370 nm) exceeds that of the x-ray (170-290 nm), our trXAS spectra correspond to a homogeneously excited volume. The sample, a mm-sized single crystal of underdoped LBCO ($x=9.5\: \%$, $T_c=32$~K) \cite{Hucker2011stripe} is cleaved to expose a fresh $ab$ surface and kept at 17~K throughout the entire experiment.

\section{Equilibrium x-ray absorption spectrum}
As shown in Fig.~\ref{fig:fig1}a, the equilibrium electronic structure of the copper oxides is mainly determined by the in-plane O $2p_{x,y}$ and Cu $3d_{x^2-y^2}$ orbitals \cite{Lee2006doping,Ogata2008the}. The $3d$ orbitals form two bands (lower and upper Hubbard bands, LHB and UHB respectively) due to the onsite Coulomb repulsion, the LHB being pushed below the oxygen $2p$ states. This band separation gives rise to a charge transfer (CT) gap, in which the lowest electronic transition occurs between oxygen (occupied) and copper (unoccupied) states. Upon hole doping, the density of states develops an additional feature at the lower end of the CT gap. This feature is generally attributed to a hybridized state between a local $3d$-hole and a $2p$-hole on the four surrounding oxygens, and is known as Zhang-Rice singlet (ZRS) \cite{Zhang1988effective,Eskes1991}. In this work, we measure absorption spectra at both the O $K$ and Cu $L_3$ edges to map the local density of states in the CuO$_2$ planes and determine the transient effective interactions.

\par The equilibrium O $K$ and Cu $L_3$ edge absorption spectra are shown in Fig.~\ref{fig:fig1}b-c. The O $K$ spectrum features two pre-edge peaks at 528.9 and 530.5 eV [see Supplementary Materials (SM) for further details]. The lower-energy transition involves Zhang-Rice singlets (ZRS) ($3d^9\underline{L}\rightarrow \underline{1s}3d^9$, with $\underline{L}$ and $\underline{1s}$ being respectively the O $2p_{x,y}$ ligand and the O $1s$ core holes) \cite{Chen1991,Eskes1991,Abbamonte2005}. The higher-energy peak corresponds  to a transition into the UHB ($3d^9\rightarrow \underline{1s}3d^{10}$), which becomes dipole-allowed due to the mixing of $3d^{10}\underline{L}$ and $3d^9$ configurations in the ground state~\cite{Chen1992}.
The Cu $L_3$ edge spectrum in Fig.~\ref{fig:fig1}c instead exhibits a main peak at 932.6~eV and a weaker shoulder at $\sim$934~eV, which correspond to transitions into the UHB ($3d^9\rightarrow \underline{2p}_{3/2}3d^{10}$) and into ligand states ($3d^9\underline{L}\rightarrow \underline{2p}_{3/2}3d^{10}\underline{L}$) \cite{Chen1992,Nuecker1995} (see SM Sec.~\ref{sec:XAS_analysis_Cu} for further details).

\begin{widetext}
    \begin{minipage}{\linewidth}
        \begin{figure}[H]
            \centering
            \includegraphics[width=\linewidth]{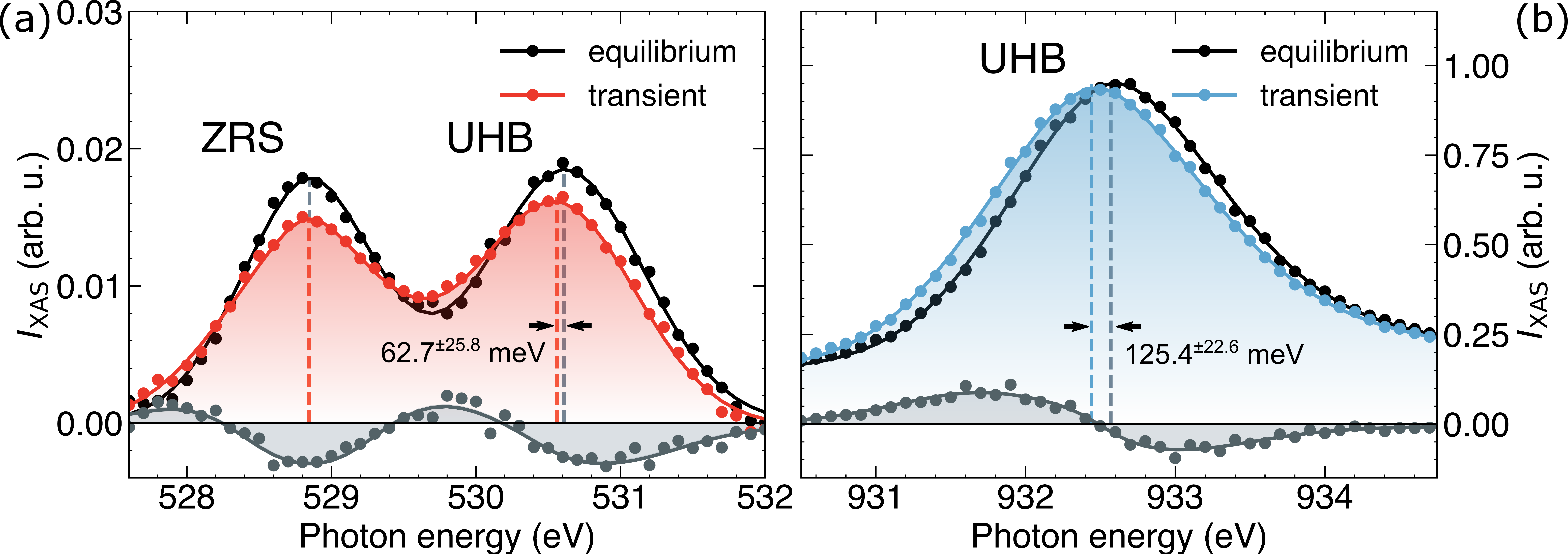}
            \caption{\label{fig:fig2} Pump-induced reshaping of the x-ray absorption. XAS spectra before ($\Delta t\sim -1$ ps) and after ($\Delta t\sim 0$ ps) the pump arrival at both O $K$ (a) and Cu $L_3$ (b) absorption edges. Symbols indicate equilibrium (black), transient (red/blue), and difference spectra (grey), while solid lines represent fits to the spectra. The main edge absorption in (a) has been subtracted for clarity. The fit peak positions for each feature are marked by dashed lines. The UHB is found to shift by the energy indicated in the panels (error bars are 95~\% confidence intervals), while the ZRS position is unchanged.}
        \end{figure}
        \vspace{1mm}
    \end{minipage}
\end{widetext}

\section{Transient x-ray absorption dynamics}
Having assigned the spectral features at both absorption edges, we are now able to track light-induced changes to the LBCO electronic structure. Figure~\ref{fig:fig2} reports our key experimental observation. By comparing equilibrium and transient XAS spectra, we observe prompt photoinduced changes on both absorption edges. At the peak of the response, the UHB peak undergoes a substantial redshift, whereas the ZRS transition energy remains at its equilibrium value. The UHB shift is especially prominent at the Cu $L_3$ edge, as its center energy decreases by ~125~meV compared to a 63~meV shift at the O $K$ edge. The integrated intensity of both peaks is unperturbed and only the ZRS transition is found to broaden by approximately 20~\%.

\begin{figure}[tbh!]
    \begin{center}
        \includegraphics[width=\columnwidth]{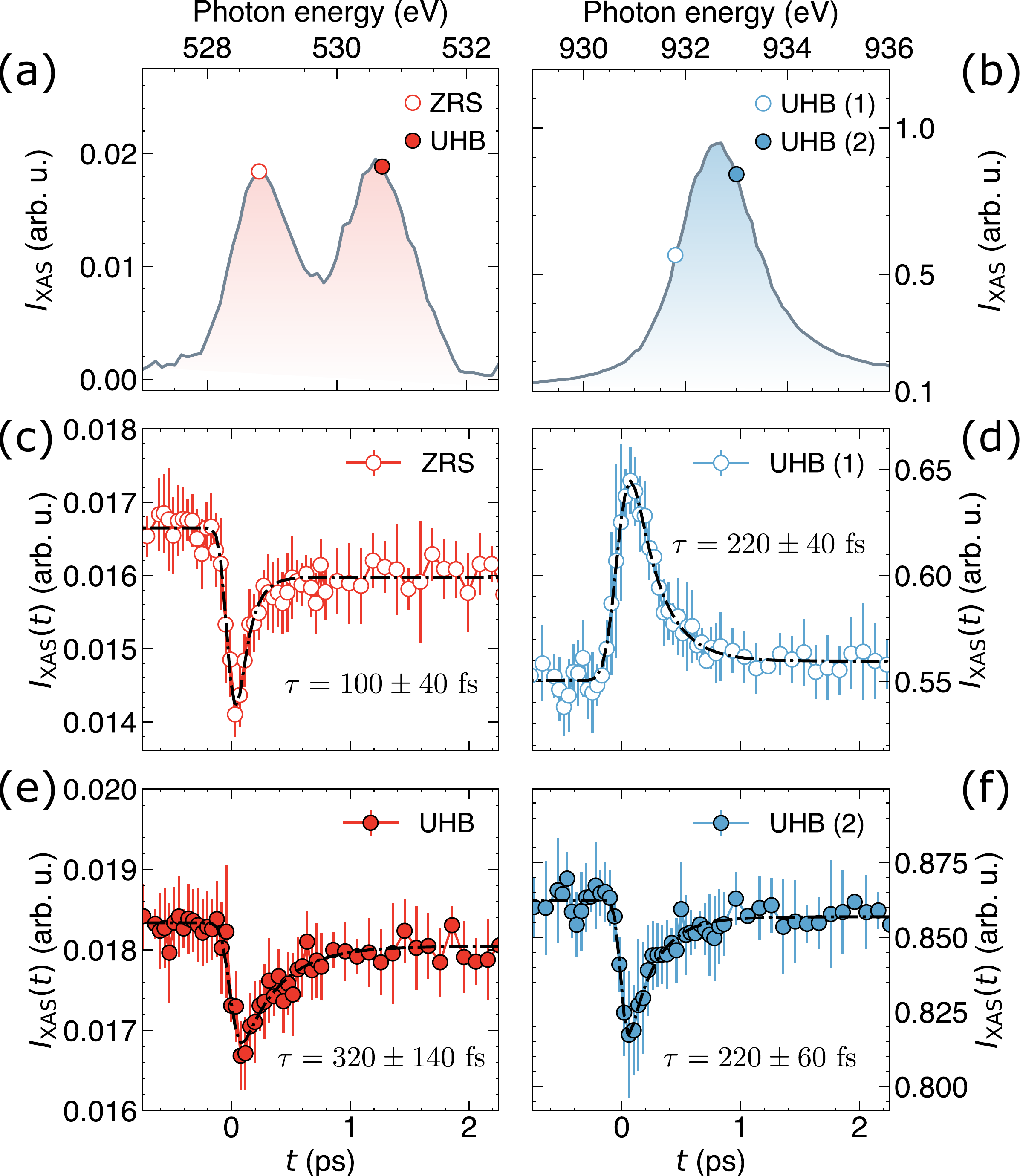}
        \caption{\label{fig:fig3} Time dependence of the pump-induced electronic dynamics. Equilibrium XAS spectra at O $K$ and Cu $L_3$ absorption edges are shown in panels (a) and (b), respectively (shaded areas). Selected energies used to measure the time-dependent recovery are indicated by filled and empty symbols. Panels (c) and (e) display the temporal evolution of ZRS and UHB peaks at the O $K$ edge, while points around the UHB peak at the the Cu $L_3$ edge are reported in panels (d) and (f). Error bars on the symbols represent standard deviations, while dash-dotted lines are fit to the data (see SM for further details).}
    \end{center}
\end{figure}
These photoinduced changes are prompt (close to the $\sim$80 fs pump-probe cross correlation) and relax to equilibrium following a single exponential behavior with a decay constant on the order of 100 fs (see Fig.~\ref{fig:fig3}). These timescales are compatible with a relaxation process mediated by a dissipative bath of optical phonons \cite{Baranov2014theory} and in agreement with previous ultrafast optical spectroscopy \cite{Okamoto2010ultrafast,Okamoto2011photoinduced,DalConte2015snapshots}, electron diffraction \cite{Carbone2008direct,Konstantinova2018} and photoemission \cite{Perfetti2007ultrafast,Dakovski2015quasiparticle,Rameau2016energy} experiments on multiple cuprate compounds.
\par Intriguingly, the UHB recovers more slowly than the ZRS peak. The UHB approaches equilibrium in 320 fs (220 fs) at the O $K$ (Cu $L_3$) edge, while the ZRS decays over $100$ fs (see SM). This is consistent with holon-doublon recombination dynamics in Mott-Hubbard and CT insulators, a mechanism also at play in other copper oxides \cite{Okamoto2010ultrafast,Okamoto2011photoinduced,Lenarcic2013ultrafast,Novelli2014witnessing,Cilento2018dynamics}, organic Mott insulators \cite{Wall2011quantum,Mitrano2014pressure}, and cold atoms in optical lattices \cite{strohmaier2010observation,Sensarma2010lifetime}. In our experiment, the 1.55 eV pump excites electrons across the CT gap, thus creating holes (holons) and double occupancies (doublons) \cite{Lenarcic2014charge}. Unlike conduction carriers within the ZRS band, holon-doublon pairs need to dissipate an energy approximately equal to the CT gap in order to recombine. This process generally requires a large number of lower-energy scattering partners (e.g. phonons, magnons, or lower energy charge fluctuations), thus making decay events increasingly rare \cite{Sensarma2010lifetime,Lenarcic2013ultrafast,Lenarcic2014charge}. As consequence, the UHB relaxation rate gets exponentially suppressed with the increasing CT gap \cite{Lenarcic2013ultrafast,Lenarcic2014charge} and becomes slower than the conduction-band quasiparticle relaxation.
\par These data cannot be rationalized in terms of pump-induced doping or heating effects. Hole doping in the closely related La$_{2-x}$Sr$_x$CuO$_4$ \cite{Chen1991,Chen1992,Nuecker1995} leads to significant intensity changes and an increased energy separation between UHB and ZRS transitions, while we observe a decrease in the energy gap between the two peaks. Electronic heating will redshift the entire absorption spectrum through the creation of unoccupied states at lower energies \cite{Stamm2007femtosecond,Higley2019femtosecond}, with the ZRS shifting more (129~meV) than the UHB peak (87 meV) at an estimated $T_e\sim2500$ K (see Appendix \ref{sec:heating} and Sec.~\ref{sec:T-dependence} of SM for electronic temperatures between $T_e=0$ and $T_e=5300$~K). In contrast, we experimentally observe a shift of the UHB while the ZRS remains fixed. Finally, the excitation of optical or acoustic phonons \cite{Perfetti2007ultrafast,Konstantinova2018,Mansart2013temperature}, which may form multiplet structures \cite{Mahan2000many} and broaden the XAS spectrum beyond the natural core hole width \cite{Hybertsen1992model,Ament2010resonant}, will primarily affect intensity and width (and not the energy) of the near-edge transitions.
Therefore, a more natural explanation for our data is a dynamic renormalization of the effective electronic interactions.

\section{Quantifying the transient Hubbard $U$}\label{sec:theory}
\par To determine which electronic interactions are affected by the pump, we calculate the transient x-ray absorption of a three-band Hubbard model with renormalized electronic parameters ~\cite{Mattheiss1987,Emery1987, Varma1987,Hybertsen1992model,Chen2013b}. The calculation is performed at $12.5\: \%$ hole doping in order to reduce the computational complexity. While the superconducting and charge-ordered phases at $12.5\: \%$ hole doping are distinct from the ones of our $x=9.5\: \%$ sample \cite{Hucker2011stripe}, the XAS spectra are similar enough in intensity and peak energy \cite{Chen1991,Chen1992} to justify this simplification. 
The theoretical spectra closely resemble the experimental x-ray absorption in Figs.~\ref{fig:fig1}-\ref{fig:fig2}. The O $K$ edge spectrum in Fig.~\ref{fig:fig4}a features one prominent ZRS peak approximately 1 eV below the edge threshold ($E_{\mathrm{edge}}$), and additional spectral peaks around 1 eV corresponding to the UHB. The Cu $L_3$ edge spectrum is instead characterized by a prominent UHB peak around $E_{\mathrm{edge}}$ and a weaker peak 1.2 eV above corresponding to the $3d^9\underline{L}$ states. All spectra are broadened only by accounting for the core-hole lifetime, which is assumed to be 0.2 eV at both the O $K$ and Cu $L_3$ edges \cite{Nicolas2012lifetime,Rossi2019experimental}. At the O $K$ edge, the separation between ZRS and UHB centroid for $U_d=8.5$ eV matches the experimental separation ($\sim1.8$ eV), while the overall UHB intensity is about 85~\% the one of the ZRS (as expected for the 12.5~\% hole-doped spectrum \cite{Chen1991}). To accurately determine the slope of the UHB peak shift, we only track the sharpest feature, which is located at slightly lower energy.
We systematically tune each parameter of the three-band Hubbard Hamiltonian and map the corresponding effect on a static XAS spectrum capturing the excited steady state.
Our observation of redshifting UHB and fixed ZRS transitions is only reproduced by changing the onsite Coulomb repulsion $U_d$ (see Fig.~\ref{fig:fig4}a,c). All other parameters ($t_{pp}$, $U_p$, and $t_{pd}$) introduce an energy shift of both peaks in tandem (see SM Sec.~\ref{sec:3band}). The shift of the sole UHB band crucially identifies the photoinduced dynamics as a transient renormalization of the Hubbard $U$ on the copper sites. By fitting the theoretical spectra \textcolor{black}{ (see SM Sec.~\ref{sec:3band})}, we subsequently convert the peak shift into an absolute change $\delta U_d$ of the Hubbard $U$. The UHB peak position is found to shift linearly for small $\delta U_d$ (see Fig.~\ref{fig:fig4}b,d), while the ZRS is almost unaffected. The slope of the UHB position as function of $U_d$ at the Cu $L_3$ edge is larger than that at the O $K$ edge, consistent with the experiment, and the observed shifts are reproduced by a Hubbard $U$ reduction $\delta U_d\sim140$ meV.

\begin{widetext}
    \begin{minipage}{\linewidth}
        \begin{figure}[H]
            \centering
            \includegraphics[width=0.9\linewidth]{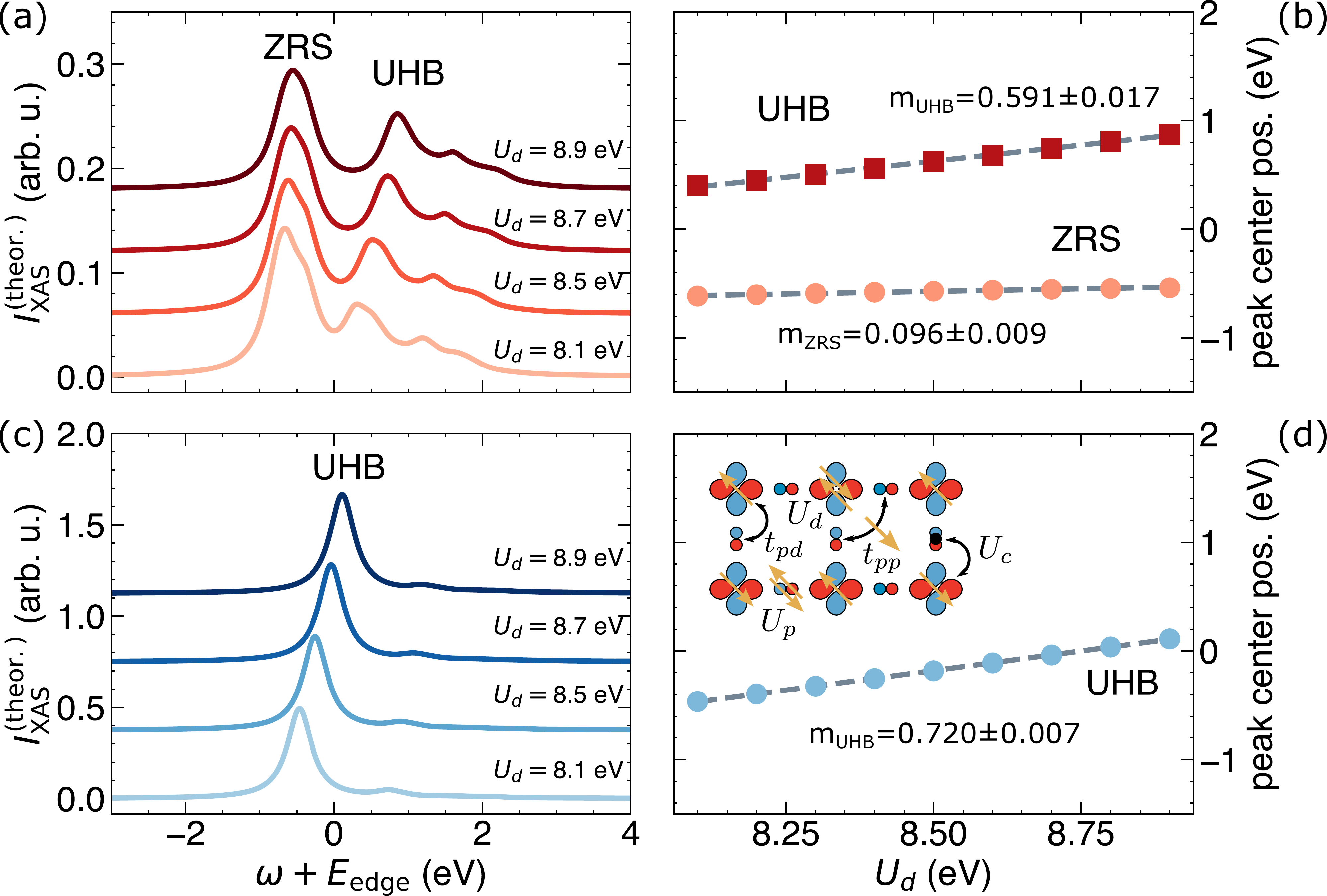}
            \caption{\label{fig:fig4} Theoretical XAS spectra of a three-band Hubbard model for variable onsite Coulomb repulsion $U_d$. Panels (a) and (c) show XAS spectra at the oxygen and copper absorption edges, respectively. The value of $U_d$ used in the calculation is reported nearby each curve, while all the other electronic parameters are kept fixed at their equilibrium values. Panels (b) and (d) show UHB and ZRS center positions a function of $U_d$. The inset in panel (d) is a cartoon of the three-band Hubbard system used in this calculation with hopping amplitudes $t_{pd}$ and $t_{pp}$, onsite Coulomb repulsions $U_d$ and $U_p$, and core-hole interaction $U_c$.}
        \end{figure}
    \end{minipage}
\end{widetext}

\section{A minimal theoretical description}
While providing quantitative information about the electronic interactions, it is not obvious \textit{a priori} that the transient dynamics can be fully captured by a quasiequilibrium three-band Hubbard model with tuned electronic parameters. Therefore, we perform a full time-dependent exact diagonalization calculation of the light-driven x-ray absorption spectrum. We downfold the three-band Hamiltonian to an effective 2D, single-band Hubbard model \cite{Hybertsen1992model,Zhang1988effective}. In this transition, the three-band UHB maintains its original character, while the three-band ZRS band becomes the single-band LHB. Non-bonding and Zhang-Rice triplet states from the three-band model are ignored and the single-band Mott gap is identified with the three-band CT gap \cite{Maekawa2004physics,Kung2016characterizing}. We include the pump pulse through the standard Peierls substitution and calculate the XAS spectra as function of pump-probe time delay, as shown in Fig.~\ref{fig:fig5}. The calculations are performed on a 12D Betts cluster with two holes in order to preserve $SU(2)$ symmetry. At equilibrium, the XAS signal features two distinct transitions into LHB and UHB, qualitatively agreeing with the experimental O $K$ edge spectra. The peak width is determined by two contributions, the intrinsic core-hole lifetime (set to 0.15~eV) and an extrinsic energy broadening introduced by the finite probe pulse (see SM Sec. \ref{sec:singlebandSM}). The UHB peak intensity is about 70~\% of the LHB peak intensity (as opposed to the equally intense peaks in Fig.~\ref{fig:fig2}a), while the energy separation is $U-U_c\sim1$~eV  \cite{Tsutsui2016incident} with $U_c$ being the core-hole attractive interaction. 
When the pump and probe pulses overlap, the peak distance shrinks by approximately40~meV, thus indicating a transient reduction of the Hubbard $U$ by 4~\%. This Mott gap closure is quantitatively consistent with the 3~\% CT gap reduction visible in Fig.~\ref{fig:fig2}a. The spectra also exhibit a general broadening and an increase of in-gap spectral weight, likely due to the creation of particle-hole excitations. The UHB shift is found to increase nonlinearly with the pump electric field, in agreement with our experimental findings (see Appendix B).
Our dynamical single-band calculation captures the experimental UHB peak shift at a quantitative level and corroborates the three-band picture of a pump-renormalized onsite Coulomb repulsion.

\section{Possible microscopic mechanisms}
\par We now discuss the possible microscopic origin of the observed Hubbard $U$ reduction. A first scenario involves the dynamical enhancement of dielectric screening \cite{Tancogne-Dejean2018,Tancogne-Dejean2020,Golez2019}. The pump pulse promotes electrons from localized levels into highly delocalized states, which increases the screening of the local Coulomb interaction. In a multi-band model, screening arises from both interband and intraband excitations, while second order ($\sim t^2/U$) doublon-holon and doublon-doublon interactions are the dominant screening modes in a single-band system.

\begin{widetext}
    \begin{minipage}{\linewidth}
        \begin{figure}[H]
            \centering
            \includegraphics[width=0.8\linewidth]{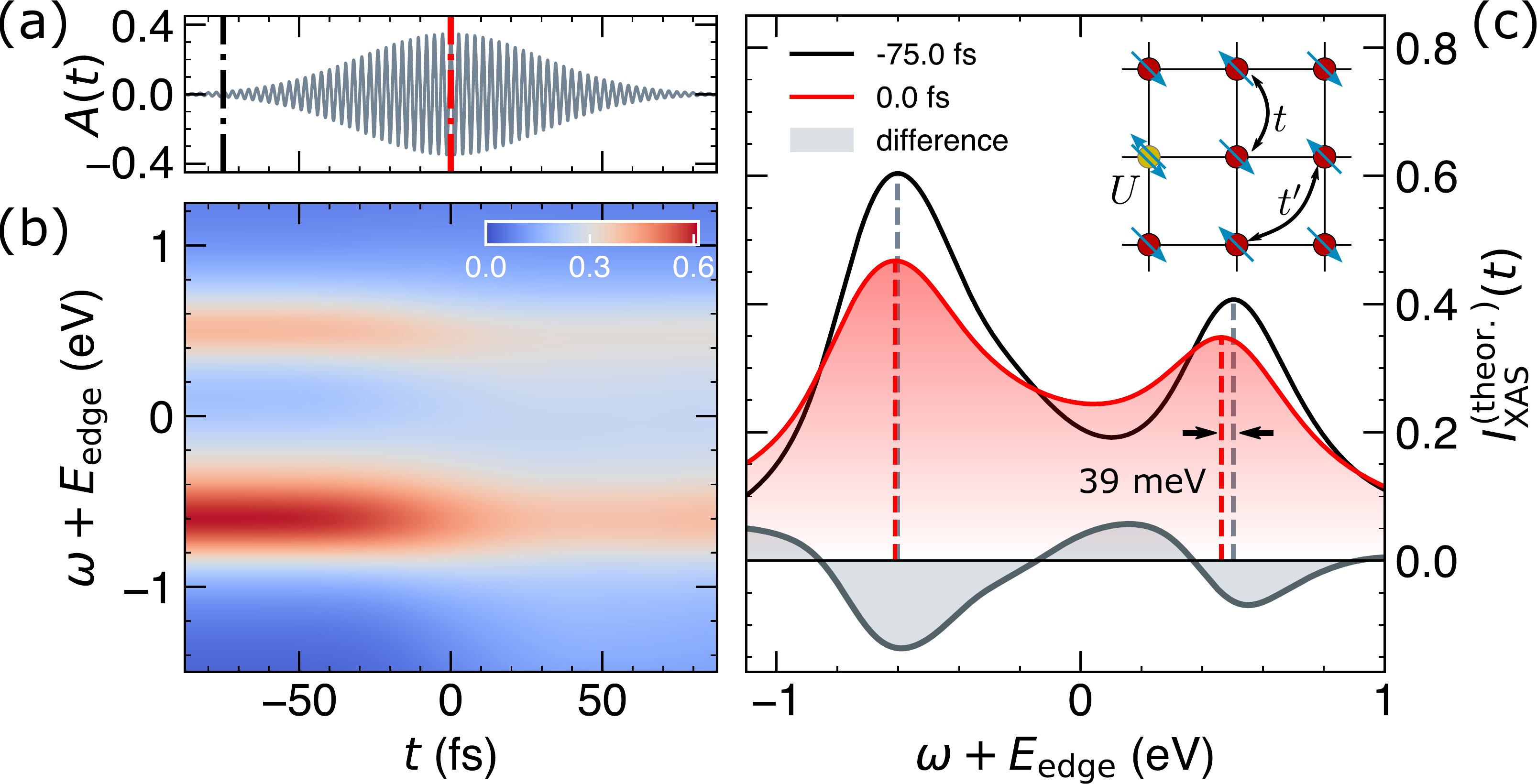}
            \caption{\label{fig:fig5} Transient XAS spectrum of the single-band Hubbard model in two dimensions. (a) Time-dependent vector potential of the pump pulse used in the time-dependent calculations. Colored dashed lines indicate selected time delays for the snapshots in panel (c). (b) Full trXAS spectrum simulation as function of time delay and photon energy (c) trXAS spectra at different pump-probe delays before (black) and during the pump (red). The peak positions of the UHB features are marked by dashed lines. The inset depicts the single-band Hubbard model with $t,t'$ hopping amplitudes, and onsite Coulomb repulsion $U$.}
        \end{figure}
    \end{minipage}
\end{widetext}

The presence of second-order charge and spin fluctuations is compatible with the scattering from antiferromagnetic spin fluctuations observed in ultrafast reflectivity experiments \cite{DalConte2015snapshots}. Intriguingly, this mechanism leads to an average dynamical reduction of the Hubbard $U$ on both $p$ and $d$ orbitals \cite{Tancogne-Dejean2018,Tancogne-Dejean2020,Golez2019}, while the fixed ZRS energy hints to an unperturbed Coulomb repulsion on the $p$ orbitals. Furthermore, a transient enhancement of dielectric screening, being proportional to the intensity instead of the electric field, would be accompanied by oscillations of the UHB position at twice the frequency of the laser field (beyond our current time resolution) \cite{Tancogne-Dejean2018,Tancogne-Dejean2020}.  
\par A second scenario involves instead a Floquet-type dressing of the effective electronic interactions \cite{Valmispild2020dynamically,Tsutsui2021antiphase}. The vector potential of the pump dynamically alters the Hubbard $U$ through both the average doublon number and their mutual interactions (of order $\sim t^2/U$). Notably, this scheme could lead to either an increase or a decrease of the Hubbard $U$ depending on the pump field strength. Based on the observation that the Hubbard $U$ renormalization lifetime is longer than the pump pulse duration (see Fig.~\ref{fig:fig3}), we argue that dynamical screening may be dominant over Floquet dressing.
Nonetheless, a final assignment will require further investigation, possibly with attosecond x-ray pulses resolving coherent dynamics at the $U$ energy scale \cite{Buades2021attosecond}. 

\section{Conclusion}
\par In summary, our experiment provides clear spectroscopic evidence for a transient and reversible renormalization of the Hubbard $U$ in a cuprate superconductor. This result has broad implications for photoinduced phase transitions and nonlinear phenomena in strongly correlated materials.
\par First, it provides a possible key to interpret the recent observation of light-enhanced superconductivity in cuprates \cite{Hu2014,Kaiser2014optically,Nicoletti2014optically,Nicoletti2018magnetic,Cremin2019photoenhanced}, which still lacks a comprehensive microscopic understanding. A 140-meV reduction of the Hubbard $U$ will close the CT gap by 70 meV, or 3~\% of its equilibrium value. Since the CT gap magnitude anticorrelates with the equilibrium superconducting $T_c$ in both theory and experiments \cite{Weber2012scaling,Ruan2016relationship}, such reduction might be favorable to the onset of out-of-equilbrium pairing. Moreover, in a spin-fluctuation-mediated picture, the CT gap closure will be accompanied by a renormalization of the pairing interaction and of the spin fluctuation spectrum \cite{Scalapino2012common,Romer2020pairing}. A lower CT gap will also increase the exchange interaction and, hence, the characteristic spin fluctuation energy \cite{LeTacon2011intense}. These effects might quantitatively explain the enhancement of superconducting correlations in photoexcited La$_{2-x}$Ba$_x$CuO$_4$ \cite{Nicoletti2014optically,Nicoletti2018magnetic} and could be resolved in future time-resolved RIXS experiments \cite{Mitrano2020probing,wang2021xray}.
\par A dynamical renormalization of the band structure has also been discussed as an essential ingredient to boost high-harmonic generation (HHG) efficiency in correlated materials, e.g. NiO  \cite{Tancogne-Dejean2018,Granas2020}. A lower Mott gap indeed reduces the energy cost of Zener tunneling and multiphoton ionization processes, thus promoting electron excitation and harmonic emission under strong light irradiation conditions.
\par Finally, achieving on-demand manipulation of the Coulomb repulsion in correlated materials may enable the observation of novel nonequilibrium states of matter, such as fragile quantum spin liquids in frustrated magnets \cite{Yang2010effective,Sahebsara2008hubbard,Shirakawa2017ground,Szasz2020chiral,Liao2017gapless,Simeng2011spin,Jiang2012identifying}, and  $\eta$-paired superconductivity in driven Mott insulators \cite{Kaneko2019photoinduced,peronaci2020enhancement,Li2020eta,Tindall2020dynamical}. In the former, a dynamical Hubbard $U$ could modify the exchange interactions and induce a transition from an antiferromagnet to an entangled spin liquid state. In the latter, tuning the Hubbard $U$ in a driven-dissipative Mott insulator will dynamically alter doublon number \cite{peronaci2020enhancement} and energy spectrum of the driven Hamiltonian \cite{Zhang1990pseudospin,Kaneko2019photoinduced}, thus paving the way to the onset of long-range staggered superconducting correlations.

\section*{Acknowledgements} We thank M. Buzzi, A. Cavalleri, M. P. M. Dean, T. P. Devereaux, M. Eckstein, D. Gole\v{z}, M. Li, B. Moritz, D. Nicoletti, A. H. Reid, A. Rubio, G. Sawatzky and M. Sentef for insightful discussions. M. M. was supported by the William F. Milton Fund at Harvard University. D. B. acknowledges support from the Swiss National Science Foundation through project P400P2\_194343.
B. J. K. was supported by IBS-R014-A2. H. J. acknowledges support by the National Research Foundation grant funded by the Korea government (MSIT) (Grant No. 2019R1F1A1060295). S. L., A. A. H., P. A. acknowledge support from DOE grant DE-FG02-06ER46285.  P. A. acknowledges support from the Gordon and Betty Moore Foundation EPiQS Initiative through grant no. GBMF9452. The trXAS experiments were performed at the SSS-RSXS endstation (proposal number: 2020-2nd-SSS-006) of the PAL-XFEL funded by the Korea government (MSIT). Work at Brookhaven National Laboratory was supported by the U.S. DOE, Office of Science, Office of Basic Energy Sciences, under Contract No. DESC0012704. This research used resources of the National Energy Research Scientific Computing Center (NERSC), a U.S.~Department of Energy Office of Science User Facility operated under Contract No.~DE-AC02-05CH11231.

\appendix
\section{Calculation of possible pump-induced heating effects on the x-ray absorption spectrum} \label{sec:heating}
In this Appendix, we consider whether the transient reshaping of the experimental x-ray absorption spectrum could be due to pump-induced sample heating. We first determine the maximum electronic temperature increase expected under the experimental pump excitation conditions and then perform an exact diagonalization calculation of the theoretical XAS spectrum of a three-band Hubbard model at such electronic temperature value.
For the first step, we follow the procedure described the Supplementary Material of Ref.~\cite{Cremin2019photoenhanced} and assume the following form for the specific heat of LBCO:
\begin{equation}
    C_s(T) = \gamma T + \beta T^3,
\end{equation}
with $\gamma=2.5$~$\mathrm{mJ\cdot mol^{-1}\cdot K^{-2}}$ and $\beta=0.25$~$\mathrm{mJ\cdot mol^{-1}\cdot K^{-4}}$ being the electronic and lattice specific heat coefficients, respectively. Assuming that the pump pulse varies on a timescale shorter than the heating-cooling dynamics, the maximum temperature increase $\Delta T=T_f-T_i$ is obtained from the absorbed fluence as:
\begin{equation}
Q_\mathrm{abs} = \int_{T_i}^{T_f} C_s(T) \mathrm{d} T.  
\end{equation}
Given the reflection coefficient of the material ($R=0.15$) and the estimated penetration depth at 800~nm ($l_\mathrm{p}\approx 370$~nm), we estimate the absorbed energy according to the relation:
\begin{equation}\label{eq:Qabs}
Q_\mathrm{abs}=\frac{F\cdot A\cdot(1-R)}{2l_\mathrm{p}}N_\mathrm{A} V_\mathrm{uc},
\end{equation}
where $F=10$~mJ/cm$^2$ denotes the laser fluence, $V_\mathrm{uc}=3.787\times 3.787 \times 13.23$~$\text{\AA}^3$ is the volume of the unit cell, and $N_\mathrm{A}$ is the Avogadro constant. The factor 2 accounts for the presence of two LBCO formula units per unit cell. Setting the initial temperature equal to the base temperature (17~K), we obtain a final temperature $T_f=120.3$~K. 
In order to get an estimate for the maximal \textit{electronic} temperature increase, $\Delta T_{e, \mathrm{max}} =T_{e, \mathrm{max}}-T_{e, 0}$, we make the following approximation of the absorbed fluence:
\begin{equation}
Q_\mathrm{abs} = \int^{T_{e, \mathrm{max}}}_{T_{e, 0}} C_e(T_e) \mathrm{d} T_e.  
\end{equation}

Here, $C_e(T_e)=\gamma T_e$ is the electronic (temperature-dependent) specific heat capacity. Using the fluence value estimated from Eq.~\eqref{eq:Qabs} and setting the initial $T_{e, 0}$ equal to the base temperature, we obtain a maximum value for $T_{e, \mathrm{max}}\approx 3240$~K. However, the pump is attenuated while traveling across the photoexcited volume, hence a more refined approximation for the maximum electronic temperature increase is \cite{Mansart2010}:
\begin{equation}
    T_{e, \mathrm{max}} = \frac{1}{l_\mathrm{p}}\int_0^{l_\mathrm{p}} \sqrt{T_{e, 0}^2+\frac{2Q_\mathrm{abs}}{\gamma}e^{-z/l_\mathrm{p}}}\mathrm{d}z.
    \label{eq:boschetto}
\end{equation}
According to Eq. \ref{eq:boschetto}, the maximum electronic temperature increase allowed within our experimental conditions is $T_{e, \mathrm{max}}\approx 2561$~K.
Having determined the maximum electronic temperature, we examine the corresponding changes in the XAS spectrum. We perform finite-temperature ED calculations within the three-band Hubbard model for temperatures ranging between $T_e=0$ and $T_e=5300$~K. In Fig.~\ref{fig:finiteTshift}, we illustrate the effect of electronic heating on the shape of the XAS signal by comparing spectra at $T_e=0$ and $T_{e, \mathrm{max}}= 2535.5$~K. Intuitively, electronic heating creates unoccupied states at lower energies and leads to a shift of the entire absorption spectrum. Both the UHB and ZRS peaks are found to shift, with the latter redshifting more than the former. Moreover, the UHB shift at the Cu $L_3$ edge is almost half the one occurring at the O $K$ edge. This is at odds with the experimental findings in Fig.~\ref{fig:fig2}, where the ZRS transition does not shift within experimental accuracy and the UHB shift at the Cu $L_3$ edge is almost twice larger than the one at the O $K$ edge. The discrepancy strongly suggests that the observed XAS spectral reshaping cannot be explained in terms of a heating-only scenario.

\section{Field dependence of the upper Hubbard band shift}
\label{sec:fluence_dep}
In this section, we compare experimental shifts of the UHB with the theoretically expected fluence-dependent trXAS spectra of a  single-band Hubbard model. The trXAS calculations are performed on a 12D Betts cluster with two holes (in order to preserve $SU(2)$ symmetry) and for variable peak electric fields between $E_0=1.75$~\mbox{MV/cm} and $E_0=17.5$~\mbox{MV/cm}. The spectra, reported in Fig.~\ref{fig:fluence}a, are calculated by using the same interaction parameters considered in Fig.~\ref{fig:fig5} and the theoretical peak positions in Fig.~\ref{fig:fluence}b are defined as the energy of the peak maximum. While the UHB is found to shift appreciably, the LHB shift is much less pronounced and discernible only for the highest field amplitudes considered here. In particular, the peak shift is nonlinear in the field amplitude, as opposed to the predicted Hubbard $U$ reduction by dynamical screening reported in Ref. \cite{Tancogne-Dejean2018}. Experimental data for the UHB peak shift at the Cu $L_3$ edge are also shown in Fig.~\ref{fig:fluence}b for two compositions, $x=0.095$ (from \ref{fig:fig2}b) and $x=0.125$ (see SM). The experimental and theoretical trends (when rescaled by the relative size of the Mott/CT gap) agree on a quantitative level and outline a nonlinear field dependence of the pump-induced Hubbard $U$ reduction. 

\begin{widetext}
    \begin{minipage}{\linewidth}
        \begin{figure}[H]
            \centering
            \includegraphics[width=\linewidth]{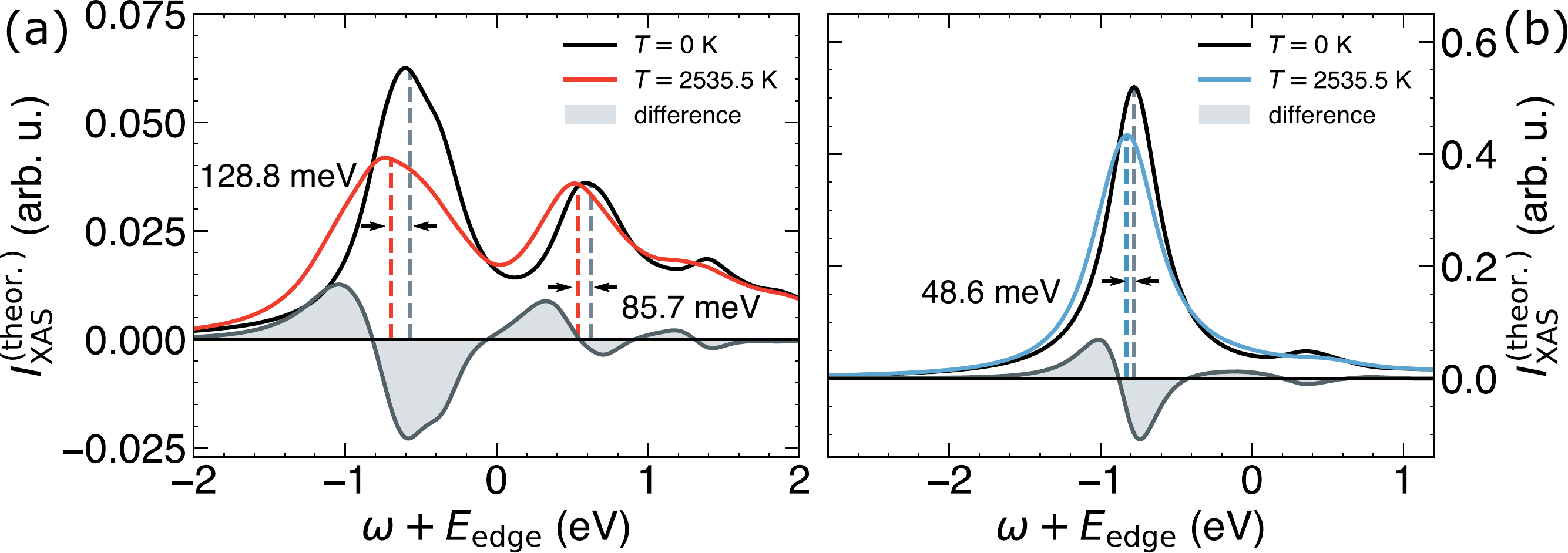}
            \caption{\label{fig:finiteTshift}Theoretical XAS spectra of the three-band Hubbard model at the O $K$ edge (panel a) and the Cu $L_3$ edge (panel b) at zero temperature (black) and at $T=2535.5$~K (color).}
        \end{figure}
        \vspace{1mm}
    \end{minipage}
\end{widetext}

\begin{widetext}
    \begin{minipage}{\linewidth}
        \begin{figure}[H]
            \centering
            \includegraphics[width=0.9\linewidth]{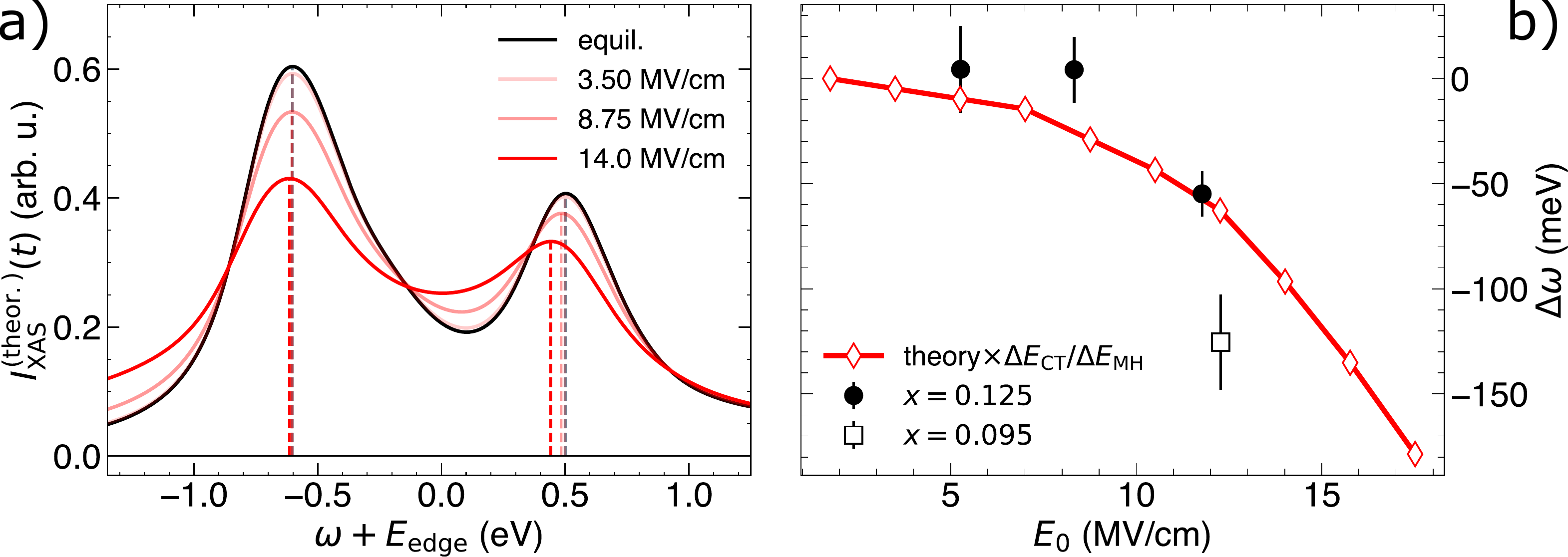}
            \caption{\label{fig:fluence} (a) Theoretical trXAS spectra of the single-band Hubbard model at equilibrium ($\Delta t=-75$~fs, black) and at $\Delta t=0$~fs for selected pump electric fields. (b) Theoretical UHB peak shift $\Delta\omega^{\mathrm{theor.}}=\omega(\mathrm{0\: fs})-\omega(\mathrm{-75\:fs})$ (red diamonds) as function of the pump peak field $E_0$ and rescaled by the ratio between CT gap and Mott-Hubbard gap. $\Delta E_{\mathrm{CT}}=1.77$ eV is the experimental separation of the equilibrium ZRS and UHB bands at the O $K$ edge, while $\Delta E_{\mathrm{MH}}=1.03$ eV is the separation between LHB and UHB in the single band Hubbard model. Experimental UHB peak shift values at the Cu $L_3$ edge for the two dopings are indicated by black symbols.}
        \end{figure}
        \vspace{1mm}
    \end{minipage}
\end{widetext}

\clearpage
\bibliography{trXAS_LBCO}

\begin{thebibliography}{109}%
\makeatletter
\providecommand \@ifxundefined [1]{%
 \@ifx{#1\undefined}
}%
\providecommand \@ifnum [1]{%
 \ifnum #1\expandafter \@firstoftwo
 \else \expandafter \@secondoftwo
 \fi
}%
\providecommand \@ifx [1]{%
 \ifx #1\expandafter \@firstoftwo
 \else \expandafter \@secondoftwo
 \fi
}%
\providecommand \natexlab [1]{#1}%
\providecommand \enquote  [1]{``#1''}%
\providecommand \bibnamefont  [1]{#1}%
\providecommand \bibfnamefont [1]{#1}%
\providecommand \citenamefont [1]{#1}%
\providecommand \href@noop [0]{\@secondoftwo}%
\providecommand \href [0]{\begingroup \@sanitize@url \@href}%
\providecommand \@href[1]{\@@startlink{#1}\@@href}%
\providecommand \@@href[1]{\endgroup#1\@@endlink}%
\providecommand \@sanitize@url [0]{\catcode `\\12\catcode `\$12\catcode
  `\&12\catcode `\#12\catcode `\^12\catcode `\_12\catcode `\%12\relax}%
\providecommand \@@startlink[1]{}%
\providecommand \@@endlink[0]{}%
\providecommand \url  [0]{\begingroup\@sanitize@url \@url }%
\providecommand \@url [1]{\endgroup\@href {#1}{\urlprefix }}%
\providecommand \urlprefix  [0]{URL }%
\providecommand \Eprint [0]{\href }%
\providecommand \doibase [0]{https://doi.org/}%
\providecommand \selectlanguage [0]{\@gobble}%
\providecommand \bibinfo  [0]{\@secondoftwo}%
\providecommand \bibfield  [0]{\@secondoftwo}%
\providecommand \translation [1]{[#1]}%
\providecommand \BibitemOpen [0]{}%
\providecommand \bibitemStop [0]{}%
\providecommand \bibitemNoStop [0]{.\EOS\space}%
\providecommand \EOS [0]{\spacefactor3000\relax}%
\providecommand \BibitemShut  [1]{\csname bibitem#1\endcsname}%
\let\auto@bib@innerbib\@empty
\bibitem [{\citenamefont {Hubbard}(1963)}]{Hubbard1963electron}%
  \BibitemOpen
  \bibfield  {author} {\bibinfo {author} {\bibfnamefont {J.}~\bibnamefont
  {Hubbard}},\ }\bibfield  {title} {\bibinfo {title} {{Electron correlations in
  narrow energy bands}},\ }\href {https://doi.org/10.1098/rspa.1963.0204}
  {\bibfield  {journal} {\bibinfo  {journal} {Proc. R. Soc. Lond. A}\ }\textbf
  {\bibinfo {volume} {276}},\ \bibinfo {pages} {238} (\bibinfo {year}
  {1963})}\BibitemShut {NoStop}%
\bibitem [{\citenamefont {Imada}\ \emph {et~al.}(1998)\citenamefont {Imada},
  \citenamefont {Fujimori},\ and\ \citenamefont {Tokura}}]{Imada1997}%
  \BibitemOpen
  \bibfield  {author} {\bibinfo {author} {\bibfnamefont {M.}~\bibnamefont
  {Imada}}, \bibinfo {author} {\bibfnamefont {A.}~\bibnamefont {Fujimori}},\
  and\ \bibinfo {author} {\bibfnamefont {Y.}~\bibnamefont {Tokura}},\
  }\bibfield  {title} {\bibinfo {title} {Metal-insulator transitions},\ }\href
  {https://doi.org/10.1103/RevModPhys.70.1039} {\bibfield  {journal} {\bibinfo
  {journal} {Rev. Mod. Phys.}\ }\textbf {\bibinfo {volume} {70}},\ \bibinfo
  {pages} {1039} (\bibinfo {year} {1998})}\BibitemShut {NoStop}%
\bibitem [{\citenamefont {Basov}\ \emph {et~al.}(2017)\citenamefont {Basov},
  \citenamefont {Averitt},\ and\ \citenamefont {Hsieh}}]{Basov2017towards}%
  \BibitemOpen
  \bibfield  {author} {\bibinfo {author} {\bibfnamefont {D.}~\bibnamefont
  {Basov}}, \bibinfo {author} {\bibfnamefont {R.}~\bibnamefont {Averitt}},\
  and\ \bibinfo {author} {\bibfnamefont {D.}~\bibnamefont {Hsieh}},\ }\bibfield
   {title} {\bibinfo {title} {Towards properties on demand in quantum
  materials},\ }\href@noop {} {\bibfield  {journal} {\bibinfo  {journal} {Nat.
  Mater.}\ }\textbf {\bibinfo {volume} {16}},\ \bibinfo {pages} {1077}
  (\bibinfo {year} {2017})}\BibitemShut {NoStop}%
\bibitem [{\citenamefont {Tokura}\ \emph {et~al.}(2017)\citenamefont {Tokura},
  \citenamefont {Kawasaki},\ and\ \citenamefont {Nagaosa}}]{Tokura2017}%
  \BibitemOpen
  \bibfield  {author} {\bibinfo {author} {\bibfnamefont {Y.}~\bibnamefont
  {Tokura}}, \bibinfo {author} {\bibfnamefont {M.}~\bibnamefont {Kawasaki}},\
  and\ \bibinfo {author} {\bibfnamefont {N.}~\bibnamefont {Nagaosa}},\
  }\bibfield  {title} {\bibinfo {title} {{Emergent functions of quantum
  materials}},\ }\href {https://doi.org/10.1038/nphys4274} {\bibfield
  {journal} {\bibinfo  {journal} {Nature Physics}\ }\textbf {\bibinfo {volume}
  {13}},\ \bibinfo {pages} {1056} (\bibinfo {year} {2017})}\BibitemShut
  {NoStop}%
\bibitem [{\citenamefont {Novelli}\ \emph {et~al.}(2013)\citenamefont
  {Novelli}, \citenamefont {Fausti}, \citenamefont {Giusti}, \citenamefont
  {Parmigiani},\ and\ \citenamefont {Hoffmann}}]{Novelli2013}%
  \BibitemOpen
  \bibfield  {author} {\bibinfo {author} {\bibfnamefont {F.}~\bibnamefont
  {Novelli}}, \bibinfo {author} {\bibfnamefont {D.}~\bibnamefont {Fausti}},
  \bibinfo {author} {\bibfnamefont {F.}~\bibnamefont {Giusti}}, \bibinfo
  {author} {\bibfnamefont {F.}~\bibnamefont {Parmigiani}},\ and\ \bibinfo
  {author} {\bibfnamefont {M.}~\bibnamefont {Hoffmann}},\ }\bibfield  {title}
  {\bibinfo {title} {{Mixed regime of light-matter interaction revealed by
  phase sensitive measurements of the dynamical Franz-Keldysh effect}},\ }\href
  {https://doi.org/10.1038/srep01227} {\bibfield  {journal} {\bibinfo
  {journal} {Scientific Reports}\ }\textbf {\bibinfo {volume} {3}},\ \bibinfo
  {pages} {1227} (\bibinfo {year} {2013})}\BibitemShut {NoStop}%
\bibitem [{\citenamefont {Schultze}\ \emph {et~al.}(2013)\citenamefont
  {Schultze}, \citenamefont {Bothschafter}, \citenamefont {Sommer},
  \citenamefont {Holzner}, \citenamefont {Schweinberger}, \citenamefont
  {Fiess}, \citenamefont {Hofstetter}, \citenamefont {Kienberger},
  \citenamefont {Apalkov}, \citenamefont {Yakovlev}, \citenamefont {Stockman},\
  and\ \citenamefont {Krausz}}]{Schultze2013}%
  \BibitemOpen
  \bibfield  {author} {\bibinfo {author} {\bibfnamefont {M.}~\bibnamefont
  {Schultze}}, \bibinfo {author} {\bibfnamefont {E.~M.}\ \bibnamefont
  {Bothschafter}}, \bibinfo {author} {\bibfnamefont {A.}~\bibnamefont
  {Sommer}}, \bibinfo {author} {\bibfnamefont {S.}~\bibnamefont {Holzner}},
  \bibinfo {author} {\bibfnamefont {W.}~\bibnamefont {Schweinberger}}, \bibinfo
  {author} {\bibfnamefont {M.}~\bibnamefont {Fiess}}, \bibinfo {author}
  {\bibfnamefont {M.}~\bibnamefont {Hofstetter}}, \bibinfo {author}
  {\bibfnamefont {R.}~\bibnamefont {Kienberger}}, \bibinfo {author}
  {\bibfnamefont {V.}~\bibnamefont {Apalkov}}, \bibinfo {author} {\bibfnamefont
  {V.~S.}\ \bibnamefont {Yakovlev}}, \bibinfo {author} {\bibfnamefont {M.~I.}\
  \bibnamefont {Stockman}},\ and\ \bibinfo {author} {\bibfnamefont
  {F.}~\bibnamefont {Krausz}},\ }\bibfield  {title} {\bibinfo {title}
  {{Controlling dielectrics with the electric field of light}},\ }\href
  {https://doi.org/10.1038/nature11720} {\bibfield  {journal} {\bibinfo
  {journal} {Nature}\ }\textbf {\bibinfo {volume} {493}},\ \bibinfo {pages}
  {75} (\bibinfo {year} {2013})}\BibitemShut {NoStop}%
\bibitem [{\citenamefont {Schultze}\ \emph {et~al.}(2014)\citenamefont
  {Schultze}, \citenamefont {Ramasesha}, \citenamefont {Pemmaraju},
  \citenamefont {Sato}, \citenamefont {Whitmore}, \citenamefont {Gandman},
  \citenamefont {Prell}, \citenamefont {Borja}, \citenamefont {Prendergast},
  \citenamefont {Yabana}, \citenamefont {Neumark},\ and\ \citenamefont
  {Leone}}]{Schultze2014}%
  \BibitemOpen
  \bibfield  {author} {\bibinfo {author} {\bibfnamefont {M.}~\bibnamefont
  {Schultze}}, \bibinfo {author} {\bibfnamefont {K.}~\bibnamefont {Ramasesha}},
  \bibinfo {author} {\bibfnamefont {C.}~\bibnamefont {Pemmaraju}}, \bibinfo
  {author} {\bibfnamefont {S.}~\bibnamefont {Sato}}, \bibinfo {author}
  {\bibfnamefont {D.}~\bibnamefont {Whitmore}}, \bibinfo {author}
  {\bibfnamefont {A.}~\bibnamefont {Gandman}}, \bibinfo {author} {\bibfnamefont
  {J.~S.}\ \bibnamefont {Prell}}, \bibinfo {author} {\bibfnamefont {L.~J.}\
  \bibnamefont {Borja}}, \bibinfo {author} {\bibfnamefont {D.}~\bibnamefont
  {Prendergast}}, \bibinfo {author} {\bibfnamefont {K.}~\bibnamefont {Yabana}},
  \bibinfo {author} {\bibfnamefont {D.~M.}\ \bibnamefont {Neumark}},\ and\
  \bibinfo {author} {\bibfnamefont {S.~R.}\ \bibnamefont {Leone}},\ }\bibfield
  {title} {\bibinfo {title} {{Attosecond band-gap dynamics in silicon}},\
  }\href {https://doi.org/10.1126/science.1260311} {\bibfield  {journal}
  {\bibinfo  {journal} {Science}\ }\textbf {\bibinfo {volume} {346}},\ \bibinfo
  {pages} {1348} (\bibinfo {year} {2014})}\BibitemShut {NoStop}%
\bibitem [{\citenamefont {Lucchini}\ \emph {et~al.}(2016)\citenamefont
  {Lucchini}, \citenamefont {Sato}, \citenamefont {Ludwig}, \citenamefont
  {Herrmann}, \citenamefont {Volkov}, \citenamefont {Kasmi}, \citenamefont
  {Shinohara}, \citenamefont {Yabana}, \citenamefont {Gallmann},\ and\
  \citenamefont {Keller}}]{Lucchini2016}%
  \BibitemOpen
  \bibfield  {author} {\bibinfo {author} {\bibfnamefont {M.}~\bibnamefont
  {Lucchini}}, \bibinfo {author} {\bibfnamefont {S.~A.}\ \bibnamefont {Sato}},
  \bibinfo {author} {\bibfnamefont {A.}~\bibnamefont {Ludwig}}, \bibinfo
  {author} {\bibfnamefont {J.}~\bibnamefont {Herrmann}}, \bibinfo {author}
  {\bibfnamefont {M.}~\bibnamefont {Volkov}}, \bibinfo {author} {\bibfnamefont
  {L.}~\bibnamefont {Kasmi}}, \bibinfo {author} {\bibfnamefont
  {Y.}~\bibnamefont {Shinohara}}, \bibinfo {author} {\bibfnamefont
  {K.}~\bibnamefont {Yabana}}, \bibinfo {author} {\bibfnamefont
  {L.}~\bibnamefont {Gallmann}},\ and\ \bibinfo {author} {\bibfnamefont
  {U.}~\bibnamefont {Keller}},\ }\bibfield  {title} {\bibinfo {title}
  {{Attosecond dynamical Franz-Keldysh effect in polycrystalline diamond}},\
  }\href {https://doi.org/10.1126/science.aag1268} {\bibfield  {journal}
  {\bibinfo  {journal} {Science}\ }\textbf {\bibinfo {volume} {353}},\ \bibinfo
  {pages} {916} (\bibinfo {year} {2016})}\BibitemShut {NoStop}%
\bibitem [{\citenamefont {Schlaepfer}\ \emph {et~al.}(2018)\citenamefont
  {Schlaepfer}, \citenamefont {Lucchini}, \citenamefont {Sato}, \citenamefont
  {Volkov}, \citenamefont {Kasmi}, \citenamefont {Hartmann}, \citenamefont
  {Rubio}, \citenamefont {Gallmann},\ and\ \citenamefont
  {Keller}}]{Schlaepfer2018}%
  \BibitemOpen
  \bibfield  {author} {\bibinfo {author} {\bibfnamefont {F.}~\bibnamefont
  {Schlaepfer}}, \bibinfo {author} {\bibfnamefont {M.}~\bibnamefont
  {Lucchini}}, \bibinfo {author} {\bibfnamefont {S.~A.}\ \bibnamefont {Sato}},
  \bibinfo {author} {\bibfnamefont {M.}~\bibnamefont {Volkov}}, \bibinfo
  {author} {\bibfnamefont {L.}~\bibnamefont {Kasmi}}, \bibinfo {author}
  {\bibfnamefont {N.}~\bibnamefont {Hartmann}}, \bibinfo {author}
  {\bibfnamefont {A.}~\bibnamefont {Rubio}}, \bibinfo {author} {\bibfnamefont
  {L.}~\bibnamefont {Gallmann}},\ and\ \bibinfo {author} {\bibfnamefont
  {U.}~\bibnamefont {Keller}},\ }\bibfield  {title} {\bibinfo {title}
  {{Attosecond optical-field-enhanced carrier injection into the GaAs
  conduction band}},\ }\href {https://doi.org/10.1038/s41567-018-0069-0}
  {\bibfield  {journal} {\bibinfo  {journal} {Nature Physics}\ }\textbf
  {\bibinfo {volume} {14}},\ \bibinfo {pages} {560} (\bibinfo {year}
  {2018})}\BibitemShut {NoStop}%
\bibitem [{\citenamefont {Gr{\aa}n{\"{a}}s}\ \emph {et~al.}(2020)\citenamefont
  {Gr{\aa}n{\"{a}}s}, \citenamefont {Vaskivskyi}, \citenamefont
  {Thunstr{\"{o}}m}, \citenamefont {Ghimire}, \citenamefont {Knut},
  \citenamefont {S{\"{o}}derstr{\"{o}}m}, \citenamefont {Kjellsson},
  \citenamefont {Turenne}, \citenamefont {Engel}, \citenamefont {Beye},
  \citenamefont {Lu}, \citenamefont {Reid}, \citenamefont {Schlotter},
  \citenamefont {Coslovich}, \citenamefont {Hoffmann}, \citenamefont {Kolesov},
  \citenamefont {Sch{\"{u}}{\ss}ler-Langeheine}, \citenamefont {Styervoyedov},
  \citenamefont {Tancogne-Dejean}, \citenamefont {Sentef}, \citenamefont
  {Reis}, \citenamefont {Rubio}, \citenamefont {Parkin}, \citenamefont {Karis},
  \citenamefont {Nordgren}, \citenamefont {Rubensson}, \citenamefont
  {Eriksson},\ and\ \citenamefont {D{\"{u}}rr}}]{Granas2020}%
  \BibitemOpen
  \bibfield  {author} {\bibinfo {author} {\bibfnamefont {O.}~\bibnamefont
  {Gr{\aa}n{\"{a}}s}}, \bibinfo {author} {\bibfnamefont {I.}~\bibnamefont
  {Vaskivskyi}}, \bibinfo {author} {\bibfnamefont {P.}~\bibnamefont
  {Thunstr{\"{o}}m}}, \bibinfo {author} {\bibfnamefont {S.}~\bibnamefont
  {Ghimire}}, \bibinfo {author} {\bibfnamefont {R.}~\bibnamefont {Knut}},
  \bibinfo {author} {\bibfnamefont {J.}~\bibnamefont {S{\"{o}}derstr{\"{o}}m}},
  \bibinfo {author} {\bibfnamefont {L.}~\bibnamefont {Kjellsson}}, \bibinfo
  {author} {\bibfnamefont {D.}~\bibnamefont {Turenne}}, \bibinfo {author}
  {\bibfnamefont {R.~Y.}\ \bibnamefont {Engel}}, \bibinfo {author}
  {\bibfnamefont {M.}~\bibnamefont {Beye}}, \bibinfo {author} {\bibfnamefont
  {J.}~\bibnamefont {Lu}}, \bibinfo {author} {\bibfnamefont {A.~H.}\
  \bibnamefont {Reid}}, \bibinfo {author} {\bibfnamefont {W.}~\bibnamefont
  {Schlotter}}, \bibinfo {author} {\bibfnamefont {G.}~\bibnamefont
  {Coslovich}}, \bibinfo {author} {\bibfnamefont {M.}~\bibnamefont {Hoffmann}},
  \bibinfo {author} {\bibfnamefont {G.}~\bibnamefont {Kolesov}}, \bibinfo
  {author} {\bibfnamefont {C.}~\bibnamefont {Sch{\"{u}}{\ss}ler-Langeheine}},
  \bibinfo {author} {\bibfnamefont {A.}~\bibnamefont {Styervoyedov}}, \bibinfo
  {author} {\bibfnamefont {N.}~\bibnamefont {Tancogne-Dejean}}, \bibinfo
  {author} {\bibfnamefont {M.~A.}\ \bibnamefont {Sentef}}, \bibinfo {author}
  {\bibfnamefont {D.~A.}\ \bibnamefont {Reis}}, \bibinfo {author}
  {\bibfnamefont {A.}~\bibnamefont {Rubio}}, \bibinfo {author} {\bibfnamefont
  {S.~S.}\ \bibnamefont {Parkin}}, \bibinfo {author} {\bibfnamefont
  {O.}~\bibnamefont {Karis}}, \bibinfo {author} {\bibfnamefont
  {J.}~\bibnamefont {Nordgren}}, \bibinfo {author} {\bibfnamefont {J.~E.}\
  \bibnamefont {Rubensson}}, \bibinfo {author} {\bibfnamefont {O.}~\bibnamefont
  {Eriksson}},\ and\ \bibinfo {author} {\bibfnamefont {H.~A.}\ \bibnamefont
  {D{\"{u}}rr}},\ }\bibfield  {title} {\bibinfo {title} {{Ultrafast
  modification of the electronic structure of a correlated insulator}},\
  }\href@noop {} {\bibfield  {journal} {\bibinfo  {journal} {arXiv}\ }\textbf
  {\bibinfo {volume} {2008.11115}},\ \bibinfo {pages} {1} (\bibinfo {year}
  {2020})},\ \Eprint {https://arxiv.org/abs/2008.11115} {arXiv:2008.11115}
  \BibitemShut {NoStop}%
\bibitem [{\citenamefont {F{\"{o}}rst}\ \emph {et~al.}(2011)\citenamefont
  {F{\"{o}}rst}, \citenamefont {Manzoni}, \citenamefont {Kaiser}, \citenamefont
  {Tomioka}, \citenamefont {Tokura}, \citenamefont {Merlin},\ and\
  \citenamefont {Cavalleri}}]{Forst2011}%
  \BibitemOpen
  \bibfield  {author} {\bibinfo {author} {\bibfnamefont {M.}~\bibnamefont
  {F{\"{o}}rst}}, \bibinfo {author} {\bibfnamefont {C.}~\bibnamefont
  {Manzoni}}, \bibinfo {author} {\bibfnamefont {S.}~\bibnamefont {Kaiser}},
  \bibinfo {author} {\bibfnamefont {Y.}~\bibnamefont {Tomioka}}, \bibinfo
  {author} {\bibfnamefont {Y.}~\bibnamefont {Tokura}}, \bibinfo {author}
  {\bibfnamefont {R.}~\bibnamefont {Merlin}},\ and\ \bibinfo {author}
  {\bibfnamefont {A.}~\bibnamefont {Cavalleri}},\ }\bibfield  {title} {\bibinfo
  {title} {{Nonlinear phononics as an ultrafast route to lattice control}},\
  }\href {https://doi.org/10.1038/nphys2055} {\bibfield  {journal} {\bibinfo
  {journal} {Nature Physics}\ }\textbf {\bibinfo {volume} {7}},\ \bibinfo
  {pages} {854} (\bibinfo {year} {2011})}\BibitemShut {NoStop}%
\bibitem [{\citenamefont {Subedi}\ \emph {et~al.}(2014)\citenamefont {Subedi},
  \citenamefont {Cavalleri},\ and\ \citenamefont {Georges}}]{Subedi2014}%
  \BibitemOpen
  \bibfield  {author} {\bibinfo {author} {\bibfnamefont {A.}~\bibnamefont
  {Subedi}}, \bibinfo {author} {\bibfnamefont {A.}~\bibnamefont {Cavalleri}},\
  and\ \bibinfo {author} {\bibfnamefont {A.}~\bibnamefont {Georges}},\
  }\bibfield  {title} {\bibinfo {title} {{Theory of nonlinear phononics for
  coherent light control of solids}},\ }\href
  {https://doi.org/10.1103/PhysRevB.89.220301} {\bibfield  {journal} {\bibinfo
  {journal} {Phys. Rev. B}\ }\textbf {\bibinfo {volume} {89}},\ \bibinfo
  {pages} {220301} (\bibinfo {year} {2014})}\BibitemShut {NoStop}%
\bibitem [{\citenamefont {Mankowsky}\ \emph {et~al.}(2014)\citenamefont
  {Mankowsky}, \citenamefont {Subedi}, \citenamefont {F{\"o}rst}, \citenamefont
  {Mariager}, \citenamefont {Chollet}, \citenamefont {Lemke}, \citenamefont
  {Robinson}, \citenamefont {Glownia}, \citenamefont {Minitti}, \citenamefont
  {Frano}, \citenamefont {Fechner}, \citenamefont {Spaldin}, \citenamefont
  {Loew}, \citenamefont {Keimer}, \citenamefont {Georges},\ and\ \citenamefont
  {Cavalleri}}]{Mankowsky2014}%
  \BibitemOpen
  \bibfield  {author} {\bibinfo {author} {\bibfnamefont {R.}~\bibnamefont
  {Mankowsky}}, \bibinfo {author} {\bibfnamefont {A.}~\bibnamefont {Subedi}},
  \bibinfo {author} {\bibfnamefont {M.}~\bibnamefont {F{\"o}rst}}, \bibinfo
  {author} {\bibfnamefont {S.~O.}\ \bibnamefont {Mariager}}, \bibinfo {author}
  {\bibfnamefont {M.}~\bibnamefont {Chollet}}, \bibinfo {author} {\bibfnamefont
  {H.~T.}\ \bibnamefont {Lemke}}, \bibinfo {author} {\bibfnamefont {J.~S.}\
  \bibnamefont {Robinson}}, \bibinfo {author} {\bibfnamefont {J.~M.}\
  \bibnamefont {Glownia}}, \bibinfo {author} {\bibfnamefont {M.~P.}\
  \bibnamefont {Minitti}}, \bibinfo {author} {\bibfnamefont {A.}~\bibnamefont
  {Frano}}, \bibinfo {author} {\bibfnamefont {M.}~\bibnamefont {Fechner}},
  \bibinfo {author} {\bibfnamefont {N.~A.}\ \bibnamefont {Spaldin}}, \bibinfo
  {author} {\bibfnamefont {T.}~\bibnamefont {Loew}}, \bibinfo {author}
  {\bibfnamefont {B.}~\bibnamefont {Keimer}}, \bibinfo {author} {\bibfnamefont
  {A.}~\bibnamefont {Georges}},\ and\ \bibinfo {author} {\bibfnamefont
  {A.}~\bibnamefont {Cavalleri}},\ }\bibfield  {title} {\bibinfo {title}
  {{Nonlinear lattice dynamics as a basis for enhanced superconductivity in
  YBa$_2$Cu$_3$O$_{6.5}$}},\ }\href {https://doi.org/10.1038/nature13875}
  {\bibfield  {journal} {\bibinfo  {journal} {Nature}\ }\textbf {\bibinfo
  {volume} {516}},\ \bibinfo {pages} {71} (\bibinfo {year} {2014})}\BibitemShut
  {NoStop}%
\bibitem [{\citenamefont {Wang}\ \emph {et~al.}(2013)\citenamefont {Wang},
  \citenamefont {Steinberg}, \citenamefont {Jarillo-Herrero},\ and\
  \citenamefont {Gedik}}]{Wang2013}%
  \BibitemOpen
  \bibfield  {author} {\bibinfo {author} {\bibfnamefont {Y.~H.}\ \bibnamefont
  {Wang}}, \bibinfo {author} {\bibfnamefont {H.}~\bibnamefont {Steinberg}},
  \bibinfo {author} {\bibfnamefont {P.}~\bibnamefont {Jarillo-Herrero}},\ and\
  \bibinfo {author} {\bibfnamefont {N.}~\bibnamefont {Gedik}},\ }\bibfield
  {title} {\bibinfo {title} {{Observation of Floquet-Bloch States on the
  Surface of a Topological Insulator}},\ }\href
  {https://doi.org/10.1126/science.1239834} {\bibfield  {journal} {\bibinfo
  {journal} {Science}\ }\textbf {\bibinfo {volume} {342}},\ \bibinfo {pages}
  {453} (\bibinfo {year} {2013})}\BibitemShut {NoStop}%
\bibitem [{\citenamefont {Mahmood}\ \emph {et~al.}(2016)\citenamefont
  {Mahmood}, \citenamefont {Chan}, \citenamefont {Alpichshev}, \citenamefont
  {Gardner}, \citenamefont {Lee}, \citenamefont {Lee},\ and\ \citenamefont
  {Gedik}}]{Mahmood2016selective}%
  \BibitemOpen
  \bibfield  {author} {\bibinfo {author} {\bibfnamefont {F.}~\bibnamefont
  {Mahmood}}, \bibinfo {author} {\bibfnamefont {C.-K.}\ \bibnamefont {Chan}},
  \bibinfo {author} {\bibfnamefont {Z.}~\bibnamefont {Alpichshev}}, \bibinfo
  {author} {\bibfnamefont {D.}~\bibnamefont {Gardner}}, \bibinfo {author}
  {\bibfnamefont {Y.}~\bibnamefont {Lee}}, \bibinfo {author} {\bibfnamefont
  {P.~A.}\ \bibnamefont {Lee}},\ and\ \bibinfo {author} {\bibfnamefont
  {N.}~\bibnamefont {Gedik}},\ }\bibfield  {title} {\bibinfo {title}
  {{Selective scattering between Floquet{\textendash}Bloch and Volkov states in
  a topological insulator}},\ }\href {https://doi.org/10.1038/nphys3609}
  {\bibfield  {journal} {\bibinfo  {journal} {Nature Physics}\ }\textbf
  {\bibinfo {volume} {12}},\ \bibinfo {pages} {306} (\bibinfo {year}
  {2016})}\BibitemShut {NoStop}%
\bibitem [{\citenamefont {McIver}\ \emph {et~al.}(2020)\citenamefont {McIver},
  \citenamefont {Schulte}, \citenamefont {Stein}, \citenamefont {Matsuyama},
  \citenamefont {Jotzu}, \citenamefont {Meier},\ and\ \citenamefont
  {Cavalleri}}]{McIver2020}%
  \BibitemOpen
  \bibfield  {author} {\bibinfo {author} {\bibfnamefont {J.~W.}\ \bibnamefont
  {McIver}}, \bibinfo {author} {\bibfnamefont {B.}~\bibnamefont {Schulte}},
  \bibinfo {author} {\bibfnamefont {F.-U.}\ \bibnamefont {Stein}}, \bibinfo
  {author} {\bibfnamefont {T.}~\bibnamefont {Matsuyama}}, \bibinfo {author}
  {\bibfnamefont {G.}~\bibnamefont {Jotzu}}, \bibinfo {author} {\bibfnamefont
  {G.}~\bibnamefont {Meier}},\ and\ \bibinfo {author} {\bibfnamefont
  {A.}~\bibnamefont {Cavalleri}},\ }\bibfield  {title} {\bibinfo {title}
  {{Light-induced anomalous Hall effect in graphene}},\ }\href
  {https://doi.org/10.1038/s41567-019-0698-y} {\bibfield  {journal} {\bibinfo
  {journal} {Nature Physics}\ }\textbf {\bibinfo {volume} {16}},\ \bibinfo
  {pages} {38} (\bibinfo {year} {2020})}\BibitemShut {NoStop}%
\bibitem [{\citenamefont {Kaiser}\ \emph
  {et~al.}(2014{\natexlab{a}})\citenamefont {Kaiser}, \citenamefont {Clark},
  \citenamefont {Nicoletti}, \citenamefont {Cotugno}, \citenamefont {Tobey},
  \citenamefont {Dean}, \citenamefont {Lupi}, \citenamefont {Okamoto},
  \citenamefont {Hasegawa}, \citenamefont {Jaksch},\ and\ \citenamefont
  {Cavalleri}}]{Kaiser2014}%
  \BibitemOpen
  \bibfield  {author} {\bibinfo {author} {\bibfnamefont {S.}~\bibnamefont
  {Kaiser}}, \bibinfo {author} {\bibfnamefont {S.~R.}\ \bibnamefont {Clark}},
  \bibinfo {author} {\bibfnamefont {D.}~\bibnamefont {Nicoletti}}, \bibinfo
  {author} {\bibfnamefont {G.}~\bibnamefont {Cotugno}}, \bibinfo {author}
  {\bibfnamefont {R.~I.}\ \bibnamefont {Tobey}}, \bibinfo {author}
  {\bibfnamefont {N.}~\bibnamefont {Dean}}, \bibinfo {author} {\bibfnamefont
  {S.}~\bibnamefont {Lupi}}, \bibinfo {author} {\bibfnamefont {H.}~\bibnamefont
  {Okamoto}}, \bibinfo {author} {\bibfnamefont {T.}~\bibnamefont {Hasegawa}},
  \bibinfo {author} {\bibfnamefont {D.}~\bibnamefont {Jaksch}},\ and\ \bibinfo
  {author} {\bibfnamefont {A.}~\bibnamefont {Cavalleri}},\ }\bibfield  {title}
  {\bibinfo {title} {{Optical Properties of a Vibrationally Modulated Solid
  State Mott Insulator}},\ }\href {https://doi.org/10.1038/srep03823}
  {\bibfield  {journal} {\bibinfo  {journal} {Scientific Reports}\ }\textbf
  {\bibinfo {volume} {4}},\ \bibinfo {pages} {3823} (\bibinfo {year}
  {2014}{\natexlab{a}})}\BibitemShut {NoStop}%
\bibitem [{\citenamefont {Singla}\ \emph {et~al.}(2015)\citenamefont {Singla},
  \citenamefont {Cotugno}, \citenamefont {Kaiser}, \citenamefont {F{\"{o}}rst},
  \citenamefont {Mitrano}, \citenamefont {Liu}, \citenamefont {Cartella},
  \citenamefont {Manzoni}, \citenamefont {Okamoto}, \citenamefont {Hasegawa},
  \citenamefont {Clark}, \citenamefont {Jaksch},\ and\ \citenamefont
  {Cavalleri}}]{Singla2015}%
  \BibitemOpen
  \bibfield  {author} {\bibinfo {author} {\bibfnamefont {R.}~\bibnamefont
  {Singla}}, \bibinfo {author} {\bibfnamefont {G.}~\bibnamefont {Cotugno}},
  \bibinfo {author} {\bibfnamefont {S.}~\bibnamefont {Kaiser}}, \bibinfo
  {author} {\bibfnamefont {M.}~\bibnamefont {F{\"{o}}rst}}, \bibinfo {author}
  {\bibfnamefont {M.}~\bibnamefont {Mitrano}}, \bibinfo {author} {\bibfnamefont
  {H.~Y.}\ \bibnamefont {Liu}}, \bibinfo {author} {\bibfnamefont
  {A.}~\bibnamefont {Cartella}}, \bibinfo {author} {\bibfnamefont
  {C.}~\bibnamefont {Manzoni}}, \bibinfo {author} {\bibfnamefont
  {H.}~\bibnamefont {Okamoto}}, \bibinfo {author} {\bibfnamefont
  {T.}~\bibnamefont {Hasegawa}}, \bibinfo {author} {\bibfnamefont {S.~R.}\
  \bibnamefont {Clark}}, \bibinfo {author} {\bibfnamefont {D.}~\bibnamefont
  {Jaksch}},\ and\ \bibinfo {author} {\bibfnamefont {A.}~\bibnamefont
  {Cavalleri}},\ }\bibfield  {title} {\bibinfo {title} {{THz-Frequency
  Modulation of the Hubbard U in an Organic Mott Insulator}},\ }\href
  {https://doi.org/10.1103/PhysRevLett.115.187401} {\bibfield  {journal}
  {\bibinfo  {journal} {Physical Review Letters}\ }\textbf {\bibinfo {volume}
  {115}},\ \bibinfo {pages} {187401} (\bibinfo {year} {2015})}\BibitemShut
  {NoStop}%
\bibitem [{\citenamefont {Buzzi}\ \emph {et~al.}(2020)\citenamefont {Buzzi},
  \citenamefont {Nicoletti}, \citenamefont {Fechner}, \citenamefont
  {Tancogne-Dejean}, \citenamefont {Sentef}, \citenamefont {Georges},
  \citenamefont {Biesner}, \citenamefont {Uykur}, \citenamefont {Dressel},
  \citenamefont {Henderson}, \citenamefont {Siegrist}, \citenamefont
  {Schlueter}, \citenamefont {Miyagawa}, \citenamefont {Kanoda}, \citenamefont
  {Nam}, \citenamefont {Ardavan}, \citenamefont {Coulthard}, \citenamefont
  {Tindall}, \citenamefont {Schlawin}, \citenamefont {Jaksch},\ and\
  \citenamefont {Cavalleri}}]{Buzzi2020}%
  \BibitemOpen
  \bibfield  {author} {\bibinfo {author} {\bibfnamefont {M.}~\bibnamefont
  {Buzzi}}, \bibinfo {author} {\bibfnamefont {D.}~\bibnamefont {Nicoletti}},
  \bibinfo {author} {\bibfnamefont {M.}~\bibnamefont {Fechner}}, \bibinfo
  {author} {\bibfnamefont {N.}~\bibnamefont {Tancogne-Dejean}}, \bibinfo
  {author} {\bibfnamefont {M.~A.}\ \bibnamefont {Sentef}}, \bibinfo {author}
  {\bibfnamefont {A.}~\bibnamefont {Georges}}, \bibinfo {author} {\bibfnamefont
  {T.}~\bibnamefont {Biesner}}, \bibinfo {author} {\bibfnamefont
  {E.}~\bibnamefont {Uykur}}, \bibinfo {author} {\bibfnamefont
  {M.}~\bibnamefont {Dressel}}, \bibinfo {author} {\bibfnamefont
  {A.}~\bibnamefont {Henderson}}, \bibinfo {author} {\bibfnamefont
  {T.}~\bibnamefont {Siegrist}}, \bibinfo {author} {\bibfnamefont {J.~A.}\
  \bibnamefont {Schlueter}}, \bibinfo {author} {\bibfnamefont {K.}~\bibnamefont
  {Miyagawa}}, \bibinfo {author} {\bibfnamefont {K.}~\bibnamefont {Kanoda}},
  \bibinfo {author} {\bibfnamefont {M.-S.}\ \bibnamefont {Nam}}, \bibinfo
  {author} {\bibfnamefont {A.}~\bibnamefont {Ardavan}}, \bibinfo {author}
  {\bibfnamefont {J.}~\bibnamefont {Coulthard}}, \bibinfo {author}
  {\bibfnamefont {J.}~\bibnamefont {Tindall}}, \bibinfo {author} {\bibfnamefont
  {F.}~\bibnamefont {Schlawin}}, \bibinfo {author} {\bibfnamefont
  {D.}~\bibnamefont {Jaksch}},\ and\ \bibinfo {author} {\bibfnamefont
  {A.}~\bibnamefont {Cavalleri}},\ }\bibfield  {title} {\bibinfo {title}
  {Photomolecular high-temperature superconductivity},\ }\href
  {https://doi.org/10.1103/PhysRevX.10.031028} {\bibfield  {journal} {\bibinfo
  {journal} {Phys. Rev. X}\ }\textbf {\bibinfo {volume} {10}},\ \bibinfo
  {pages} {031028} (\bibinfo {year} {2020})}\BibitemShut {NoStop}%
\bibitem [{\citenamefont {Kawakami}\ \emph {et~al.}(2009)\citenamefont
  {Kawakami}, \citenamefont {Iwai}, \citenamefont {Fukatsu}, \citenamefont
  {Miura}, \citenamefont {Yoneyama}, \citenamefont {Sasaki},\ and\
  \citenamefont {Kobayashi}}]{Kawakami2009}%
  \BibitemOpen
  \bibfield  {author} {\bibinfo {author} {\bibfnamefont {Y.}~\bibnamefont
  {Kawakami}}, \bibinfo {author} {\bibfnamefont {S.}~\bibnamefont {Iwai}},
  \bibinfo {author} {\bibfnamefont {T.}~\bibnamefont {Fukatsu}}, \bibinfo
  {author} {\bibfnamefont {M.}~\bibnamefont {Miura}}, \bibinfo {author}
  {\bibfnamefont {N.}~\bibnamefont {Yoneyama}}, \bibinfo {author}
  {\bibfnamefont {T.}~\bibnamefont {Sasaki}},\ and\ \bibinfo {author}
  {\bibfnamefont {N.}~\bibnamefont {Kobayashi}},\ }\bibfield  {title} {\bibinfo
  {title} {{Optical Modulation of Effective On-Site Coulomb Energy for the Mott
  Transition in an Organic Dimer Insulator}},\ }\href
  {https://doi.org/10.1103/PhysRevLett.103.066403} {\bibfield  {journal}
  {\bibinfo  {journal} {Phys. Rev. Lett.}\ }\textbf {\bibinfo {volume} {103}},\
  \bibinfo {pages} {66403} (\bibinfo {year} {2009})}\BibitemShut {NoStop}%
\bibitem [{\citenamefont {Tancogne-Dejean}\ \emph {et~al.}(2018)\citenamefont
  {Tancogne-Dejean}, \citenamefont {Sentef},\ and\ \citenamefont
  {Rubio}}]{Tancogne-Dejean2018}%
  \BibitemOpen
  \bibfield  {author} {\bibinfo {author} {\bibfnamefont {N.}~\bibnamefont
  {Tancogne-Dejean}}, \bibinfo {author} {\bibfnamefont {M.~A.}\ \bibnamefont
  {Sentef}},\ and\ \bibinfo {author} {\bibfnamefont {A.}~\bibnamefont
  {Rubio}},\ }\bibfield  {title} {\bibinfo {title} {{Ultrafast Modification of
  Hubbard U in a Strongly Correlated Material : Ab initio High-Harmonic
  Generation in NiO}},\ }\href {https://doi.org/10.1103/PhysRevLett.121.097402}
  {\bibfield  {journal} {\bibinfo  {journal} {Physical Review Letters}\
  }\textbf {\bibinfo {volume} {121}},\ \bibinfo {pages} {97402} (\bibinfo
  {year} {2018})}\BibitemShut {NoStop}%
\bibitem [{\citenamefont {Topp}\ \emph {et~al.}(2018)\citenamefont {Topp},
  \citenamefont {Tancogne-Dejean}, \citenamefont {Kemper}, \citenamefont
  {Rubio},\ and\ \citenamefont {Sentef}}]{Topp2018}%
  \BibitemOpen
  \bibfield  {author} {\bibinfo {author} {\bibfnamefont {G.~E.}\ \bibnamefont
  {Topp}}, \bibinfo {author} {\bibfnamefont {N.}~\bibnamefont
  {Tancogne-Dejean}}, \bibinfo {author} {\bibfnamefont {A.~F.}\ \bibnamefont
  {Kemper}}, \bibinfo {author} {\bibfnamefont {A.}~\bibnamefont {Rubio}},\ and\
  \bibinfo {author} {\bibfnamefont {M.~A.}\ \bibnamefont {Sentef}},\ }\bibfield
   {title} {\bibinfo {title} {{All-optical nonequilibrium pathway to
  stabilising magnetic Weyl semimetals in pyrochlore iridates}},\ }\href
  {https://doi.org/10.1038/s41467-018-06991-8} {\bibfield  {journal} {\bibinfo
  {journal} {Nature Communications}\ }\textbf {\bibinfo {volume} {9}},\
  \bibinfo {pages} {4452} (\bibinfo {year} {2018})}\BibitemShut {NoStop}%
\bibitem [{\citenamefont {Gole\ifmmode~\check{z}\else \v{z}\fi{}}\ \emph
  {et~al.}(2019)\citenamefont {Gole\ifmmode~\check{z}\else \v{z}\fi{}},
  \citenamefont {Eckstein},\ and\ \citenamefont {Werner}}]{Golez2019}%
  \BibitemOpen
  \bibfield  {author} {\bibinfo {author} {\bibfnamefont {D.}~\bibnamefont
  {Gole\ifmmode~\check{z}\else \v{z}\fi{}}}, \bibinfo {author} {\bibfnamefont
  {M.}~\bibnamefont {Eckstein}},\ and\ \bibinfo {author} {\bibfnamefont
  {P.}~\bibnamefont {Werner}},\ }\bibfield  {title} {\bibinfo {title}
  {Multiband nonequilibrium {$GW+\text{EDMFT}$} formalism for correlated
  insulators},\ }\href {https://doi.org/10.1103/PhysRevB.100.235117} {\bibfield
   {journal} {\bibinfo  {journal} {Phys. Rev. B}\ }\textbf {\bibinfo {volume}
  {100}},\ \bibinfo {pages} {235117} (\bibinfo {year} {2019})}\BibitemShut
  {NoStop}%
\bibitem [{\citenamefont {Tancogne-Dejean}\ \emph {et~al.}(2020)\citenamefont
  {Tancogne-Dejean}, \citenamefont {Sentef},\ and\ \citenamefont
  {Rubio}}]{Tancogne-Dejean2020}%
  \BibitemOpen
  \bibfield  {author} {\bibinfo {author} {\bibfnamefont {N.}~\bibnamefont
  {Tancogne-Dejean}}, \bibinfo {author} {\bibfnamefont {M.~A.}\ \bibnamefont
  {Sentef}},\ and\ \bibinfo {author} {\bibfnamefont {A.}~\bibnamefont
  {Rubio}},\ }\bibfield  {title} {\bibinfo {title} {{Ultrafast transient
  absorption spectroscopy of the charge-transfer insulator NiO: Beyond the
  dynamical Franz-Keldysh effect}},\ }\href
  {https://doi.org/10.1103/PhysRevB.102.115106} {\bibfield  {journal} {\bibinfo
   {journal} {Physical Review B}\ }\textbf {\bibinfo {volume} {102}},\ \bibinfo
  {pages} {115106} (\bibinfo {year} {2020})}\BibitemShut {NoStop}%
\bibitem [{\citenamefont {Beaulieu}\ \emph {et~al.}(2021)\citenamefont
  {Beaulieu}, \citenamefont {Dong}, \citenamefont {Tancogne-Dejean},
  \citenamefont {Dendzik}, \citenamefont {Pincelli}, \citenamefont {Maklar},
  \citenamefont {Xian}, \citenamefont {Sentef}, \citenamefont {Wolf},
  \citenamefont {Rubio}, \citenamefont {Rettig},\ and\ \citenamefont
  {Ernstorfer}}]{Beaulieu2020}%
  \BibitemOpen
  \bibfield  {author} {\bibinfo {author} {\bibfnamefont {S.}~\bibnamefont
  {Beaulieu}}, \bibinfo {author} {\bibfnamefont {S.}~\bibnamefont {Dong}},
  \bibinfo {author} {\bibfnamefont {N.}~\bibnamefont {Tancogne-Dejean}},
  \bibinfo {author} {\bibfnamefont {M.}~\bibnamefont {Dendzik}}, \bibinfo
  {author} {\bibfnamefont {T.}~\bibnamefont {Pincelli}}, \bibinfo {author}
  {\bibfnamefont {J.}~\bibnamefont {Maklar}}, \bibinfo {author} {\bibfnamefont
  {R.~P.}\ \bibnamefont {Xian}}, \bibinfo {author} {\bibfnamefont {M.~A.}\
  \bibnamefont {Sentef}}, \bibinfo {author} {\bibfnamefont {M.}~\bibnamefont
  {Wolf}}, \bibinfo {author} {\bibfnamefont {A.}~\bibnamefont {Rubio}},
  \bibinfo {author} {\bibfnamefont {L.}~\bibnamefont {Rettig}},\ and\ \bibinfo
  {author} {\bibfnamefont {R.}~\bibnamefont {Ernstorfer}},\ }\bibfield  {title}
  {\bibinfo {title} {{Ultrafast dynamical Lifshitz transition}},\ }\href
  {https://doi.org/10.1126/sciadv.abd9275} {\bibfield  {journal} {\bibinfo
  {journal} {Science Advances}\ }\textbf {\bibinfo {volume} {7}},\ \bibinfo
  {pages} {eabd9275} (\bibinfo {year} {2021})}\BibitemShut {NoStop}%
\bibitem [{\citenamefont {Valmispild}\ \emph {et~al.}(2020)\citenamefont
  {Valmispild}, \citenamefont {Dutreix}, \citenamefont {Eckstein},
  \citenamefont {Katsnelson}, \citenamefont {Lichtenstein},\ and\ \citenamefont
  {Stepanov}}]{Valmispild2020dynamically}%
  \BibitemOpen
  \bibfield  {author} {\bibinfo {author} {\bibfnamefont {V.~N.}\ \bibnamefont
  {Valmispild}}, \bibinfo {author} {\bibfnamefont {C.}~\bibnamefont {Dutreix}},
  \bibinfo {author} {\bibfnamefont {M.}~\bibnamefont {Eckstein}}, \bibinfo
  {author} {\bibfnamefont {M.~I.}\ \bibnamefont {Katsnelson}}, \bibinfo
  {author} {\bibfnamefont {A.~I.}\ \bibnamefont {Lichtenstein}},\ and\ \bibinfo
  {author} {\bibfnamefont {E.~A.}\ \bibnamefont {Stepanov}},\ }\bibfield
  {title} {\bibinfo {title} {{Dynamically induced doublon repulsion in the
  Fermi-Hubbard model probed by a single-particle density of states}},\ }\href
  {https://doi.org/10.1103/PhysRevB.102.220301} {\bibfield  {journal} {\bibinfo
   {journal} {Phys. Rev. B}\ }\textbf {\bibinfo {volume} {102}},\ \bibinfo
  {pages} {220301} (\bibinfo {year} {2020})}\BibitemShut {NoStop}%
\bibitem [{\citenamefont {Zhang}\ and\ \citenamefont
  {Averitt}(2014)}]{Zhang2014dynamics}%
  \BibitemOpen
  \bibfield  {author} {\bibinfo {author} {\bibfnamefont {J.}~\bibnamefont
  {Zhang}}\ and\ \bibinfo {author} {\bibfnamefont {R.}~\bibnamefont
  {Averitt}},\ }\bibfield  {title} {\bibinfo {title} {Dynamics and control in
  complex transition metal oxides},\ }\href
  {https://doi.org/10.1146/annurev-matsci-070813-113258} {\bibfield  {journal}
  {\bibinfo  {journal} {Annual Review of Materials Research}\ }\textbf
  {\bibinfo {volume} {44}},\ \bibinfo {pages} {19} (\bibinfo {year}
  {2014})}\BibitemShut {NoStop}%
\bibitem [{\citenamefont {Ahn}\ \emph {et~al.}(2021)\citenamefont {Ahn},
  \citenamefont {Cavalleri}, \citenamefont {Georges}, \citenamefont
  {Ismail-Beigi}, \citenamefont {Millis},\ and\ \citenamefont
  {Triscone}}]{Ahn2021designing}%
  \BibitemOpen
  \bibfield  {author} {\bibinfo {author} {\bibfnamefont {C.}~\bibnamefont
  {Ahn}}, \bibinfo {author} {\bibfnamefont {A.}~\bibnamefont {Cavalleri}},
  \bibinfo {author} {\bibfnamefont {A.}~\bibnamefont {Georges}}, \bibinfo
  {author} {\bibfnamefont {S.}~\bibnamefont {Ismail-Beigi}}, \bibinfo {author}
  {\bibfnamefont {A.~J.}\ \bibnamefont {Millis}},\ and\ \bibinfo {author}
  {\bibfnamefont {J.-M.}\ \bibnamefont {Triscone}},\ }\bibfield  {title}
  {\bibinfo {title} {Designing and controlling the properties of transition
  metal oxide quantum materials},\ }\bibfield  {journal} {\bibinfo  {journal}
  {Nature Materials}\ }\href {https://doi.org/10.1038/s41563-021-00989-2}
  {10.1038/s41563-021-00989-2} (\bibinfo {year} {2021})\BibitemShut {NoStop}%
\bibitem [{\citenamefont {Lee}\ \emph {et~al.}(2006)\citenamefont {Lee},
  \citenamefont {Nagaosa},\ and\ \citenamefont {Wen}}]{Lee2006doping}%
  \BibitemOpen
  \bibfield  {author} {\bibinfo {author} {\bibfnamefont {P.~A.}\ \bibnamefont
  {Lee}}, \bibinfo {author} {\bibfnamefont {N.}~\bibnamefont {Nagaosa}},\ and\
  \bibinfo {author} {\bibfnamefont {X.-G.}\ \bibnamefont {Wen}},\ }\bibfield
  {title} {\bibinfo {title} {{Doping a Mott insulator: Physics of
  high-temperature superconductivity}},\ }\href
  {https://doi.org/10.1103/RevModPhys.78.17} {\bibfield  {journal} {\bibinfo
  {journal} {Rev. Mod. Phys.}\ }\textbf {\bibinfo {volume} {78}},\ \bibinfo
  {pages} {17} (\bibinfo {year} {2006})}\BibitemShut {NoStop}%
\bibitem [{\citenamefont {Keimer}\ \emph {et~al.}(2015)\citenamefont {Keimer},
  \citenamefont {Kivelson}, \citenamefont {Norman}, \citenamefont {Uchida},\
  and\ \citenamefont {Zaanen}}]{Keimer2015}%
  \BibitemOpen
  \bibfield  {author} {\bibinfo {author} {\bibfnamefont {B.}~\bibnamefont
  {Keimer}}, \bibinfo {author} {\bibfnamefont {S.~A.}\ \bibnamefont
  {Kivelson}}, \bibinfo {author} {\bibfnamefont {M.~R.}\ \bibnamefont
  {Norman}}, \bibinfo {author} {\bibfnamefont {S.}~\bibnamefont {Uchida}},\
  and\ \bibinfo {author} {\bibfnamefont {J.}~\bibnamefont {Zaanen}},\
  }\bibfield  {title} {\bibinfo {title} {From quantum matter to
  high-temperature superconductivity in copper oxides},\ }\href
  {https://doi.org/10.1038/nature14165} {\bibfield  {journal} {\bibinfo
  {journal} {Nature}\ }\textbf {\bibinfo {volume} {518}},\ \bibinfo {pages}
  {179} (\bibinfo {year} {2015})}\BibitemShut {NoStop}%
\bibitem [{\citenamefont {Georges}\ \emph {et~al.}(1996)\citenamefont
  {Georges}, \citenamefont {Kotliar}, \citenamefont {Krauth},\ and\
  \citenamefont {Rozenberg}}]{Georges1996dynamical}%
  \BibitemOpen
  \bibfield  {author} {\bibinfo {author} {\bibfnamefont {A.}~\bibnamefont
  {Georges}}, \bibinfo {author} {\bibfnamefont {G.}~\bibnamefont {Kotliar}},
  \bibinfo {author} {\bibfnamefont {W.}~\bibnamefont {Krauth}},\ and\ \bibinfo
  {author} {\bibfnamefont {M.~J.}\ \bibnamefont {Rozenberg}},\ }\bibfield
  {title} {\bibinfo {title} {Dynamical mean-field theory of strongly correlated
  fermion systems and the limit of infinite dimensions},\ }\href
  {https://doi.org/10.1103/RevModPhys.68.13} {\bibfield  {journal} {\bibinfo
  {journal} {Rev. Mod. Phys.}\ }\textbf {\bibinfo {volume} {68}},\ \bibinfo
  {pages} {13} (\bibinfo {year} {1996})}\BibitemShut {NoStop}%
\bibitem [{\citenamefont {Rossi}\ \emph {et~al.}(2020)\citenamefont {Rossi},
  \citenamefont {{\v{S}}imkovic},\ and\ \citenamefont
  {Ferrero}}]{Rossi2020renormalized}%
  \BibitemOpen
  \bibfield  {author} {\bibinfo {author} {\bibfnamefont {R.}~\bibnamefont
  {Rossi}}, \bibinfo {author} {\bibfnamefont {F.}~\bibnamefont
  {{\v{S}}imkovic}},\ and\ \bibinfo {author} {\bibfnamefont {M.}~\bibnamefont
  {Ferrero}},\ }\bibfield  {title} {\bibinfo {title} {Renormalized perturbation
  theory at large expansion orders},\ }\href
  {https://doi.org/10.1209/0295-5075/132/11001} {\bibfield  {journal} {\bibinfo
   {journal} {{EPL} (Europhysics Letters)}\ }\textbf {\bibinfo {volume}
  {132}},\ \bibinfo {pages} {11001} (\bibinfo {year} {2020})}\BibitemShut
  {NoStop}%
\bibitem [{\citenamefont {Scalapino}(2012)}]{Scalapino2012common}%
  \BibitemOpen
  \bibfield  {author} {\bibinfo {author} {\bibfnamefont {D.~J.}\ \bibnamefont
  {Scalapino}},\ }\bibfield  {title} {\bibinfo {title} {A common thread: The
  pairing interaction for unconventional superconductors},\ }\href
  {https://doi.org/10.1103/RevModPhys.84.1383} {\bibfield  {journal} {\bibinfo
  {journal} {Rev. Mod. Phys.}\ }\textbf {\bibinfo {volume} {84}},\ \bibinfo
  {pages} {1383} (\bibinfo {year} {2012})}\BibitemShut {NoStop}%
\bibitem [{\citenamefont {Nilsson}\ \emph {et~al.}(2019)\citenamefont
  {Nilsson}, \citenamefont {Karlsson},\ and\ \citenamefont
  {Aryasetiawan}}]{Nilsson2019dynamically}%
  \BibitemOpen
  \bibfield  {author} {\bibinfo {author} {\bibfnamefont {F.}~\bibnamefont
  {Nilsson}}, \bibinfo {author} {\bibfnamefont {K.}~\bibnamefont {Karlsson}},\
  and\ \bibinfo {author} {\bibfnamefont {F.}~\bibnamefont {Aryasetiawan}},\
  }\bibfield  {title} {\bibinfo {title} {Dynamically screened coulomb
  interaction in the parent compounds of hole-doped cuprates: Trends and
  exceptions},\ }\href {https://doi.org/10.1103/PhysRevB.99.075135} {\bibfield
  {journal} {\bibinfo  {journal} {Phys. Rev. B}\ }\textbf {\bibinfo {volume}
  {99}},\ \bibinfo {pages} {075135} (\bibinfo {year} {2019})}\BibitemShut
  {NoStop}%
\bibitem [{\citenamefont {Abbamonte}\ \emph {et~al.}(2005)\citenamefont
  {Abbamonte}, \citenamefont {Rusydi}, \citenamefont {Smadici}, \citenamefont
  {Gu}, \citenamefont {Sawatzky},\ and\ \citenamefont {Feng}}]{Abbamonte2005}%
  \BibitemOpen
  \bibfield  {author} {\bibinfo {author} {\bibfnamefont {P.}~\bibnamefont
  {Abbamonte}}, \bibinfo {author} {\bibfnamefont {A.}~\bibnamefont {Rusydi}},
  \bibinfo {author} {\bibfnamefont {S.}~\bibnamefont {Smadici}}, \bibinfo
  {author} {\bibfnamefont {G.~D.}\ \bibnamefont {Gu}}, \bibinfo {author}
  {\bibfnamefont {G.~A.}\ \bibnamefont {Sawatzky}},\ and\ \bibinfo {author}
  {\bibfnamefont {D.~L.}\ \bibnamefont {Feng}},\ }\bibfield  {title} {\bibinfo
  {title} {{Spatially modulated 'Mottness' in
  {$\mathrm{La}_{2-x}\mathrm{Ba}_x\mathrm{CuO}_4$}}},\ }\href
  {https://doi.org/10.1038/nphys178} {\bibfield  {journal} {\bibinfo  {journal}
  {Nature Physics}\ }\textbf {\bibinfo {volume} {1}},\ \bibinfo {pages} {155}
  (\bibinfo {year} {2005})}\BibitemShut {NoStop}%
\bibitem [{\citenamefont {Tranquada}\ \emph {et~al.}(2004)\citenamefont
  {Tranquada}, \citenamefont {Woo}, \citenamefont {Perring}, \citenamefont
  {Goka}, \citenamefont {Gu}, \citenamefont {Xu}, \citenamefont {Fujita},\ and\
  \citenamefont {Yamada}}]{Tranquada2004}%
  \BibitemOpen
  \bibfield  {author} {\bibinfo {author} {\bibfnamefont {J.~M.}\ \bibnamefont
  {Tranquada}}, \bibinfo {author} {\bibfnamefont {H.}~\bibnamefont {Woo}},
  \bibinfo {author} {\bibfnamefont {T.~G.}\ \bibnamefont {Perring}}, \bibinfo
  {author} {\bibfnamefont {H.}~\bibnamefont {Goka}}, \bibinfo {author}
  {\bibfnamefont {G.~D.}\ \bibnamefont {Gu}}, \bibinfo {author} {\bibfnamefont
  {G.}~\bibnamefont {Xu}}, \bibinfo {author} {\bibfnamefont {M.}~\bibnamefont
  {Fujita}},\ and\ \bibinfo {author} {\bibfnamefont {K.}~\bibnamefont
  {Yamada}},\ }\bibfield  {title} {\bibinfo {title} {{Quantum magnetic
  excitations from stripes in copper oxide superconductors}},\ }\href
  {https://doi.org/10.1038/nature02574} {\bibfield  {journal} {\bibinfo
  {journal} {Nature}\ }\textbf {\bibinfo {volume} {429}},\ \bibinfo {pages}
  {534} (\bibinfo {year} {2004})}\BibitemShut {NoStop}%
\bibitem [{\citenamefont {H\"ucker}\ \emph {et~al.}(2011)\citenamefont
  {H\"ucker}, \citenamefont {v.~Zimmermann}, \citenamefont {Gu}, \citenamefont
  {Xu}, \citenamefont {Wen}, \citenamefont {Xu}, \citenamefont {Kang},
  \citenamefont {Zheludev},\ and\ \citenamefont
  {Tranquada}}]{Hucker2011stripe}%
  \BibitemOpen
  \bibfield  {author} {\bibinfo {author} {\bibfnamefont {M.}~\bibnamefont
  {H\"ucker}}, \bibinfo {author} {\bibfnamefont {M.}~\bibnamefont
  {v.~Zimmermann}}, \bibinfo {author} {\bibfnamefont {G.~D.}\ \bibnamefont
  {Gu}}, \bibinfo {author} {\bibfnamefont {Z.~J.}\ \bibnamefont {Xu}}, \bibinfo
  {author} {\bibfnamefont {J.~S.}\ \bibnamefont {Wen}}, \bibinfo {author}
  {\bibfnamefont {G.}~\bibnamefont {Xu}}, \bibinfo {author} {\bibfnamefont
  {H.~J.}\ \bibnamefont {Kang}}, \bibinfo {author} {\bibfnamefont
  {A.}~\bibnamefont {Zheludev}},\ and\ \bibinfo {author} {\bibfnamefont
  {J.~M.}\ \bibnamefont {Tranquada}},\ }\bibfield  {title} {\bibinfo {title}
  {Stripe order in superconducting
  {La${}_{2\ensuremath{-}x}$Ba${}_{x}$CuO${}_{4}$
  ($0.095\ensuremath{\leqslant}x\ensuremath{\leqslant}0.155$})},\ }\href
  {https://doi.org/10.1103/PhysRevB.83.104506} {\bibfield  {journal} {\bibinfo
  {journal} {Phys. Rev. B}\ }\textbf {\bibinfo {volume} {83}},\ \bibinfo
  {pages} {104506} (\bibinfo {year} {2011})}\BibitemShut {NoStop}%
\bibitem [{\citenamefont {Cavalleri}\ \emph {et~al.}(2005)\citenamefont
  {Cavalleri}, \citenamefont {Rini}, \citenamefont {Chong}, \citenamefont
  {Fourmaux}, \citenamefont {Glover}, \citenamefont {Heimann}, \citenamefont
  {Kieffer},\ and\ \citenamefont {Schoenlein}}]{Cavalleri2005band}%
  \BibitemOpen
  \bibfield  {author} {\bibinfo {author} {\bibfnamefont {A.}~\bibnamefont
  {Cavalleri}}, \bibinfo {author} {\bibfnamefont {M.}~\bibnamefont {Rini}},
  \bibinfo {author} {\bibfnamefont {H.~H.~W.}\ \bibnamefont {Chong}}, \bibinfo
  {author} {\bibfnamefont {S.}~\bibnamefont {Fourmaux}}, \bibinfo {author}
  {\bibfnamefont {T.~E.}\ \bibnamefont {Glover}}, \bibinfo {author}
  {\bibfnamefont {P.~A.}\ \bibnamefont {Heimann}}, \bibinfo {author}
  {\bibfnamefont {J.~C.}\ \bibnamefont {Kieffer}},\ and\ \bibinfo {author}
  {\bibfnamefont {R.~W.}\ \bibnamefont {Schoenlein}},\ }\bibfield  {title}
  {\bibinfo {title} {{Band-Selective Measurements of Electron Dynamics in
  ${\mathrm{VO}}_{2}$ Using Femtosecond Near-Edge X-Ray Absorption}},\ }\href
  {https://doi.org/10.1103/PhysRevLett.95.067405} {\bibfield  {journal}
  {\bibinfo  {journal} {Phys. Rev. Lett.}\ }\textbf {\bibinfo {volume} {95}},\
  \bibinfo {pages} {067405} (\bibinfo {year} {2005})}\BibitemShut {NoStop}%
\bibitem [{\citenamefont {Stamm}\ \emph {et~al.}(2007)\citenamefont {Stamm},
  \citenamefont {Kachel}, \citenamefont {Pontius}, \citenamefont {Mitzner},
  \citenamefont {Quast}, \citenamefont {Holldack}, \citenamefont {Khan},
  \citenamefont {Lupulescu}, \citenamefont {Aziz}, \citenamefont {Wietstruk},
  \citenamefont {D{\"{u}}rr},\ and\ \citenamefont
  {Eberhardt}}]{Stamm2007femtosecond}%
  \BibitemOpen
  \bibfield  {author} {\bibinfo {author} {\bibfnamefont {C.}~\bibnamefont
  {Stamm}}, \bibinfo {author} {\bibfnamefont {T.}~\bibnamefont {Kachel}},
  \bibinfo {author} {\bibfnamefont {N.}~\bibnamefont {Pontius}}, \bibinfo
  {author} {\bibfnamefont {R.}~\bibnamefont {Mitzner}}, \bibinfo {author}
  {\bibfnamefont {T.}~\bibnamefont {Quast}}, \bibinfo {author} {\bibfnamefont
  {K.}~\bibnamefont {Holldack}}, \bibinfo {author} {\bibfnamefont
  {S.}~\bibnamefont {Khan}}, \bibinfo {author} {\bibfnamefont {C.}~\bibnamefont
  {Lupulescu}}, \bibinfo {author} {\bibfnamefont {E.~F.}\ \bibnamefont {Aziz}},
  \bibinfo {author} {\bibfnamefont {M.}~\bibnamefont {Wietstruk}}, \bibinfo
  {author} {\bibfnamefont {H.~A.}\ \bibnamefont {D{\"{u}}rr}},\ and\ \bibinfo
  {author} {\bibfnamefont {W.}~\bibnamefont {Eberhardt}},\ }\bibfield  {title}
  {\bibinfo {title} {{Femtosecond modification of electron localization and
  transfer of angular momentum in nickel}},\ }\href
  {https://doi.org/10.1038/nmat1985} {\bibfield  {journal} {\bibinfo  {journal}
  {Nature Materials}\ }\textbf {\bibinfo {volume} {6}},\ \bibinfo {pages} {740}
  (\bibinfo {year} {2007})}\BibitemShut {NoStop}%
\bibitem [{\citenamefont {Moulet}\ \emph {et~al.}(2017)\citenamefont {Moulet},
  \citenamefont {Bertrand}, \citenamefont {Klostermann}, \citenamefont
  {Guggenmos}, \citenamefont {Karpowicz},\ and\ \citenamefont
  {Goulielmakis}}]{Moulet2017}%
  \BibitemOpen
  \bibfield  {author} {\bibinfo {author} {\bibfnamefont {A.}~\bibnamefont
  {Moulet}}, \bibinfo {author} {\bibfnamefont {J.~B.}\ \bibnamefont
  {Bertrand}}, \bibinfo {author} {\bibfnamefont {T.}~\bibnamefont
  {Klostermann}}, \bibinfo {author} {\bibfnamefont {A.}~\bibnamefont
  {Guggenmos}}, \bibinfo {author} {\bibfnamefont {N.}~\bibnamefont
  {Karpowicz}},\ and\ \bibinfo {author} {\bibfnamefont {E.}~\bibnamefont
  {Goulielmakis}},\ }\bibfield  {title} {\bibinfo {title} {Soft x-ray
  excitonics},\ }\href {https://doi.org/10.1126/science.aan4737} {\bibfield
  {journal} {\bibinfo  {journal} {Science}\ }\textbf {\bibinfo {volume}
  {357}},\ \bibinfo {pages} {1134} (\bibinfo {year} {2017})}\BibitemShut
  {NoStop}%
\bibitem [{\citenamefont {Carneiro}\ \emph {et~al.}(2017)\citenamefont
  {Carneiro}, \citenamefont {Cushing}, \citenamefont {Liu}, \citenamefont {Su},
  \citenamefont {Yang}, \citenamefont {Alivisatos},\ and\ \citenamefont
  {Leone}}]{Carneiro2017}%
  \BibitemOpen
  \bibfield  {author} {\bibinfo {author} {\bibfnamefont {L.~M.}\ \bibnamefont
  {Carneiro}}, \bibinfo {author} {\bibfnamefont {S.~K.}\ \bibnamefont
  {Cushing}}, \bibinfo {author} {\bibfnamefont {C.}~\bibnamefont {Liu}},
  \bibinfo {author} {\bibfnamefont {Y.}~\bibnamefont {Su}}, \bibinfo {author}
  {\bibfnamefont {P.}~\bibnamefont {Yang}}, \bibinfo {author} {\bibfnamefont
  {A.~P.}\ \bibnamefont {Alivisatos}},\ and\ \bibinfo {author} {\bibfnamefont
  {S.~R.}\ \bibnamefont {Leone}},\ }\bibfield  {title} {\bibinfo {title}
  {{Excitation-wavelength-dependent small polaron trapping of photoexcited
  carriers in {$\ensuremath{\alpha}$}-Fe2O3}},\ }\href
  {https://doi.org/10.1038/nmat4936} {\bibfield  {journal} {\bibinfo  {journal}
  {Nature Materials}\ }\textbf {\bibinfo {volume} {16}},\ \bibinfo {pages}
  {819} (\bibinfo {year} {2017})}\BibitemShut {NoStop}%
\bibitem [{\citenamefont {Z{\"{u}}rch}\ \emph {et~al.}(2017)\citenamefont
  {Z{\"{u}}rch}, \citenamefont {Chang}, \citenamefont {Borja}, \citenamefont
  {Kraus}, \citenamefont {Cushing}, \citenamefont {Gandman}, \citenamefont
  {Kaplan}, \citenamefont {Oh}, \citenamefont {Prell}, \citenamefont
  {Prendergast}, \citenamefont {Pemmaraju}, \citenamefont {Neumark},\ and\
  \citenamefont {Leone}}]{Zurch2017a}%
  \BibitemOpen
  \bibfield  {author} {\bibinfo {author} {\bibfnamefont {M.}~\bibnamefont
  {Z{\"{u}}rch}}, \bibinfo {author} {\bibfnamefont {H.-T.}\ \bibnamefont
  {Chang}}, \bibinfo {author} {\bibfnamefont {L.~J.}\ \bibnamefont {Borja}},
  \bibinfo {author} {\bibfnamefont {P.~M.}\ \bibnamefont {Kraus}}, \bibinfo
  {author} {\bibfnamefont {S.~K.}\ \bibnamefont {Cushing}}, \bibinfo {author}
  {\bibfnamefont {A.}~\bibnamefont {Gandman}}, \bibinfo {author} {\bibfnamefont
  {C.~J.}\ \bibnamefont {Kaplan}}, \bibinfo {author} {\bibfnamefont {M.~H.}\
  \bibnamefont {Oh}}, \bibinfo {author} {\bibfnamefont {J.~S.}\ \bibnamefont
  {Prell}}, \bibinfo {author} {\bibfnamefont {D.}~\bibnamefont {Prendergast}},
  \bibinfo {author} {\bibfnamefont {C.~D.}\ \bibnamefont {Pemmaraju}}, \bibinfo
  {author} {\bibfnamefont {D.~M.}\ \bibnamefont {Neumark}},\ and\ \bibinfo
  {author} {\bibfnamefont {S.~R.}\ \bibnamefont {Leone}},\ }\bibfield  {title}
  {\bibinfo {title} {{Direct and simultaneous observation of ultrafast electron
  and hole dynamics in germanium}},\ }\href
  {https://doi.org/10.1038/ncomms15734} {\bibfield  {journal} {\bibinfo
  {journal} {Nature Communications}\ }\textbf {\bibinfo {volume} {8}},\
  \bibinfo {pages} {15734} (\bibinfo {year} {2017})}\BibitemShut {NoStop}%
\bibitem [{\citenamefont {Attar}\ \emph {et~al.}(2020)\citenamefont {Attar},
  \citenamefont {Chang}, \citenamefont {Britz}, \citenamefont {Zhang},
  \citenamefont {Lin}, \citenamefont {Krishnamoorthy}, \citenamefont {Linker},
  \citenamefont {Fritz}, \citenamefont {Neumark}, \citenamefont {Kalia},
  \citenamefont {Nakano}, \citenamefont {Ajayan}, \citenamefont {Vashishta},
  \citenamefont {Bergmann},\ and\ \citenamefont {Leone}}]{Attar2020}%
  \BibitemOpen
  \bibfield  {author} {\bibinfo {author} {\bibfnamefont {A.~R.}\ \bibnamefont
  {Attar}}, \bibinfo {author} {\bibfnamefont {H.-T.}\ \bibnamefont {Chang}},
  \bibinfo {author} {\bibfnamefont {A.}~\bibnamefont {Britz}}, \bibinfo
  {author} {\bibfnamefont {X.}~\bibnamefont {Zhang}}, \bibinfo {author}
  {\bibfnamefont {M.-F.}\ \bibnamefont {Lin}}, \bibinfo {author} {\bibfnamefont
  {A.}~\bibnamefont {Krishnamoorthy}}, \bibinfo {author} {\bibfnamefont
  {T.}~\bibnamefont {Linker}}, \bibinfo {author} {\bibfnamefont
  {D.}~\bibnamefont {Fritz}}, \bibinfo {author} {\bibfnamefont {D.~M.}\
  \bibnamefont {Neumark}}, \bibinfo {author} {\bibfnamefont {R.~K.}\
  \bibnamefont {Kalia}}, \bibinfo {author} {\bibfnamefont {A.}~\bibnamefont
  {Nakano}}, \bibinfo {author} {\bibfnamefont {P.}~\bibnamefont {Ajayan}},
  \bibinfo {author} {\bibfnamefont {P.}~\bibnamefont {Vashishta}}, \bibinfo
  {author} {\bibfnamefont {U.}~\bibnamefont {Bergmann}},\ and\ \bibinfo
  {author} {\bibfnamefont {S.~R.}\ \bibnamefont {Leone}},\ }\bibfield  {title}
  {\bibinfo {title} {{Simultaneous Observation of Carrier-Specific
  Redistribution and Coherent Lattice Dynamics in {2H-MoTe{$_2$}} with
  Femtosecond Core-Level Spectroscopy}},\ }\href
  {https://doi.org/10.1021/acsnano.0c06988} {\bibfield  {journal} {\bibinfo
  {journal} {ACS Nano}\ }\textbf {\bibinfo {volume} {14}},\ \bibinfo {pages}
  {15829} (\bibinfo {year} {2020})}\BibitemShut {NoStop}%
\bibitem [{\citenamefont {Britz}\ \emph {et~al.}(2021)\citenamefont {Britz},
  \citenamefont {Attar}, \citenamefont {Zhang}, \citenamefont {Chang},
  \citenamefont {Nyby}, \citenamefont {Krishnamoorthy}, \citenamefont {Park},
  \citenamefont {Kwon}, \citenamefont {Kim}, \citenamefont {Nordlund},
  \citenamefont {Sainio}, \citenamefont {Heinz}, \citenamefont {Leone},
  \citenamefont {Lindenberg}, \citenamefont {Nakano}, \citenamefont {Ajayan},
  \citenamefont {Vashishta}, \citenamefont {Fritz}, \citenamefont {Lin},\ and\
  \citenamefont {Bergmann}}]{Britz2021}%
  \BibitemOpen
  \bibfield  {author} {\bibinfo {author} {\bibfnamefont {A.}~\bibnamefont
  {Britz}}, \bibinfo {author} {\bibfnamefont {A.~R.}\ \bibnamefont {Attar}},
  \bibinfo {author} {\bibfnamefont {X.}~\bibnamefont {Zhang}}, \bibinfo
  {author} {\bibfnamefont {H.-T.}\ \bibnamefont {Chang}}, \bibinfo {author}
  {\bibfnamefont {C.}~\bibnamefont {Nyby}}, \bibinfo {author} {\bibfnamefont
  {A.}~\bibnamefont {Krishnamoorthy}}, \bibinfo {author} {\bibfnamefont
  {S.~H.}\ \bibnamefont {Park}}, \bibinfo {author} {\bibfnamefont
  {S.}~\bibnamefont {Kwon}}, \bibinfo {author} {\bibfnamefont {M.}~\bibnamefont
  {Kim}}, \bibinfo {author} {\bibfnamefont {D.}~\bibnamefont {Nordlund}},
  \bibinfo {author} {\bibfnamefont {S.}~\bibnamefont {Sainio}}, \bibinfo
  {author} {\bibfnamefont {T.~F.}\ \bibnamefont {Heinz}}, \bibinfo {author}
  {\bibfnamefont {S.~R.}\ \bibnamefont {Leone}}, \bibinfo {author}
  {\bibfnamefont {A.~M.}\ \bibnamefont {Lindenberg}}, \bibinfo {author}
  {\bibfnamefont {A.}~\bibnamefont {Nakano}}, \bibinfo {author} {\bibfnamefont
  {P.}~\bibnamefont {Ajayan}}, \bibinfo {author} {\bibfnamefont
  {P.}~\bibnamefont {Vashishta}}, \bibinfo {author} {\bibfnamefont
  {D.}~\bibnamefont {Fritz}}, \bibinfo {author} {\bibfnamefont {M.-F.}\
  \bibnamefont {Lin}},\ and\ \bibinfo {author} {\bibfnamefont {U.}~\bibnamefont
  {Bergmann}},\ }\bibfield  {title} {\bibinfo {title} {Carrier-specific
  dynamics in {2H-MoTe{$_2$}} observed by femtosecond soft x-ray absorption
  spectroscopy using an x-ray free-electron laser},\ }\href
  {https://doi.org/10.1063/4.0000048} {\bibfield  {journal} {\bibinfo
  {journal} {Structural Dynamics}\ }\textbf {\bibinfo {volume} {8}},\ \bibinfo
  {pages} {014501} (\bibinfo {year} {2021})}\BibitemShut {NoStop}%
\bibitem [{\citenamefont {Mardegan}\ \emph {et~al.}(2021)\citenamefont
  {Mardegan}, \citenamefont {Zerdane}, \citenamefont {Mancini}, \citenamefont
  {Esposito}, \citenamefont {Rouxel}, \citenamefont {Mankowsky}, \citenamefont
  {Svetina}, \citenamefont {Gurung}, \citenamefont {Parchenko}, \citenamefont
  {Porer}, \citenamefont {Burganov}, \citenamefont {Deng}, \citenamefont
  {Beaud}, \citenamefont {Ingold}, \citenamefont {Pedrini}, \citenamefont
  {Arrell}, \citenamefont {Erny}, \citenamefont {Dax}, \citenamefont {Lemke},
  \citenamefont {Decker}, \citenamefont {Ortiz}, \citenamefont {Milne},
  \citenamefont {Smolentsev}, \citenamefont {Maurel}, \citenamefont {Johnson},
  \citenamefont {Mitsuda}, \citenamefont {Wada}, \citenamefont {Yokoyama},
  \citenamefont {Wadati},\ and\ \citenamefont {Staub}}]{Mardegan2021ultrafast}%
  \BibitemOpen
  \bibfield  {author} {\bibinfo {author} {\bibfnamefont {J.~R.~L.}\
  \bibnamefont {Mardegan}}, \bibinfo {author} {\bibfnamefont {S.}~\bibnamefont
  {Zerdane}}, \bibinfo {author} {\bibfnamefont {G.}~\bibnamefont {Mancini}},
  \bibinfo {author} {\bibfnamefont {V.}~\bibnamefont {Esposito}}, \bibinfo
  {author} {\bibfnamefont {J.~R.}\ \bibnamefont {Rouxel}}, \bibinfo {author}
  {\bibfnamefont {R.}~\bibnamefont {Mankowsky}}, \bibinfo {author}
  {\bibfnamefont {C.}~\bibnamefont {Svetina}}, \bibinfo {author} {\bibfnamefont
  {N.}~\bibnamefont {Gurung}}, \bibinfo {author} {\bibfnamefont
  {S.}~\bibnamefont {Parchenko}}, \bibinfo {author} {\bibfnamefont
  {M.}~\bibnamefont {Porer}}, \bibinfo {author} {\bibfnamefont
  {B.}~\bibnamefont {Burganov}}, \bibinfo {author} {\bibfnamefont
  {Y.}~\bibnamefont {Deng}}, \bibinfo {author} {\bibfnamefont {P.}~\bibnamefont
  {Beaud}}, \bibinfo {author} {\bibfnamefont {G.}~\bibnamefont {Ingold}},
  \bibinfo {author} {\bibfnamefont {B.}~\bibnamefont {Pedrini}}, \bibinfo
  {author} {\bibfnamefont {C.}~\bibnamefont {Arrell}}, \bibinfo {author}
  {\bibfnamefont {C.}~\bibnamefont {Erny}}, \bibinfo {author} {\bibfnamefont
  {A.}~\bibnamefont {Dax}}, \bibinfo {author} {\bibfnamefont {H.}~\bibnamefont
  {Lemke}}, \bibinfo {author} {\bibfnamefont {M.}~\bibnamefont {Decker}},
  \bibinfo {author} {\bibfnamefont {N.}~\bibnamefont {Ortiz}}, \bibinfo
  {author} {\bibfnamefont {C.}~\bibnamefont {Milne}}, \bibinfo {author}
  {\bibfnamefont {G.}~\bibnamefont {Smolentsev}}, \bibinfo {author}
  {\bibfnamefont {L.}~\bibnamefont {Maurel}}, \bibinfo {author} {\bibfnamefont
  {S.~L.}\ \bibnamefont {Johnson}}, \bibinfo {author} {\bibfnamefont
  {A.}~\bibnamefont {Mitsuda}}, \bibinfo {author} {\bibfnamefont
  {H.}~\bibnamefont {Wada}}, \bibinfo {author} {\bibfnamefont {Y.}~\bibnamefont
  {Yokoyama}}, \bibinfo {author} {\bibfnamefont {H.}~\bibnamefont {Wadati}},\
  and\ \bibinfo {author} {\bibfnamefont {U.}~\bibnamefont {Staub}},\ }\bibfield
   {title} {\bibinfo {title} {Ultrafast electron localization in the
  {$\mathrm{Eu}{\mathrm{Ni}}_{2}{({\mathrm{Si}}_{0.21}{\mathrm{Ge}}_{0.79})}_{2}$}
  correlated metal},\ }\href {https://doi.org/10.1103/PhysRevResearch.3.033211}
  {\bibfield  {journal} {\bibinfo  {journal} {Phys. Rev. Research}\ }\textbf
  {\bibinfo {volume} {3}},\ \bibinfo {pages} {033211} (\bibinfo {year}
  {2021})}\BibitemShut {NoStop}%
\bibitem [{\citenamefont {Jang}\ \emph {et~al.}(2020)\citenamefont {Jang},
  \citenamefont {Kim}, \citenamefont {Kim}, \citenamefont {Park}, \citenamefont
  {Kwon}, \citenamefont {Lee}, \citenamefont {Park}, \citenamefont {Park},
  \citenamefont {Kim}, \citenamefont {Hyun}, \citenamefont {Hwang},
  \citenamefont {Lee}, \citenamefont {Lim}, \citenamefont {Gang}, \citenamefont
  {Kim}, \citenamefont {Heo}, \citenamefont {Kim}, \citenamefont {Jung},
  \citenamefont {Kim}, \citenamefont {Park}, \citenamefont {Kim}, \citenamefont
  {Shin}, \citenamefont {Park}, \citenamefont {Koo}, \citenamefont {Shin},
  \citenamefont {Heo}, \citenamefont {Kim}, \citenamefont {Min}, \citenamefont
  {Han}, \citenamefont {Kang}, \citenamefont {Lee}, \citenamefont {Kim},
  \citenamefont {Eom},\ and\ \citenamefont {Rah}}]{Jang2020}%
  \BibitemOpen
  \bibfield  {author} {\bibinfo {author} {\bibfnamefont {H.}~\bibnamefont
  {Jang}}, \bibinfo {author} {\bibfnamefont {H.-D.}\ \bibnamefont {Kim}},
  \bibinfo {author} {\bibfnamefont {M.}~\bibnamefont {Kim}}, \bibinfo {author}
  {\bibfnamefont {S.~H.}\ \bibnamefont {Park}}, \bibinfo {author}
  {\bibfnamefont {S.}~\bibnamefont {Kwon}}, \bibinfo {author} {\bibfnamefont
  {J.~Y.}\ \bibnamefont {Lee}}, \bibinfo {author} {\bibfnamefont {S.-Y.}\
  \bibnamefont {Park}}, \bibinfo {author} {\bibfnamefont {G.}~\bibnamefont
  {Park}}, \bibinfo {author} {\bibfnamefont {S.}~\bibnamefont {Kim}}, \bibinfo
  {author} {\bibfnamefont {H.}~\bibnamefont {Hyun}}, \bibinfo {author}
  {\bibfnamefont {S.}~\bibnamefont {Hwang}}, \bibinfo {author} {\bibfnamefont
  {C.-S.}\ \bibnamefont {Lee}}, \bibinfo {author} {\bibfnamefont {C.-Y.}\
  \bibnamefont {Lim}}, \bibinfo {author} {\bibfnamefont {W.}~\bibnamefont
  {Gang}}, \bibinfo {author} {\bibfnamefont {M.}~\bibnamefont {Kim}}, \bibinfo
  {author} {\bibfnamefont {S.}~\bibnamefont {Heo}}, \bibinfo {author}
  {\bibfnamefont {J.}~\bibnamefont {Kim}}, \bibinfo {author} {\bibfnamefont
  {G.}~\bibnamefont {Jung}}, \bibinfo {author} {\bibfnamefont {S.}~\bibnamefont
  {Kim}}, \bibinfo {author} {\bibfnamefont {J.}~\bibnamefont {Park}}, \bibinfo
  {author} {\bibfnamefont {J.}~\bibnamefont {Kim}}, \bibinfo {author}
  {\bibfnamefont {H.}~\bibnamefont {Shin}}, \bibinfo {author} {\bibfnamefont
  {J.}~\bibnamefont {Park}}, \bibinfo {author} {\bibfnamefont {T.-Y.}\
  \bibnamefont {Koo}}, \bibinfo {author} {\bibfnamefont {H.-J.}\ \bibnamefont
  {Shin}}, \bibinfo {author} {\bibfnamefont {H.}~\bibnamefont {Heo}}, \bibinfo
  {author} {\bibfnamefont {C.}~\bibnamefont {Kim}}, \bibinfo {author}
  {\bibfnamefont {C.-K.}\ \bibnamefont {Min}}, \bibinfo {author} {\bibfnamefont
  {J.-H.}\ \bibnamefont {Han}}, \bibinfo {author} {\bibfnamefont {H.-S.}\
  \bibnamefont {Kang}}, \bibinfo {author} {\bibfnamefont {H.-S.}\ \bibnamefont
  {Lee}}, \bibinfo {author} {\bibfnamefont {K.~S.}\ \bibnamefont {Kim}},
  \bibinfo {author} {\bibfnamefont {I.}~\bibnamefont {Eom}},\ and\ \bibinfo
  {author} {\bibfnamefont {S.}~\bibnamefont {Rah}},\ }\bibfield  {title}
  {\bibinfo {title} {{Time-resolved resonant elastic soft x-ray scattering at
  Pohang Accelerator Laboratory X-ray Free Electron Laser}},\ }\href
  {https://doi.org/10.1063/5.0016414} {\bibfield  {journal} {\bibinfo
  {journal} {Review of Scientific Instruments}\ }\textbf {\bibinfo {volume}
  {91}},\ \bibinfo {pages} {083904} (\bibinfo {year} {2020})}\BibitemShut
  {NoStop}%
\bibitem [{\citenamefont {Ogata}\ and\ \citenamefont
  {Fukuyama}(2008)}]{Ogata2008the}%
  \BibitemOpen
  \bibfield  {author} {\bibinfo {author} {\bibfnamefont {M.}~\bibnamefont
  {Ogata}}\ and\ \bibinfo {author} {\bibfnamefont {H.}~\bibnamefont
  {Fukuyama}},\ }\bibfield  {title} {\bibinfo {title} {{The t{\textendash}J
  model for the oxide high-T$_c$ superconductors}},\ }\href
  {https://doi.org/10.1088/0034-4885/71/3/036501} {\bibfield  {journal}
  {\bibinfo  {journal} {Reports on Progress in Physics}\ }\textbf {\bibinfo
  {volume} {71}},\ \bibinfo {pages} {036501} (\bibinfo {year}
  {2008})}\BibitemShut {NoStop}%
\bibitem [{\citenamefont {Zhang}\ and\ \citenamefont
  {Rice}(1988)}]{Zhang1988effective}%
  \BibitemOpen
  \bibfield  {author} {\bibinfo {author} {\bibfnamefont {F.~C.}\ \bibnamefont
  {Zhang}}\ and\ \bibinfo {author} {\bibfnamefont {T.~M.}\ \bibnamefont
  {Rice}},\ }\bibfield  {title} {\bibinfo {title} {Effective hamiltonian for
  the superconducting cu oxides},\ }\href
  {https://doi.org/10.1103/PhysRevB.37.3759} {\bibfield  {journal} {\bibinfo
  {journal} {Phys. Rev. B}\ }\textbf {\bibinfo {volume} {37}},\ \bibinfo
  {pages} {3759} (\bibinfo {year} {1988})}\BibitemShut {NoStop}%
\bibitem [{\citenamefont {Eskes}\ \emph {et~al.}(1991)\citenamefont {Eskes},
  \citenamefont {Meinders},\ and\ \citenamefont {Sawatzky}}]{Eskes1991}%
  \BibitemOpen
  \bibfield  {author} {\bibinfo {author} {\bibfnamefont {H.}~\bibnamefont
  {Eskes}}, \bibinfo {author} {\bibfnamefont {M.~B.~J.}\ \bibnamefont
  {Meinders}},\ and\ \bibinfo {author} {\bibfnamefont {G.~A.}\ \bibnamefont
  {Sawatzky}},\ }\bibfield  {title} {\bibinfo {title} {{Anomalous transfer of
  spectral weight in doped strongly correlated systems}},\ }\href
  {https://doi.org/10.1103/PhysRevLett.67.1035} {\bibfield  {journal} {\bibinfo
   {journal} {Phys. Rev. Lett.}\ }\textbf {\bibinfo {volume} {67}},\ \bibinfo
  {pages} {1035} (\bibinfo {year} {1991})}\BibitemShut {NoStop}%
\bibitem [{\citenamefont {Chen}\ \emph {et~al.}(1991)\citenamefont {Chen},
  \citenamefont {Sette}, \citenamefont {Ma}, \citenamefont {Hybertsen},
  \citenamefont {Stechel}, \citenamefont {Foulkes}, \citenamefont {Schulter},
  \citenamefont {Cheong}, \citenamefont {Cooper}, \citenamefont {Rupp},
  \citenamefont {Batlogg}, \citenamefont {Soo}, \citenamefont {Ming},
  \citenamefont {Krol},\ and\ \citenamefont {Kao}}]{Chen1991}%
  \BibitemOpen
  \bibfield  {author} {\bibinfo {author} {\bibfnamefont {C.~T.}\ \bibnamefont
  {Chen}}, \bibinfo {author} {\bibfnamefont {F.}~\bibnamefont {Sette}},
  \bibinfo {author} {\bibfnamefont {Y.}~\bibnamefont {Ma}}, \bibinfo {author}
  {\bibfnamefont {M.~S.}\ \bibnamefont {Hybertsen}}, \bibinfo {author}
  {\bibfnamefont {E.~B.}\ \bibnamefont {Stechel}}, \bibinfo {author}
  {\bibfnamefont {W.~M.~C.}\ \bibnamefont {Foulkes}}, \bibinfo {author}
  {\bibfnamefont {M.}~\bibnamefont {Schulter}}, \bibinfo {author}
  {\bibfnamefont {S.-W.}\ \bibnamefont {Cheong}}, \bibinfo {author}
  {\bibfnamefont {A.~S.}\ \bibnamefont {Cooper}}, \bibinfo {author}
  {\bibfnamefont {L.~W.}\ \bibnamefont {Rupp}}, \bibinfo {author}
  {\bibfnamefont {B.}~\bibnamefont {Batlogg}}, \bibinfo {author} {\bibfnamefont
  {Y.~L.}\ \bibnamefont {Soo}}, \bibinfo {author} {\bibfnamefont {Z.~H.}\
  \bibnamefont {Ming}}, \bibinfo {author} {\bibfnamefont {A.}~\bibnamefont
  {Krol}},\ and\ \bibinfo {author} {\bibfnamefont {Y.~H.}\ \bibnamefont
  {Kao}},\ }\bibfield  {title} {\bibinfo {title} {{Electronic states in
  ${\mathrm{La}}_{2\mathrm{\ensuremath{-}}\mathit{x}}$${\mathrm{Sr}}_{\mathit{x}}$${\mathrm{CuO}}_{4+\mathrm{\ensuremath{\delta}}}$
  probed by soft-x-ray absorption}},\ }\href
  {https://doi.org/10.1103/PhysRevLett.66.104} {\bibfield  {journal} {\bibinfo
  {journal} {Phys. Rev. Lett.}\ }\textbf {\bibinfo {volume} {66}},\ \bibinfo
  {pages} {104} (\bibinfo {year} {1991})}\BibitemShut {NoStop}%
\bibitem [{\citenamefont {Chen}\ \emph {et~al.}(1992)\citenamefont {Chen},
  \citenamefont {Tjeng}, \citenamefont {Kwo}, \citenamefont {Kao},
  \citenamefont {Rudolf}, \citenamefont {Sette},\ and\ \citenamefont
  {Fleming}}]{Chen1992}%
  \BibitemOpen
  \bibfield  {author} {\bibinfo {author} {\bibfnamefont {C.~T.}\ \bibnamefont
  {Chen}}, \bibinfo {author} {\bibfnamefont {L.~H.}\ \bibnamefont {Tjeng}},
  \bibinfo {author} {\bibfnamefont {J.}~\bibnamefont {Kwo}}, \bibinfo {author}
  {\bibfnamefont {H.~L.}\ \bibnamefont {Kao}}, \bibinfo {author} {\bibfnamefont
  {P.}~\bibnamefont {Rudolf}}, \bibinfo {author} {\bibfnamefont
  {F.}~\bibnamefont {Sette}},\ and\ \bibinfo {author} {\bibfnamefont {R.~M.}\
  \bibnamefont {Fleming}},\ }\bibfield  {title} {\bibinfo {title}
  {{Out-of-plane orbital characters of intrinsic and doped holes in
  ${\mathrm{La}}_{2\mathrm{\ensuremath{-}}\mathit{x}}$${\mathrm{Sr}}_{\mathit{x}}$${\mathrm{CuO}}_{4}$}},\
  }\href {https://doi.org/10.1103/PhysRevLett.68.2543} {\bibfield  {journal}
  {\bibinfo  {journal} {Phys. Rev. Lett.}\ }\textbf {\bibinfo {volume} {68}},\
  \bibinfo {pages} {2543} (\bibinfo {year} {1992})}\BibitemShut {NoStop}%
\bibitem [{\citenamefont {N\"ucker}\ \emph {et~al.}(1995)\citenamefont
  {N\"ucker}, \citenamefont {Pellegrin}, \citenamefont {Schweiss},
  \citenamefont {Fink}, \citenamefont {Molodtsov}, \citenamefont {Simmons},
  \citenamefont {Kaindl}, \citenamefont {Frentrup}, \citenamefont {Erb},\ and\
  \citenamefont {M\"uller-Vogt}}]{Nuecker1995}%
  \BibitemOpen
  \bibfield  {author} {\bibinfo {author} {\bibfnamefont {N.}~\bibnamefont
  {N\"ucker}}, \bibinfo {author} {\bibfnamefont {E.}~\bibnamefont {Pellegrin}},
  \bibinfo {author} {\bibfnamefont {P.}~\bibnamefont {Schweiss}}, \bibinfo
  {author} {\bibfnamefont {J.}~\bibnamefont {Fink}}, \bibinfo {author}
  {\bibfnamefont {S.~L.}\ \bibnamefont {Molodtsov}}, \bibinfo {author}
  {\bibfnamefont {C.~T.}\ \bibnamefont {Simmons}}, \bibinfo {author}
  {\bibfnamefont {G.}~\bibnamefont {Kaindl}}, \bibinfo {author} {\bibfnamefont
  {W.}~\bibnamefont {Frentrup}}, \bibinfo {author} {\bibfnamefont
  {A.}~\bibnamefont {Erb}},\ and\ \bibinfo {author} {\bibfnamefont
  {G.}~\bibnamefont {M\"uller-Vogt}},\ }\bibfield  {title} {\bibinfo {title}
  {Site-specific and doping-dependent electronic structure of
  {${\mathrm{YBa}}_{2}$${\mathrm{Cu}}_{3}$${\mathrm{O}}_{\mathit{x}}$} probed
  by {O} 1s and {Cu} 2p x-ray-absorption spectroscopy},\ }\href
  {https://doi.org/10.1103/PhysRevB.51.8529} {\bibfield  {journal} {\bibinfo
  {journal} {Phys. Rev. B}\ }\textbf {\bibinfo {volume} {51}},\ \bibinfo
  {pages} {8529} (\bibinfo {year} {1995})}\BibitemShut {NoStop}%
\bibitem [{\citenamefont {Baranov}\ and\ \citenamefont
  {Kabanov}(2014)}]{Baranov2014theory}%
  \BibitemOpen
  \bibfield  {author} {\bibinfo {author} {\bibfnamefont {V.~V.}\ \bibnamefont
  {Baranov}}\ and\ \bibinfo {author} {\bibfnamefont {V.~V.}\ \bibnamefont
  {Kabanov}},\ }\bibfield  {title} {\bibinfo {title} {Theory of electronic
  relaxation in a metal excited by an ultrashort optical pump},\ }\href
  {https://doi.org/10.1103/PhysRevB.89.125102} {\bibfield  {journal} {\bibinfo
  {journal} {Phys. Rev. B}\ }\textbf {\bibinfo {volume} {89}},\ \bibinfo
  {pages} {125102} (\bibinfo {year} {2014})}\BibitemShut {NoStop}%
\bibitem [{\citenamefont {Okamoto}\ \emph {et~al.}(2010)\citenamefont
  {Okamoto}, \citenamefont {Miyagoe}, \citenamefont {Kobayashi}, \citenamefont
  {Uemura}, \citenamefont {Nishioka}, \citenamefont {Matsuzaki}, \citenamefont
  {Sawa},\ and\ \citenamefont {Tokura}}]{Okamoto2010ultrafast}%
  \BibitemOpen
  \bibfield  {author} {\bibinfo {author} {\bibfnamefont {H.}~\bibnamefont
  {Okamoto}}, \bibinfo {author} {\bibfnamefont {T.}~\bibnamefont {Miyagoe}},
  \bibinfo {author} {\bibfnamefont {K.}~\bibnamefont {Kobayashi}}, \bibinfo
  {author} {\bibfnamefont {H.}~\bibnamefont {Uemura}}, \bibinfo {author}
  {\bibfnamefont {H.}~\bibnamefont {Nishioka}}, \bibinfo {author}
  {\bibfnamefont {H.}~\bibnamefont {Matsuzaki}}, \bibinfo {author}
  {\bibfnamefont {A.}~\bibnamefont {Sawa}},\ and\ \bibinfo {author}
  {\bibfnamefont {Y.}~\bibnamefont {Tokura}},\ }\bibfield  {title} {\bibinfo
  {title} {Ultrafast charge dynamics in photoexcited
  {${\text{Nd}}_{2}{\text{CuO}}_{4}$} and {${\text{La}}_{2}{\text{CuO}}_{4}$}
  cuprate compounds investigated by femtosecond absorption spectroscopy},\
  }\href {https://doi.org/10.1103/PhysRevB.82.060513} {\bibfield  {journal}
  {\bibinfo  {journal} {Phys. Rev. B}\ }\textbf {\bibinfo {volume} {82}},\
  \bibinfo {pages} {060513} (\bibinfo {year} {2010})}\BibitemShut {NoStop}%
\bibitem [{\citenamefont {Okamoto}\ \emph {et~al.}(2011)\citenamefont
  {Okamoto}, \citenamefont {Miyagoe}, \citenamefont {Kobayashi}, \citenamefont
  {Uemura}, \citenamefont {Nishioka}, \citenamefont {Matsuzaki}, \citenamefont
  {Sawa},\ and\ \citenamefont {Tokura}}]{Okamoto2011photoinduced}%
  \BibitemOpen
  \bibfield  {author} {\bibinfo {author} {\bibfnamefont {H.}~\bibnamefont
  {Okamoto}}, \bibinfo {author} {\bibfnamefont {T.}~\bibnamefont {Miyagoe}},
  \bibinfo {author} {\bibfnamefont {K.}~\bibnamefont {Kobayashi}}, \bibinfo
  {author} {\bibfnamefont {H.}~\bibnamefont {Uemura}}, \bibinfo {author}
  {\bibfnamefont {H.}~\bibnamefont {Nishioka}}, \bibinfo {author}
  {\bibfnamefont {H.}~\bibnamefont {Matsuzaki}}, \bibinfo {author}
  {\bibfnamefont {A.}~\bibnamefont {Sawa}},\ and\ \bibinfo {author}
  {\bibfnamefont {Y.}~\bibnamefont {Tokura}},\ }\bibfield  {title} {\bibinfo
  {title} {{Photoinduced transition from Mott insulator to metal in the undoped
  cuprates {${\text{Nd}}_{2}{\text{CuO}}_{4}$ and
  ${\text{La}}_{2}{\text{CuO}}_{4}$}}},\ }\href
  {https://doi.org/10.1103/PhysRevB.83.125102} {\bibfield  {journal} {\bibinfo
  {journal} {Phys. Rev. B}\ }\textbf {\bibinfo {volume} {83}},\ \bibinfo
  {pages} {125102} (\bibinfo {year} {2011})}\BibitemShut {NoStop}%
\bibitem [{\citenamefont {{Dal Conte}}\ \emph {et~al.}(2015)\citenamefont {{Dal
  Conte}}, \citenamefont {Vidmar}, \citenamefont {Gole{\v{z}}}, \citenamefont
  {Mierzejewski}, \citenamefont {Soavi}, \citenamefont {Peli}, \citenamefont
  {Banfi}, \citenamefont {Ferrini}, \citenamefont {Comin}, \citenamefont
  {Ludbrook}, \citenamefont {Chauviere}, \citenamefont {Zhigadlo},
  \citenamefont {Eisaki}, \citenamefont {Greven}, \citenamefont {Lupi},
  \citenamefont {Damascelli}, \citenamefont {Brida}, \citenamefont {Capone},
  \citenamefont {Bon{\v{c}}a}, \citenamefont {Cerullo},\ and\ \citenamefont
  {Giannetti}}]{DalConte2015snapshots}%
  \BibitemOpen
  \bibfield  {author} {\bibinfo {author} {\bibfnamefont {S.}~\bibnamefont {{Dal
  Conte}}}, \bibinfo {author} {\bibfnamefont {L.}~\bibnamefont {Vidmar}},
  \bibinfo {author} {\bibfnamefont {D.}~\bibnamefont {Gole{\v{z}}}}, \bibinfo
  {author} {\bibfnamefont {M.}~\bibnamefont {Mierzejewski}}, \bibinfo {author}
  {\bibfnamefont {G.}~\bibnamefont {Soavi}}, \bibinfo {author} {\bibfnamefont
  {S.}~\bibnamefont {Peli}}, \bibinfo {author} {\bibfnamefont {F.}~\bibnamefont
  {Banfi}}, \bibinfo {author} {\bibfnamefont {G.}~\bibnamefont {Ferrini}},
  \bibinfo {author} {\bibfnamefont {R.}~\bibnamefont {Comin}}, \bibinfo
  {author} {\bibfnamefont {B.~M.}\ \bibnamefont {Ludbrook}}, \bibinfo {author}
  {\bibfnamefont {L.}~\bibnamefont {Chauviere}}, \bibinfo {author}
  {\bibfnamefont {N.~D.}\ \bibnamefont {Zhigadlo}}, \bibinfo {author}
  {\bibfnamefont {H.}~\bibnamefont {Eisaki}}, \bibinfo {author} {\bibfnamefont
  {M.}~\bibnamefont {Greven}}, \bibinfo {author} {\bibfnamefont
  {S.}~\bibnamefont {Lupi}}, \bibinfo {author} {\bibfnamefont {A.}~\bibnamefont
  {Damascelli}}, \bibinfo {author} {\bibfnamefont {D.}~\bibnamefont {Brida}},
  \bibinfo {author} {\bibfnamefont {M.}~\bibnamefont {Capone}}, \bibinfo
  {author} {\bibfnamefont {J.}~\bibnamefont {Bon{\v{c}}a}}, \bibinfo {author}
  {\bibfnamefont {G.}~\bibnamefont {Cerullo}},\ and\ \bibinfo {author}
  {\bibfnamefont {C.}~\bibnamefont {Giannetti}},\ }\bibfield  {title} {\bibinfo
  {title} {{Snapshots of the retarded interaction of charge carriers with
  ultrafast fluctuations in cuprates}},\ }\href
  {https://doi.org/10.1038/nphys3265} {\bibfield  {journal} {\bibinfo
  {journal} {Nature Physics}\ }\textbf {\bibinfo {volume} {11}},\ \bibinfo
  {pages} {421} (\bibinfo {year} {2015})}\BibitemShut {NoStop}%
\bibitem [{\citenamefont {Carbone}\ \emph {et~al.}(2008)\citenamefont
  {Carbone}, \citenamefont {Yang}, \citenamefont {Giannini},\ and\
  \citenamefont {Zewail}}]{Carbone2008direct}%
  \BibitemOpen
  \bibfield  {author} {\bibinfo {author} {\bibfnamefont {F.}~\bibnamefont
  {Carbone}}, \bibinfo {author} {\bibfnamefont {D.-S.}\ \bibnamefont {Yang}},
  \bibinfo {author} {\bibfnamefont {E.}~\bibnamefont {Giannini}},\ and\
  \bibinfo {author} {\bibfnamefont {A.~H.}\ \bibnamefont {Zewail}},\ }\bibfield
   {title} {\bibinfo {title} {Direct role of structural dynamics in
  electron-lattice coupling of superconducting cuprates},\ }\href
  {https://doi.org/10.1073/pnas.0811335106} {\bibfield  {journal} {\bibinfo
  {journal} {Proceedings of the National Academy of Sciences}\ }\textbf
  {\bibinfo {volume} {105}},\ \bibinfo {pages} {20161} (\bibinfo {year}
  {2008})}\BibitemShut {NoStop}%
\bibitem [{\citenamefont {Konstantinova}\ \emph {et~al.}(2018)\citenamefont
  {Konstantinova}, \citenamefont {Rameau}, \citenamefont {Reid}, \citenamefont
  {Abdurazakov}, \citenamefont {Wu}, \citenamefont {Li}, \citenamefont {Shen},
  \citenamefont {Gu}, \citenamefont {Huang}, \citenamefont {Rettig},
  \citenamefont {Avigo}, \citenamefont {Ligges}, \citenamefont {Freericks},
  \citenamefont {Kemper}, \citenamefont {Dürr}, \citenamefont {Bovensiepen},
  \citenamefont {Johnson}, \citenamefont {Wang},\ and\ \citenamefont
  {Zhu}}]{Konstantinova2018}%
  \BibitemOpen
  \bibfield  {author} {\bibinfo {author} {\bibfnamefont {T.}~\bibnamefont
  {Konstantinova}}, \bibinfo {author} {\bibfnamefont {J.~D.}\ \bibnamefont
  {Rameau}}, \bibinfo {author} {\bibfnamefont {A.~H.}\ \bibnamefont {Reid}},
  \bibinfo {author} {\bibfnamefont {O.}~\bibnamefont {Abdurazakov}}, \bibinfo
  {author} {\bibfnamefont {L.}~\bibnamefont {Wu}}, \bibinfo {author}
  {\bibfnamefont {R.}~\bibnamefont {Li}}, \bibinfo {author} {\bibfnamefont
  {X.}~\bibnamefont {Shen}}, \bibinfo {author} {\bibfnamefont {G.}~\bibnamefont
  {Gu}}, \bibinfo {author} {\bibfnamefont {Y.}~\bibnamefont {Huang}}, \bibinfo
  {author} {\bibfnamefont {L.}~\bibnamefont {Rettig}}, \bibinfo {author}
  {\bibfnamefont {I.}~\bibnamefont {Avigo}}, \bibinfo {author} {\bibfnamefont
  {M.}~\bibnamefont {Ligges}}, \bibinfo {author} {\bibfnamefont {J.~K.}\
  \bibnamefont {Freericks}}, \bibinfo {author} {\bibfnamefont {A.~F.}\
  \bibnamefont {Kemper}}, \bibinfo {author} {\bibfnamefont {H.~A.}\
  \bibnamefont {Dürr}}, \bibinfo {author} {\bibfnamefont {U.}~\bibnamefont
  {Bovensiepen}}, \bibinfo {author} {\bibfnamefont {P.~D.}\ \bibnamefont
  {Johnson}}, \bibinfo {author} {\bibfnamefont {X.}~\bibnamefont {Wang}},\ and\
  \bibinfo {author} {\bibfnamefont {Y.}~\bibnamefont {Zhu}},\ }\bibfield
  {title} {\bibinfo {title} {{Nonequilibrium electron and lattice dynamics of
  strongly correlated
  ${\mathrm{{B}i}}_{2}{\mathrm{Sr}}_{2}{\mathrm{CaCu}}_{2}{\mathrm{O}}_{8+\ensuremath{\delta}}$
  single crystals}},\ }\href {https://doi.org/10.1126/sciadv.aap7427}
  {\bibfield  {journal} {\bibinfo  {journal} {Science Advances}\ }\textbf
  {\bibinfo {volume} {4}},\ \bibinfo {pages} {eaap7427} (\bibinfo {year}
  {2018})}\BibitemShut {NoStop}%
\bibitem [{\citenamefont {Perfetti}\ \emph {et~al.}(2007)\citenamefont
  {Perfetti}, \citenamefont {Loukakos}, \citenamefont {Lisowski}, \citenamefont
  {Bovensiepen}, \citenamefont {Eisaki},\ and\ \citenamefont
  {Wolf}}]{Perfetti2007ultrafast}%
  \BibitemOpen
  \bibfield  {author} {\bibinfo {author} {\bibfnamefont {L.}~\bibnamefont
  {Perfetti}}, \bibinfo {author} {\bibfnamefont {P.~A.}\ \bibnamefont
  {Loukakos}}, \bibinfo {author} {\bibfnamefont {M.}~\bibnamefont {Lisowski}},
  \bibinfo {author} {\bibfnamefont {U.}~\bibnamefont {Bovensiepen}}, \bibinfo
  {author} {\bibfnamefont {H.}~\bibnamefont {Eisaki}},\ and\ \bibinfo {author}
  {\bibfnamefont {M.}~\bibnamefont {Wolf}},\ }\bibfield  {title} {\bibinfo
  {title} {Ultrafast electron relaxation in superconducting
  {${{\mathrm{Bi}}_{2}{\mathrm{Sr}}_{2}{\mathrm{CaCu}}_{2}{\mathrm{O}}_{8+\ensuremath{\delta}}}$}
  by time-resolved photoelectron spectroscopy},\ }\href
  {https://doi.org/10.1103/PhysRevLett.99.197001} {\bibfield  {journal}
  {\bibinfo  {journal} {Phys. Rev. Lett.}\ }\textbf {\bibinfo {volume} {99}},\
  \bibinfo {pages} {197001} (\bibinfo {year} {2007})}\BibitemShut {NoStop}%
\bibitem [{\citenamefont {Dakovski}\ \emph {et~al.}(2015)\citenamefont
  {Dakovski}, \citenamefont {Durakiewicz}, \citenamefont {Zhu}, \citenamefont
  {Riseborough}, \citenamefont {Gu}, \citenamefont {Gilbertson}, \citenamefont
  {Taylor},\ and\ \citenamefont {Rodriguez}}]{Dakovski2015quasiparticle}%
  \BibitemOpen
  \bibfield  {author} {\bibinfo {author} {\bibfnamefont {G.~L.}\ \bibnamefont
  {Dakovski}}, \bibinfo {author} {\bibfnamefont {T.}~\bibnamefont
  {Durakiewicz}}, \bibinfo {author} {\bibfnamefont {J.-X.}\ \bibnamefont
  {Zhu}}, \bibinfo {author} {\bibfnamefont {P.~S.}\ \bibnamefont
  {Riseborough}}, \bibinfo {author} {\bibfnamefont {G.}~\bibnamefont {Gu}},
  \bibinfo {author} {\bibfnamefont {S.~M.}\ \bibnamefont {Gilbertson}},
  \bibinfo {author} {\bibfnamefont {A.}~\bibnamefont {Taylor}},\ and\ \bibinfo
  {author} {\bibfnamefont {G.}~\bibnamefont {Rodriguez}},\ }\bibfield  {title}
  {\bibinfo {title} {{Quasiparticle dynamics across the full Brillouin zone of
  Bi$_2$Sr$_2$CaCu$_2$O$_{8+\delta}$ traced with ultrafast time and
  angle-resolved photoemission spectroscopy}},\ }\href
  {https://doi.org/10.1063/1.4933133} {\bibfield  {journal} {\bibinfo
  {journal} {Structural Dynamics}\ }\textbf {\bibinfo {volume} {2}},\ \bibinfo
  {pages} {054501} (\bibinfo {year} {2015})}\BibitemShut {NoStop}%
\bibitem [{\citenamefont {Rameau}\ \emph {et~al.}(2016)\citenamefont {Rameau},
  \citenamefont {Freutel}, \citenamefont {Kemper}, \citenamefont {Sentef},
  \citenamefont {Freericks}, \citenamefont {Avigo}, \citenamefont {Ligges},
  \citenamefont {Rettig}, \citenamefont {Yoshida}, \citenamefont {Eisaki},
  \citenamefont {Schneeloch}, \citenamefont {Zhong}, \citenamefont {Xu},
  \citenamefont {Gu}, \citenamefont {Johnson},\ and\ \citenamefont
  {Bovensiepen}}]{Rameau2016energy}%
  \BibitemOpen
  \bibfield  {author} {\bibinfo {author} {\bibfnamefont {J.~D.}\ \bibnamefont
  {Rameau}}, \bibinfo {author} {\bibfnamefont {S.}~\bibnamefont {Freutel}},
  \bibinfo {author} {\bibfnamefont {A.~F.}\ \bibnamefont {Kemper}}, \bibinfo
  {author} {\bibfnamefont {M.~A.}\ \bibnamefont {Sentef}}, \bibinfo {author}
  {\bibfnamefont {J.~K.}\ \bibnamefont {Freericks}}, \bibinfo {author}
  {\bibfnamefont {I.}~\bibnamefont {Avigo}}, \bibinfo {author} {\bibfnamefont
  {M.}~\bibnamefont {Ligges}}, \bibinfo {author} {\bibfnamefont
  {L.}~\bibnamefont {Rettig}}, \bibinfo {author} {\bibfnamefont
  {Y.}~\bibnamefont {Yoshida}}, \bibinfo {author} {\bibfnamefont
  {H.}~\bibnamefont {Eisaki}}, \bibinfo {author} {\bibfnamefont
  {J.}~\bibnamefont {Schneeloch}}, \bibinfo {author} {\bibfnamefont {R.~D.}\
  \bibnamefont {Zhong}}, \bibinfo {author} {\bibfnamefont {Z.~J.}\ \bibnamefont
  {Xu}}, \bibinfo {author} {\bibfnamefont {G.~D.}\ \bibnamefont {Gu}}, \bibinfo
  {author} {\bibfnamefont {P.~D.}\ \bibnamefont {Johnson}},\ and\ \bibinfo
  {author} {\bibfnamefont {U.}~\bibnamefont {Bovensiepen}},\ }\bibfield
  {title} {\bibinfo {title} {{Energy dissipation from a correlated system
  driven out of equilibrium}},\ }\href {https://doi.org/10.1038/ncomms13761}
  {\bibfield  {journal} {\bibinfo  {journal} {Nature Communications}\ }\textbf
  {\bibinfo {volume} {7}},\ \bibinfo {pages} {13761} (\bibinfo {year}
  {2016})}\BibitemShut {NoStop}%
\bibitem [{\citenamefont {Lenar\ifmmode \check{c}\else
  \v{c}\fi{}i\ifmmode~\check{c}\else \v{c}\fi{}}\ and\ \citenamefont
  {Prelov\ifmmode~\check{s}\else \v{s}\fi{}ek}(2013)}]{Lenarcic2013ultrafast}%
  \BibitemOpen
  \bibfield  {author} {\bibinfo {author} {\bibfnamefont {Z.}~\bibnamefont
  {Lenar\ifmmode \check{c}\else \v{c}\fi{}i\ifmmode~\check{c}\else
  \v{c}\fi{}}}\ and\ \bibinfo {author} {\bibfnamefont {P.}~\bibnamefont
  {Prelov\ifmmode~\check{s}\else \v{s}\fi{}ek}},\ }\bibfield  {title} {\bibinfo
  {title} {{Ultrafast Charge Recombination in a Photoexcited Mott-Hubbard
  Insulator}},\ }\href {https://doi.org/10.1103/PhysRevLett.111.016401}
  {\bibfield  {journal} {\bibinfo  {journal} {Phys. Rev. Lett.}\ }\textbf
  {\bibinfo {volume} {111}},\ \bibinfo {pages} {016401} (\bibinfo {year}
  {2013})}\BibitemShut {NoStop}%
\bibitem [{\citenamefont {Novelli}\ \emph {et~al.}(2014)\citenamefont
  {Novelli}, \citenamefont {De~Filippis}, \citenamefont {Cataudella},
  \citenamefont {Esposito}, \citenamefont {Vergara}, \citenamefont {Cilento},
  \citenamefont {Sindici}, \citenamefont {Amaricci}, \citenamefont {Giannetti},
  \citenamefont {Prabhakaran}, \citenamefont {Wall}, \citenamefont {Perucchi},
  \citenamefont {Dal~Conte}, \citenamefont {Cerullo}, \citenamefont {Capone},
  \citenamefont {Mishchenko}, \citenamefont {Gr\:{u}ninger}, \citenamefont
  {Nagaosa}, \citenamefont {Parmigiani},\ and\ \citenamefont
  {Fausti}}]{Novelli2014witnessing}%
  \BibitemOpen
  \bibfield  {author} {\bibinfo {author} {\bibfnamefont {F.}~\bibnamefont
  {Novelli}}, \bibinfo {author} {\bibfnamefont {G.}~\bibnamefont
  {De~Filippis}}, \bibinfo {author} {\bibfnamefont {V.}~\bibnamefont
  {Cataudella}}, \bibinfo {author} {\bibfnamefont {M.}~\bibnamefont
  {Esposito}}, \bibinfo {author} {\bibfnamefont {I.}~\bibnamefont {Vergara}},
  \bibinfo {author} {\bibfnamefont {F.}~\bibnamefont {Cilento}}, \bibinfo
  {author} {\bibfnamefont {E.}~\bibnamefont {Sindici}}, \bibinfo {author}
  {\bibfnamefont {A.}~\bibnamefont {Amaricci}}, \bibinfo {author}
  {\bibfnamefont {C.}~\bibnamefont {Giannetti}}, \bibinfo {author}
  {\bibfnamefont {D.}~\bibnamefont {Prabhakaran}}, \bibinfo {author}
  {\bibfnamefont {S.}~\bibnamefont {Wall}}, \bibinfo {author} {\bibfnamefont
  {A.}~\bibnamefont {Perucchi}}, \bibinfo {author} {\bibfnamefont
  {S.}~\bibnamefont {Dal~Conte}}, \bibinfo {author} {\bibfnamefont
  {G.}~\bibnamefont {Cerullo}}, \bibinfo {author} {\bibfnamefont
  {M.}~\bibnamefont {Capone}}, \bibinfo {author} {\bibfnamefont
  {A.}~\bibnamefont {Mishchenko}}, \bibinfo {author} {\bibfnamefont
  {M.}~\bibnamefont {Gr\:{u}ninger}}, \bibinfo {author} {\bibfnamefont
  {N.}~\bibnamefont {Nagaosa}}, \bibinfo {author} {\bibfnamefont
  {F.}~\bibnamefont {Parmigiani}},\ and\ \bibinfo {author} {\bibfnamefont
  {D.}~\bibnamefont {Fausti}},\ }\bibfield  {title} {\bibinfo {title}
  {Witnessing the formation and relaxation of dressed quasi-particles in a
  strongly correlated electron system},\ }\href
  {https://doi.org/10.1038/ncomms6112} {\bibfield  {journal} {\bibinfo
  {journal} {Nature Communications}\ }\textbf {\bibinfo {volume} {5}},\
  \bibinfo {pages} {5112} (\bibinfo {year} {2014})}\BibitemShut {NoStop}%
\bibitem [{\citenamefont {Cilento}\ \emph {et~al.}(2018)\citenamefont
  {Cilento}, \citenamefont {Manzoni}, \citenamefont {Sterzi}, \citenamefont
  {Peli}, \citenamefont {Ronchi}, \citenamefont {Crepaldi}, \citenamefont
  {Boschini}, \citenamefont {Cacho}, \citenamefont {Chapman}, \citenamefont
  {Springate}, \citenamefont {Eisaki}, \citenamefont {Greven}, \citenamefont
  {Berciu}, \citenamefont {Kemper}, \citenamefont {Damascelli}, \citenamefont
  {Capone}, \citenamefont {Giannetti},\ and\ \citenamefont
  {Parmigiani}}]{Cilento2018dynamics}%
  \BibitemOpen
  \bibfield  {author} {\bibinfo {author} {\bibfnamefont {F.}~\bibnamefont
  {Cilento}}, \bibinfo {author} {\bibfnamefont {G.}~\bibnamefont {Manzoni}},
  \bibinfo {author} {\bibfnamefont {A.}~\bibnamefont {Sterzi}}, \bibinfo
  {author} {\bibfnamefont {S.}~\bibnamefont {Peli}}, \bibinfo {author}
  {\bibfnamefont {A.}~\bibnamefont {Ronchi}}, \bibinfo {author} {\bibfnamefont
  {A.}~\bibnamefont {Crepaldi}}, \bibinfo {author} {\bibfnamefont
  {F.}~\bibnamefont {Boschini}}, \bibinfo {author} {\bibfnamefont
  {C.}~\bibnamefont {Cacho}}, \bibinfo {author} {\bibfnamefont
  {R.}~\bibnamefont {Chapman}}, \bibinfo {author} {\bibfnamefont
  {E.}~\bibnamefont {Springate}}, \bibinfo {author} {\bibfnamefont
  {H.}~\bibnamefont {Eisaki}}, \bibinfo {author} {\bibfnamefont
  {M.}~\bibnamefont {Greven}}, \bibinfo {author} {\bibfnamefont
  {M.}~\bibnamefont {Berciu}}, \bibinfo {author} {\bibfnamefont {A.~F.}\
  \bibnamefont {Kemper}}, \bibinfo {author} {\bibfnamefont {A.}~\bibnamefont
  {Damascelli}}, \bibinfo {author} {\bibfnamefont {M.}~\bibnamefont {Capone}},
  \bibinfo {author} {\bibfnamefont {C.}~\bibnamefont {Giannetti}},\ and\
  \bibinfo {author} {\bibfnamefont {F.}~\bibnamefont {Parmigiani}},\ }\bibfield
   {title} {\bibinfo {title} {Dynamics of correlation-frozen antinodal
  quasiparticles in superconducting cuprates},\ }\bibfield  {journal} {\bibinfo
   {journal} {Science Advances}\ }\textbf {\bibinfo {volume} {4}},\ \href
  {https://doi.org/10.1126/sciadv.aar1998} {10.1126/sciadv.aar1998} (\bibinfo
  {year} {2018})\BibitemShut {NoStop}%
\bibitem [{\citenamefont {Wall}\ \emph {et~al.}(2011)\citenamefont {Wall},
  \citenamefont {Brida}, \citenamefont {Clark}, \citenamefont {Ehrke},
  \citenamefont {Jaksch}, \citenamefont {Ardavan}, \citenamefont {Bonora},
  \citenamefont {Uemura}, \citenamefont {Takahashi}, \citenamefont {Hasegawa},
  \citenamefont {Okamoto}, \citenamefont {Cerullo},\ and\ \citenamefont
  {Cavalleri}}]{Wall2011quantum}%
  \BibitemOpen
  \bibfield  {author} {\bibinfo {author} {\bibfnamefont {S.}~\bibnamefont
  {Wall}}, \bibinfo {author} {\bibfnamefont {D.}~\bibnamefont {Brida}},
  \bibinfo {author} {\bibfnamefont {S.}~\bibnamefont {Clark}}, \bibinfo
  {author} {\bibfnamefont {H.~P.}\ \bibnamefont {Ehrke}}, \bibinfo {author}
  {\bibfnamefont {D.}~\bibnamefont {Jaksch}}, \bibinfo {author} {\bibfnamefont
  {A.}~\bibnamefont {Ardavan}}, \bibinfo {author} {\bibfnamefont
  {S.}~\bibnamefont {Bonora}}, \bibinfo {author} {\bibfnamefont
  {H.}~\bibnamefont {Uemura}}, \bibinfo {author} {\bibfnamefont
  {Y.}~\bibnamefont {Takahashi}}, \bibinfo {author} {\bibfnamefont
  {T.}~\bibnamefont {Hasegawa}}, \bibinfo {author} {\bibfnamefont
  {H.}~\bibnamefont {Okamoto}}, \bibinfo {author} {\bibfnamefont
  {G.}~\bibnamefont {Cerullo}},\ and\ \bibinfo {author} {\bibfnamefont
  {A.}~\bibnamefont {Cavalleri}},\ }\bibfield  {title} {\bibinfo {title}
  {{Quantum interference between charge excitation paths in a solid-state Mott
  insulator}},\ }\href {https://doi.org/10.1038/nphys1831} {\bibfield
  {journal} {\bibinfo  {journal} {Nature Physics}\ }\textbf {\bibinfo {volume}
  {7}},\ \bibinfo {pages} {114} (\bibinfo {year} {2011})}\BibitemShut {NoStop}%
\bibitem [{\citenamefont {Mitrano}\ \emph {et~al.}(2014)\citenamefont
  {Mitrano}, \citenamefont {Cotugno}, \citenamefont {Clark}, \citenamefont
  {Singla}, \citenamefont {Kaiser}, \citenamefont {St{\"{a}}hler},
  \citenamefont {Beyer}, \citenamefont {Dressel}, \citenamefont {Baldassarre},
  \citenamefont {Nicoletti}, \citenamefont {Perucchi}, \citenamefont
  {Hasegawa}, \citenamefont {Okamoto}, \citenamefont {Jaksch},\ and\
  \citenamefont {Cavalleri}}]{Mitrano2014pressure}%
  \BibitemOpen
  \bibfield  {author} {\bibinfo {author} {\bibfnamefont {M.}~\bibnamefont
  {Mitrano}}, \bibinfo {author} {\bibfnamefont {G.}~\bibnamefont {Cotugno}},
  \bibinfo {author} {\bibfnamefont {S.~R.}\ \bibnamefont {Clark}}, \bibinfo
  {author} {\bibfnamefont {R.}~\bibnamefont {Singla}}, \bibinfo {author}
  {\bibfnamefont {S.}~\bibnamefont {Kaiser}}, \bibinfo {author} {\bibfnamefont
  {J.}~\bibnamefont {St{\"{a}}hler}}, \bibinfo {author} {\bibfnamefont
  {R.}~\bibnamefont {Beyer}}, \bibinfo {author} {\bibfnamefont
  {M.}~\bibnamefont {Dressel}}, \bibinfo {author} {\bibfnamefont
  {L.}~\bibnamefont {Baldassarre}}, \bibinfo {author} {\bibfnamefont
  {D.}~\bibnamefont {Nicoletti}}, \bibinfo {author} {\bibfnamefont
  {A.}~\bibnamefont {Perucchi}}, \bibinfo {author} {\bibfnamefont
  {T.}~\bibnamefont {Hasegawa}}, \bibinfo {author} {\bibfnamefont
  {H.}~\bibnamefont {Okamoto}}, \bibinfo {author} {\bibfnamefont
  {D.}~\bibnamefont {Jaksch}},\ and\ \bibinfo {author} {\bibfnamefont
  {A.}~\bibnamefont {Cavalleri}},\ }\bibfield  {title} {\bibinfo {title}
  {{Pressure-Dependent Relaxation in the Photoexcited Mott Insulator
  {$\mathrm{ET}-{\mathrm{F}}_{2}\mathrm{TCNQ}$}: Influence of Hopping and
  Correlations on Quasiparticle Recombination Rates}},\ }\href
  {https://doi.org/10.1103/PhysRevLett.112.117801} {\bibfield  {journal}
  {\bibinfo  {journal} {Phys. Rev. Lett.}\ }\textbf {\bibinfo {volume} {112}},\
  \bibinfo {pages} {117801} (\bibinfo {year} {2014})}\BibitemShut {NoStop}%
\bibitem [{\citenamefont {Strohmaier}\ \emph {et~al.}(2010)\citenamefont
  {Strohmaier}, \citenamefont {Greif}, \citenamefont {J\"ordens}, \citenamefont
  {Tarruell}, \citenamefont {Moritz}, \citenamefont {Esslinger}, \citenamefont
  {Sensarma}, \citenamefont {Pekker}, \citenamefont {Altman},\ and\
  \citenamefont {Demler}}]{strohmaier2010observation}%
  \BibitemOpen
  \bibfield  {author} {\bibinfo {author} {\bibfnamefont {N.}~\bibnamefont
  {Strohmaier}}, \bibinfo {author} {\bibfnamefont {D.}~\bibnamefont {Greif}},
  \bibinfo {author} {\bibfnamefont {R.}~\bibnamefont {J\"ordens}}, \bibinfo
  {author} {\bibfnamefont {L.}~\bibnamefont {Tarruell}}, \bibinfo {author}
  {\bibfnamefont {H.}~\bibnamefont {Moritz}}, \bibinfo {author} {\bibfnamefont
  {T.}~\bibnamefont {Esslinger}}, \bibinfo {author} {\bibfnamefont
  {R.}~\bibnamefont {Sensarma}}, \bibinfo {author} {\bibfnamefont
  {D.}~\bibnamefont {Pekker}}, \bibinfo {author} {\bibfnamefont
  {E.}~\bibnamefont {Altman}},\ and\ \bibinfo {author} {\bibfnamefont
  {E.}~\bibnamefont {Demler}},\ }\bibfield  {title} {\bibinfo {title}
  {{Observation of Elastic Doublon Decay in the Fermi-Hubbard Model}},\ }\href
  {https://doi.org/10.1103/PhysRevLett.104.080401} {\bibfield  {journal}
  {\bibinfo  {journal} {Phys. Rev. Lett.}\ }\textbf {\bibinfo {volume} {104}},\
  \bibinfo {pages} {080401} (\bibinfo {year} {2010})}\BibitemShut {NoStop}%
\bibitem [{\citenamefont {Sensarma}\ \emph {et~al.}(2010)\citenamefont
  {Sensarma}, \citenamefont {Pekker}, \citenamefont {Altman}, \citenamefont
  {Demler}, \citenamefont {Strohmaier}, \citenamefont {Greif}, \citenamefont
  {J\"ordens}, \citenamefont {Tarruell}, \citenamefont {Moritz},\ and\
  \citenamefont {Esslinger}}]{Sensarma2010lifetime}%
  \BibitemOpen
  \bibfield  {author} {\bibinfo {author} {\bibfnamefont {R.}~\bibnamefont
  {Sensarma}}, \bibinfo {author} {\bibfnamefont {D.}~\bibnamefont {Pekker}},
  \bibinfo {author} {\bibfnamefont {E.}~\bibnamefont {Altman}}, \bibinfo
  {author} {\bibfnamefont {E.}~\bibnamefont {Demler}}, \bibinfo {author}
  {\bibfnamefont {N.}~\bibnamefont {Strohmaier}}, \bibinfo {author}
  {\bibfnamefont {D.}~\bibnamefont {Greif}}, \bibinfo {author} {\bibfnamefont
  {R.}~\bibnamefont {J\"ordens}}, \bibinfo {author} {\bibfnamefont
  {L.}~\bibnamefont {Tarruell}}, \bibinfo {author} {\bibfnamefont
  {H.}~\bibnamefont {Moritz}},\ and\ \bibinfo {author} {\bibfnamefont
  {T.}~\bibnamefont {Esslinger}},\ }\bibfield  {title} {\bibinfo {title}
  {Lifetime of double occupancies in the {Fermi-Hubbard} model},\ }\href
  {https://doi.org/10.1103/PhysRevB.82.224302} {\bibfield  {journal} {\bibinfo
  {journal} {Phys. Rev. B}\ }\textbf {\bibinfo {volume} {82}},\ \bibinfo
  {pages} {224302} (\bibinfo {year} {2010})}\BibitemShut {NoStop}%
\bibitem [{\citenamefont {Lenar\ifmmode \check{c}\else
  \v{c}\fi{}i\ifmmode~\check{c}\else \v{c}\fi{}}\ and\ \citenamefont
  {Prelov\ifmmode~\check{s}\else \v{s}\fi{}ek}(2014)}]{Lenarcic2014charge}%
  \BibitemOpen
  \bibfield  {author} {\bibinfo {author} {\bibfnamefont {Z.}~\bibnamefont
  {Lenar\ifmmode \check{c}\else \v{c}\fi{}i\ifmmode~\check{c}\else
  \v{c}\fi{}}}\ and\ \bibinfo {author} {\bibfnamefont {P.}~\bibnamefont
  {Prelov\ifmmode~\check{s}\else \v{s}\fi{}ek}},\ }\bibfield  {title} {\bibinfo
  {title} {Charge recombination in undoped cuprates},\ }\href
  {https://doi.org/10.1103/PhysRevB.90.235136} {\bibfield  {journal} {\bibinfo
  {journal} {Phys. Rev. B}\ }\textbf {\bibinfo {volume} {90}},\ \bibinfo
  {pages} {235136} (\bibinfo {year} {2014})}\BibitemShut {NoStop}%
\bibitem [{\citenamefont {Higley}\ \emph {et~al.}(2019)\citenamefont {Higley},
  \citenamefont {Reid}, \citenamefont {Chen}, \citenamefont {Guyader},
  \citenamefont {Hellwig}, \citenamefont {Lutman}, \citenamefont {Liu},
  \citenamefont {Shafer}, \citenamefont {Chase}, \citenamefont {Dakovski},
  \citenamefont {Mitra}, \citenamefont {Yuan}, \citenamefont {Schlappa},
  \citenamefont {D\"{u}rr}, \citenamefont {Schlotter},\ and\ \citenamefont
  {St\"{o}hr}}]{Higley2019femtosecond}%
  \BibitemOpen
  \bibfield  {author} {\bibinfo {author} {\bibfnamefont {D.~J.}\ \bibnamefont
  {Higley}}, \bibinfo {author} {\bibfnamefont {A.~H.}\ \bibnamefont {Reid}},
  \bibinfo {author} {\bibfnamefont {Z.}~\bibnamefont {Chen}}, \bibinfo {author}
  {\bibfnamefont {L.~L.}\ \bibnamefont {Guyader}}, \bibinfo {author}
  {\bibfnamefont {O.}~\bibnamefont {Hellwig}}, \bibinfo {author} {\bibfnamefont
  {A.~A.}\ \bibnamefont {Lutman}}, \bibinfo {author} {\bibfnamefont
  {T.}~\bibnamefont {Liu}}, \bibinfo {author} {\bibfnamefont {P.}~\bibnamefont
  {Shafer}}, \bibinfo {author} {\bibfnamefont {T.}~\bibnamefont {Chase}},
  \bibinfo {author} {\bibfnamefont {G.~L.}\ \bibnamefont {Dakovski}}, \bibinfo
  {author} {\bibfnamefont {A.}~\bibnamefont {Mitra}}, \bibinfo {author}
  {\bibfnamefont {E.}~\bibnamefont {Yuan}}, \bibinfo {author} {\bibfnamefont
  {J.}~\bibnamefont {Schlappa}}, \bibinfo {author} {\bibfnamefont {H.~A.}\
  \bibnamefont {D\"{u}rr}}, \bibinfo {author} {\bibfnamefont {W.~F.}\
  \bibnamefont {Schlotter}},\ and\ \bibinfo {author} {\bibfnamefont
  {J.}~\bibnamefont {St\"{o}hr}},\ }\bibfield  {title} {\bibinfo {title}
  {Femtosecond x-ray induced changes of the electronic and magnetic response of
  solids from electron redistribution},\ }\href
  {https://doi.org/10.1038/s41467-019-13272-5} {\bibfield  {journal} {\bibinfo
  {journal} {Nature Communications}\ }\textbf {\bibinfo {volume} {10}},\
  \bibinfo {pages} {5289} (\bibinfo {year} {2019})}\BibitemShut {NoStop}%
\bibitem [{\citenamefont {Mansart}\ \emph {et~al.}(2013)\citenamefont
  {Mansart}, \citenamefont {Cottet}, \citenamefont {Mancini}, \citenamefont
  {Jarlborg}, \citenamefont {Dugdale}, \citenamefont {Johnson}, \citenamefont
  {Mariager}, \citenamefont {Milne}, \citenamefont {Beaud}, \citenamefont
  {Gr\"ubel}, \citenamefont {Johnson}, \citenamefont {Kubacka}, \citenamefont
  {Ingold}, \citenamefont {Prsa}, \citenamefont {R\o{}nnow}, \citenamefont
  {Conder}, \citenamefont {Pomjakushina}, \citenamefont {Chergui},\ and\
  \citenamefont {Carbone}}]{Mansart2013temperature}%
  \BibitemOpen
  \bibfield  {author} {\bibinfo {author} {\bibfnamefont {B.}~\bibnamefont
  {Mansart}}, \bibinfo {author} {\bibfnamefont {M.~J.~G.}\ \bibnamefont
  {Cottet}}, \bibinfo {author} {\bibfnamefont {G.~F.}\ \bibnamefont {Mancini}},
  \bibinfo {author} {\bibfnamefont {T.}~\bibnamefont {Jarlborg}}, \bibinfo
  {author} {\bibfnamefont {S.~B.}\ \bibnamefont {Dugdale}}, \bibinfo {author}
  {\bibfnamefont {S.~L.}\ \bibnamefont {Johnson}}, \bibinfo {author}
  {\bibfnamefont {S.~O.}\ \bibnamefont {Mariager}}, \bibinfo {author}
  {\bibfnamefont {C.~J.}\ \bibnamefont {Milne}}, \bibinfo {author}
  {\bibfnamefont {P.}~\bibnamefont {Beaud}}, \bibinfo {author} {\bibfnamefont
  {S.}~\bibnamefont {Gr\"ubel}}, \bibinfo {author} {\bibfnamefont {J.~A.}\
  \bibnamefont {Johnson}}, \bibinfo {author} {\bibfnamefont {T.}~\bibnamefont
  {Kubacka}}, \bibinfo {author} {\bibfnamefont {G.}~\bibnamefont {Ingold}},
  \bibinfo {author} {\bibfnamefont {K.}~\bibnamefont {Prsa}}, \bibinfo {author}
  {\bibfnamefont {H.~M.}\ \bibnamefont {R\o{}nnow}}, \bibinfo {author}
  {\bibfnamefont {K.}~\bibnamefont {Conder}}, \bibinfo {author} {\bibfnamefont
  {E.}~\bibnamefont {Pomjakushina}}, \bibinfo {author} {\bibfnamefont
  {M.}~\bibnamefont {Chergui}},\ and\ \bibinfo {author} {\bibfnamefont
  {F.}~\bibnamefont {Carbone}},\ }\bibfield  {title} {\bibinfo {title}
  {Temperature-dependent electron-phonon coupling in
  {La${}_{2\ensuremath{-}x}$Sr${}_{x}$CuO${}_{4}$} probed by femtosecond x-ray
  diffraction},\ }\href {https://doi.org/10.1103/PhysRevB.88.054507} {\bibfield
   {journal} {\bibinfo  {journal} {Phys. Rev. B}\ }\textbf {\bibinfo {volume}
  {88}},\ \bibinfo {pages} {054507} (\bibinfo {year} {2013})}\BibitemShut
  {NoStop}%
\bibitem [{\citenamefont {Mahan}(2000)}]{Mahan2000many}%
  \BibitemOpen
  \bibfield  {author} {\bibinfo {author} {\bibfnamefont {G.~D.}\ \bibnamefont
  {Mahan}},\ }\href@noop {} {\emph {\bibinfo {title} {{Many-Particle
  Physics}}}},\ \bibinfo {edition} {3rd}\ ed.\ (\bibinfo  {publisher}
  {Plenum},\ \bibinfo {address} {New York, N.Y.},\ \bibinfo {year}
  {2000})\BibitemShut {NoStop}%
\bibitem [{\citenamefont {Hybertsen}\ \emph {et~al.}(1992)\citenamefont
  {Hybertsen}, \citenamefont {Stechel}, \citenamefont {Foulkes},\ and\
  \citenamefont {Schl\"uter}}]{Hybertsen1992model}%
  \BibitemOpen
  \bibfield  {author} {\bibinfo {author} {\bibfnamefont {M.~S.}\ \bibnamefont
  {Hybertsen}}, \bibinfo {author} {\bibfnamefont {E.~B.}\ \bibnamefont
  {Stechel}}, \bibinfo {author} {\bibfnamefont {W.~M.~C.}\ \bibnamefont
  {Foulkes}},\ and\ \bibinfo {author} {\bibfnamefont {M.}~\bibnamefont
  {Schl\"uter}},\ }\bibfield  {title} {\bibinfo {title} {Model for low-energy
  electronic states probed by x-ray absorption in
  {high-${\mathit{T}}_{\mathit{c}}$} cuprates},\ }\href
  {https://doi.org/10.1103/PhysRevB.45.10032} {\bibfield  {journal} {\bibinfo
  {journal} {Phys. Rev. B}\ }\textbf {\bibinfo {volume} {45}},\ \bibinfo
  {pages} {10032} (\bibinfo {year} {1992})}\BibitemShut {NoStop}%
\bibitem [{\citenamefont {Ament}(2010)}]{Ament2010resonant}%
  \BibitemOpen
  \bibfield  {author} {\bibinfo {author} {\bibfnamefont {L.~J.~P.}\
  \bibnamefont {Ament}},\ }\emph {\bibinfo {title} {{Resonant inelastic x-ray
  scattering studies of elementary excitations }}},\ \href@noop {} {Ph.D.
  thesis},\ \bibinfo  {school} {Leiden University} (\bibinfo {year}
  {2010})\BibitemShut {NoStop}%
\bibitem [{\citenamefont {Mattheiss}(1987)}]{Mattheiss1987}%
  \BibitemOpen
  \bibfield  {author} {\bibinfo {author} {\bibfnamefont {L.~F.}\ \bibnamefont
  {Mattheiss}},\ }\bibfield  {title} {\bibinfo {title} {{Electronic band
  properties and superconductivity in
  ${\mathrm{La}}_{2\mathrm{\ensuremath{-}}\mathrm{y}}$${\mathrm{X}}_{\mathrm{y}}$${\mathrm{CuO}}_{4}$}},\
  }\href {https://doi.org/10.1103/PhysRevLett.58.1028} {\bibfield  {journal}
  {\bibinfo  {journal} {Phys. Rev. Lett.}\ }\textbf {\bibinfo {volume} {58}},\
  \bibinfo {pages} {1028} (\bibinfo {year} {1987})}\BibitemShut {NoStop}%
\bibitem [{\citenamefont {Emery}(1987)}]{Emery1987}%
  \BibitemOpen
  \bibfield  {author} {\bibinfo {author} {\bibfnamefont {V.~J.}\ \bibnamefont
  {Emery}},\ }\bibfield  {title} {\bibinfo {title} {{Theory of
  high-${\mathrm{T}}_{\mathrm{c}}$ superconductivity in oxides}},\ }\href
  {https://doi.org/10.1103/PhysRevLett.58.2794} {\bibfield  {journal} {\bibinfo
   {journal} {Phys. Rev. Lett.}\ }\textbf {\bibinfo {volume} {58}},\ \bibinfo
  {pages} {2794} (\bibinfo {year} {1987})}\BibitemShut {NoStop}%
\bibitem [{\citenamefont {Varma}\ \emph {et~al.}(1987)\citenamefont {Varma},
  \citenamefont {Schmitt-Rink},\ and\ \citenamefont {Abrahams}}]{Varma1987}%
  \BibitemOpen
  \bibfield  {author} {\bibinfo {author} {\bibfnamefont {C.~M.}\ \bibnamefont
  {Varma}}, \bibinfo {author} {\bibfnamefont {S.}~\bibnamefont
  {Schmitt-Rink}},\ and\ \bibinfo {author} {\bibfnamefont {E.}~\bibnamefont
  {Abrahams}},\ }\bibfield  {title} {\bibinfo {title} {{Charge transfer
  excitations and superconductivity in “ionic” metals}},\ }\href
  {https://doi.org/https://doi.org/10.1016/0038-1098(87)90407-8} {\bibfield
  {journal} {\bibinfo  {journal} {Solid State Communications}\ }\textbf
  {\bibinfo {volume} {62}},\ \bibinfo {pages} {681} (\bibinfo {year}
  {1987})}\BibitemShut {NoStop}%
\bibitem [{\citenamefont {Chen}\ \emph {et~al.}(2013)\citenamefont {Chen},
  \citenamefont {Sentef}, \citenamefont {Kung}, \citenamefont {Jia},
  \citenamefont {Thomale}, \citenamefont {Moritz}, \citenamefont {Kampf},\ and\
  \citenamefont {Devereaux}}]{Chen2013b}%
  \BibitemOpen
  \bibfield  {author} {\bibinfo {author} {\bibfnamefont {C.-C.}\ \bibnamefont
  {Chen}}, \bibinfo {author} {\bibfnamefont {M.}~\bibnamefont {Sentef}},
  \bibinfo {author} {\bibfnamefont {Y.~F.}\ \bibnamefont {Kung}}, \bibinfo
  {author} {\bibfnamefont {C.~J.}\ \bibnamefont {Jia}}, \bibinfo {author}
  {\bibfnamefont {R.}~\bibnamefont {Thomale}}, \bibinfo {author} {\bibfnamefont
  {B.}~\bibnamefont {Moritz}}, \bibinfo {author} {\bibfnamefont {A.~P.}\
  \bibnamefont {Kampf}},\ and\ \bibinfo {author} {\bibfnamefont {T.~P.}\
  \bibnamefont {Devereaux}},\ }\bibfield  {title} {\bibinfo {title} {{Doping
  evolution of the oxygen $K$-edge x-ray absorption spectra of cuprate
  superconductors using a three-orbital Hubbard model}},\ }\href
  {https://doi.org/10.1103/PhysRevB.87.165144} {\bibfield  {journal} {\bibinfo
  {journal} {Physical Review B}\ }\textbf {\bibinfo {volume} {87}},\ \bibinfo
  {pages} {165144} (\bibinfo {year} {2013})}\BibitemShut {NoStop}%
\bibitem [{\citenamefont {Nicolas}\ and\ \citenamefont
  {Miron}(2012)}]{Nicolas2012lifetime}%
  \BibitemOpen
  \bibfield  {author} {\bibinfo {author} {\bibfnamefont {C.}~\bibnamefont
  {Nicolas}}\ and\ \bibinfo {author} {\bibfnamefont {C.}~\bibnamefont
  {Miron}},\ }\bibfield  {title} {\bibinfo {title} {Lifetime broadening of
  core-excited and -ionized states},\ }\href
  {https://doi.org/https://doi.org/10.1016/j.elspec.2012.05.008} {\bibfield
  {journal} {\bibinfo  {journal} {Journal of Electron Spectroscopy and Related
  Phenomena}\ }\textbf {\bibinfo {volume} {185}},\ \bibinfo {pages} {267}
  (\bibinfo {year} {2012})},\ \bibinfo {note} {special Issue in honor of Prof.
  T. Darrah Thomas: High-Resolution Spectroscopy of Isolated
  Species}\BibitemShut {NoStop}%
\bibitem [{\citenamefont {Rossi}\ \emph {et~al.}(2019)\citenamefont {Rossi},
  \citenamefont {Arpaia}, \citenamefont {Fumagalli}, \citenamefont
  {Moretti~Sala}, \citenamefont {Betto}, \citenamefont {Kummer}, \citenamefont
  {De~Luca}, \citenamefont {van~den Brink}, \citenamefont {Salluzzo},
  \citenamefont {Brookes}, \citenamefont {Braicovich},\ and\ \citenamefont
  {Ghiringhelli}}]{Rossi2019experimental}%
  \BibitemOpen
  \bibfield  {author} {\bibinfo {author} {\bibfnamefont {M.}~\bibnamefont
  {Rossi}}, \bibinfo {author} {\bibfnamefont {R.}~\bibnamefont {Arpaia}},
  \bibinfo {author} {\bibfnamefont {R.}~\bibnamefont {Fumagalli}}, \bibinfo
  {author} {\bibfnamefont {M.}~\bibnamefont {Moretti~Sala}}, \bibinfo {author}
  {\bibfnamefont {D.}~\bibnamefont {Betto}}, \bibinfo {author} {\bibfnamefont
  {K.}~\bibnamefont {Kummer}}, \bibinfo {author} {\bibfnamefont {G.~M.}\
  \bibnamefont {De~Luca}}, \bibinfo {author} {\bibfnamefont {J.}~\bibnamefont
  {van~den Brink}}, \bibinfo {author} {\bibfnamefont {M.}~\bibnamefont
  {Salluzzo}}, \bibinfo {author} {\bibfnamefont {N.~B.}\ \bibnamefont
  {Brookes}}, \bibinfo {author} {\bibfnamefont {L.}~\bibnamefont
  {Braicovich}},\ and\ \bibinfo {author} {\bibfnamefont {G.}~\bibnamefont
  {Ghiringhelli}},\ }\bibfield  {title} {\bibinfo {title} {Experimental
  determination of momentum-resolved electron-phonon coupling},\ }\href
  {https://doi.org/10.1103/PhysRevLett.123.027001} {\bibfield  {journal}
  {\bibinfo  {journal} {Phys. Rev. Lett.}\ }\textbf {\bibinfo {volume} {123}},\
  \bibinfo {pages} {027001} (\bibinfo {year} {2019})}\BibitemShut {NoStop}%
\bibitem [{\citenamefont {Maekawa}\ \emph {et~al.}(2004)\citenamefont
  {Maekawa}, \citenamefont {Tohyama}, \citenamefont {Barnes}, \citenamefont
  {Ishihara}, \citenamefont {Koshibae},\ and\ \citenamefont
  {Khaliullin}}]{Maekawa2004physics}%
  \BibitemOpen
  \bibfield  {author} {\bibinfo {author} {\bibfnamefont {S.}~\bibnamefont
  {Maekawa}}, \bibinfo {author} {\bibfnamefont {T.}~\bibnamefont {Tohyama}},
  \bibinfo {author} {\bibfnamefont {S.~E.}\ \bibnamefont {Barnes}}, \bibinfo
  {author} {\bibfnamefont {S.}~\bibnamefont {Ishihara}}, \bibinfo {author}
  {\bibfnamefont {W.}~\bibnamefont {Koshibae}},\ and\ \bibinfo {author}
  {\bibfnamefont {G.}~\bibnamefont {Khaliullin}},\ }\href@noop {} {\emph
  {\bibinfo {title} {{Physics of Transition Metal Oxides}}}},\ \bibinfo
  {edition} {1st}\ ed.\ (\bibinfo  {publisher} {Springer, Berlin, Heidelberg},\
  \bibinfo {address} {New York, N.Y.},\ \bibinfo {year} {2004})\BibitemShut
  {NoStop}%
\bibitem [{\citenamefont {Kung}\ \emph {et~al.}(2016)\citenamefont {Kung},
  \citenamefont {Chen}, \citenamefont {Wang}, \citenamefont {Huang},
  \citenamefont {Nowadnick}, \citenamefont {Moritz}, \citenamefont {Scalettar},
  \citenamefont {Johnston},\ and\ \citenamefont
  {Devereaux}}]{Kung2016characterizing}%
  \BibitemOpen
  \bibfield  {author} {\bibinfo {author} {\bibfnamefont {Y.~F.}\ \bibnamefont
  {Kung}}, \bibinfo {author} {\bibfnamefont {C.-C.}\ \bibnamefont {Chen}},
  \bibinfo {author} {\bibfnamefont {Y.}~\bibnamefont {Wang}}, \bibinfo {author}
  {\bibfnamefont {E.~W.}\ \bibnamefont {Huang}}, \bibinfo {author}
  {\bibfnamefont {E.~A.}\ \bibnamefont {Nowadnick}}, \bibinfo {author}
  {\bibfnamefont {B.}~\bibnamefont {Moritz}}, \bibinfo {author} {\bibfnamefont
  {R.~T.}\ \bibnamefont {Scalettar}}, \bibinfo {author} {\bibfnamefont
  {S.}~\bibnamefont {Johnston}},\ and\ \bibinfo {author} {\bibfnamefont
  {T.~P.}\ \bibnamefont {Devereaux}},\ }\bibfield  {title} {\bibinfo {title}
  {{Characterizing the three-orbital Hubbard model with determinant quantum
  Monte Carlo}},\ }\href {https://doi.org/10.1103/PhysRevB.93.155166}
  {\bibfield  {journal} {\bibinfo  {journal} {Phys. Rev. B}\ }\textbf {\bibinfo
  {volume} {93}},\ \bibinfo {pages} {155166} (\bibinfo {year}
  {2016})}\BibitemShut {NoStop}%
\bibitem [{\citenamefont {Tsutsui}\ and\ \citenamefont
  {Tohyama}(2016)}]{Tsutsui2016incident}%
  \BibitemOpen
  \bibfield  {author} {\bibinfo {author} {\bibfnamefont {K.}~\bibnamefont
  {Tsutsui}}\ and\ \bibinfo {author} {\bibfnamefont {T.}~\bibnamefont
  {Tohyama}},\ }\bibfield  {title} {\bibinfo {title} {Incident-energy-dependent
  spectral weight of resonant inelastic x-ray scattering in doped cuprates},\
  }\href {https://doi.org/10.1103/PhysRevB.94.085144} {\bibfield  {journal}
  {\bibinfo  {journal} {Phys. Rev. B}\ }\textbf {\bibinfo {volume} {94}},\
  \bibinfo {pages} {085144} (\bibinfo {year} {2016})}\BibitemShut {NoStop}%
\bibitem [{\citenamefont {Tsutsui}\ \emph {et~al.}(2021)\citenamefont
  {Tsutsui}, \citenamefont {Shinjo},\ and\ \citenamefont
  {Tohyama}}]{Tsutsui2021antiphase}%
  \BibitemOpen
  \bibfield  {author} {\bibinfo {author} {\bibfnamefont {K.}~\bibnamefont
  {Tsutsui}}, \bibinfo {author} {\bibfnamefont {K.}~\bibnamefont {Shinjo}},\
  and\ \bibinfo {author} {\bibfnamefont {T.}~\bibnamefont {Tohyama}},\
  }\bibfield  {title} {\bibinfo {title} {{Antiphase Oscillations in the
  Time-Resolved Spin Structure Factor of a Photoexcited Mott Insulator}},\
  }\href {https://doi.org/10.1103/PhysRevLett.126.127404} {\bibfield  {journal}
  {\bibinfo  {journal} {Phys. Rev. Lett.}\ }\textbf {\bibinfo {volume} {126}},\
  \bibinfo {pages} {127404} (\bibinfo {year} {2021})}\BibitemShut {NoStop}%
\bibitem [{\citenamefont {Buades}\ \emph {et~al.}(2021)\citenamefont {Buades},
  \citenamefont {Picón}, \citenamefont {Berger}, \citenamefont {León},
  \citenamefont {Di~Palo}, \citenamefont {Cousin}, \citenamefont {Cocchi},
  \citenamefont {Pellegrin}, \citenamefont {Martin}, \citenamefont
  {Mañas-Valero}, \citenamefont {Coronado}, \citenamefont {Danz},
  \citenamefont {Draxl}, \citenamefont {Uemoto}, \citenamefont {Yabana},
  \citenamefont {Schultze}, \citenamefont {Wall}, \citenamefont {Zürch},\ and\
  \citenamefont {Biegert}}]{Buades2021attosecond}%
  \BibitemOpen
  \bibfield  {author} {\bibinfo {author} {\bibfnamefont {B.}~\bibnamefont
  {Buades}}, \bibinfo {author} {\bibfnamefont {A.}~\bibnamefont {Picón}},
  \bibinfo {author} {\bibfnamefont {E.}~\bibnamefont {Berger}}, \bibinfo
  {author} {\bibfnamefont {I.}~\bibnamefont {León}}, \bibinfo {author}
  {\bibfnamefont {N.}~\bibnamefont {Di~Palo}}, \bibinfo {author} {\bibfnamefont
  {S.~L.}\ \bibnamefont {Cousin}}, \bibinfo {author} {\bibfnamefont
  {C.}~\bibnamefont {Cocchi}}, \bibinfo {author} {\bibfnamefont
  {E.}~\bibnamefont {Pellegrin}}, \bibinfo {author} {\bibfnamefont {J.~H.}\
  \bibnamefont {Martin}}, \bibinfo {author} {\bibfnamefont {S.}~\bibnamefont
  {Mañas-Valero}}, \bibinfo {author} {\bibfnamefont {E.}~\bibnamefont
  {Coronado}}, \bibinfo {author} {\bibfnamefont {T.}~\bibnamefont {Danz}},
  \bibinfo {author} {\bibfnamefont {C.}~\bibnamefont {Draxl}}, \bibinfo
  {author} {\bibfnamefont {M.}~\bibnamefont {Uemoto}}, \bibinfo {author}
  {\bibfnamefont {K.}~\bibnamefont {Yabana}}, \bibinfo {author} {\bibfnamefont
  {M.}~\bibnamefont {Schultze}}, \bibinfo {author} {\bibfnamefont
  {S.}~\bibnamefont {Wall}}, \bibinfo {author} {\bibfnamefont {M.}~\bibnamefont
  {Zürch}},\ and\ \bibinfo {author} {\bibfnamefont {J.}~\bibnamefont
  {Biegert}},\ }\bibfield  {title} {\bibinfo {title} {Attosecond state-resolved
  carrier motion in quantum materials probed by soft x-ray {XANES}},\ }\href
  {https://doi.org/10.1063/5.0020649} {\bibfield  {journal} {\bibinfo
  {journal} {Applied Physics Reviews}\ }\textbf {\bibinfo {volume} {8}},\
  \bibinfo {pages} {011408} (\bibinfo {year} {2021})}\BibitemShut {NoStop}%
\bibitem [{\citenamefont {Hu}\ \emph {et~al.}(2014)\citenamefont {Hu},
  \citenamefont {Kaiser}, \citenamefont {Nicoletti}, \citenamefont {Hunt},
  \citenamefont {Gierz}, \citenamefont {Hoffmann}, \citenamefont {Le~Tacon},
  \citenamefont {Loew}, \citenamefont {Keimer},\ and\ \citenamefont
  {Cavalleri}}]{Hu2014}%
  \BibitemOpen
  \bibfield  {author} {\bibinfo {author} {\bibfnamefont {W.}~\bibnamefont
  {Hu}}, \bibinfo {author} {\bibfnamefont {S.}~\bibnamefont {Kaiser}}, \bibinfo
  {author} {\bibfnamefont {D.}~\bibnamefont {Nicoletti}}, \bibinfo {author}
  {\bibfnamefont {C.~R.}\ \bibnamefont {Hunt}}, \bibinfo {author}
  {\bibfnamefont {I.}~\bibnamefont {Gierz}}, \bibinfo {author} {\bibfnamefont
  {M.~C.}\ \bibnamefont {Hoffmann}}, \bibinfo {author} {\bibfnamefont
  {M.}~\bibnamefont {Le~Tacon}}, \bibinfo {author} {\bibfnamefont
  {T.}~\bibnamefont {Loew}}, \bibinfo {author} {\bibfnamefont {B.}~\bibnamefont
  {Keimer}},\ and\ \bibinfo {author} {\bibfnamefont {A.}~\bibnamefont
  {Cavalleri}},\ }\bibfield  {title} {\bibinfo {title} {{Optically enhanced
  coherent transport in {YBa$_2$Cu$_3$O$_{6.5}$} by ultrafast redistribution of
  interlayer coupling}},\ }\href {https://doi.org/10.1038/nmat3963} {\bibfield
  {journal} {\bibinfo  {journal} {Nature Materials}\ }\textbf {\bibinfo
  {volume} {13}},\ \bibinfo {pages} {705} (\bibinfo {year} {2014})}\BibitemShut
  {NoStop}%
\bibitem [{\citenamefont {Kaiser}\ \emph
  {et~al.}(2014{\natexlab{b}})\citenamefont {Kaiser}, \citenamefont {Hunt},
  \citenamefont {Nicoletti}, \citenamefont {Hu}, \citenamefont {Gierz},
  \citenamefont {Liu}, \citenamefont {Le~Tacon}, \citenamefont {Loew},
  \citenamefont {Haug}, \citenamefont {Keimer},\ and\ \citenamefont
  {Cavalleri}}]{Kaiser2014optically}%
  \BibitemOpen
  \bibfield  {author} {\bibinfo {author} {\bibfnamefont {S.}~\bibnamefont
  {Kaiser}}, \bibinfo {author} {\bibfnamefont {C.~R.}\ \bibnamefont {Hunt}},
  \bibinfo {author} {\bibfnamefont {D.}~\bibnamefont {Nicoletti}}, \bibinfo
  {author} {\bibfnamefont {W.}~\bibnamefont {Hu}}, \bibinfo {author}
  {\bibfnamefont {I.}~\bibnamefont {Gierz}}, \bibinfo {author} {\bibfnamefont
  {H.~Y.}\ \bibnamefont {Liu}}, \bibinfo {author} {\bibfnamefont
  {M.}~\bibnamefont {Le~Tacon}}, \bibinfo {author} {\bibfnamefont
  {T.}~\bibnamefont {Loew}}, \bibinfo {author} {\bibfnamefont {D.}~\bibnamefont
  {Haug}}, \bibinfo {author} {\bibfnamefont {B.}~\bibnamefont {Keimer}},\ and\
  \bibinfo {author} {\bibfnamefont {A.}~\bibnamefont {Cavalleri}},\ }\bibfield
  {title} {\bibinfo {title} {Optically induced coherent transport far above
  ${T}_{c}$ in underdoped
  {${\mathrm{YBa}}_{2}{\mathrm{Cu}}_{3}{\mathrm{O}}_{6+\ensuremath{\delta}}$}},\
  }\href {https://doi.org/10.1103/PhysRevB.89.184516} {\bibfield  {journal}
  {\bibinfo  {journal} {Phys. Rev. B}\ }\textbf {\bibinfo {volume} {89}},\
  \bibinfo {pages} {184516} (\bibinfo {year} {2014}{\natexlab{b}})}\BibitemShut
  {NoStop}%
\bibitem [{\citenamefont {Nicoletti}\ \emph {et~al.}(2014)\citenamefont
  {Nicoletti}, \citenamefont {Casandruc}, \citenamefont {Laplace},
  \citenamefont {Khanna}, \citenamefont {Hunt}, \citenamefont {Kaiser},
  \citenamefont {Dhesi}, \citenamefont {Gu}, \citenamefont {Hill},\ and\
  \citenamefont {Cavalleri}}]{Nicoletti2014optically}%
  \BibitemOpen
  \bibfield  {author} {\bibinfo {author} {\bibfnamefont {D.}~\bibnamefont
  {Nicoletti}}, \bibinfo {author} {\bibfnamefont {E.}~\bibnamefont
  {Casandruc}}, \bibinfo {author} {\bibfnamefont {Y.}~\bibnamefont {Laplace}},
  \bibinfo {author} {\bibfnamefont {V.}~\bibnamefont {Khanna}}, \bibinfo
  {author} {\bibfnamefont {C.~R.}\ \bibnamefont {Hunt}}, \bibinfo {author}
  {\bibfnamefont {S.}~\bibnamefont {Kaiser}}, \bibinfo {author} {\bibfnamefont
  {S.~S.}\ \bibnamefont {Dhesi}}, \bibinfo {author} {\bibfnamefont {G.~D.}\
  \bibnamefont {Gu}}, \bibinfo {author} {\bibfnamefont {J.~P.}\ \bibnamefont
  {Hill}},\ and\ \bibinfo {author} {\bibfnamefont {A.}~\bibnamefont
  {Cavalleri}},\ }\bibfield  {title} {\bibinfo {title} {Optically induced
  superconductivity in striped
  {${\mathrm{La}}_{2\ensuremath{-}x}{\mathrm{Ba}}_{x}{\mathrm{CuO}}_{4}$} by
  polarization-selective excitation in the near infrared},\ }\href
  {https://doi.org/10.1103/PhysRevB.90.100503} {\bibfield  {journal} {\bibinfo
  {journal} {Phys. Rev. B}\ }\textbf {\bibinfo {volume} {90}},\ \bibinfo
  {pages} {100503} (\bibinfo {year} {2014})}\BibitemShut {NoStop}%
\bibitem [{\citenamefont {Nicoletti}\ \emph {et~al.}(2018)\citenamefont
  {Nicoletti}, \citenamefont {Fu}, \citenamefont {Mehio}, \citenamefont
  {Moore}, \citenamefont {Disa}, \citenamefont {Gu},\ and\ \citenamefont
  {Cavalleri}}]{Nicoletti2018magnetic}%
  \BibitemOpen
  \bibfield  {author} {\bibinfo {author} {\bibfnamefont {D.}~\bibnamefont
  {Nicoletti}}, \bibinfo {author} {\bibfnamefont {D.}~\bibnamefont {Fu}},
  \bibinfo {author} {\bibfnamefont {O.}~\bibnamefont {Mehio}}, \bibinfo
  {author} {\bibfnamefont {S.}~\bibnamefont {Moore}}, \bibinfo {author}
  {\bibfnamefont {A.~S.}\ \bibnamefont {Disa}}, \bibinfo {author}
  {\bibfnamefont {G.~D.}\ \bibnamefont {Gu}},\ and\ \bibinfo {author}
  {\bibfnamefont {A.}~\bibnamefont {Cavalleri}},\ }\bibfield  {title} {\bibinfo
  {title} {Magnetic-field tuning of light-induced superconductivity in striped
  {${\mathrm{La}}_{2\ensuremath{-}x}{\mathrm{Ba}}_{x}{\mathrm{CuO}}_{4}$}},\
  }\href {https://doi.org/10.1103/PhysRevLett.121.267003} {\bibfield  {journal}
  {\bibinfo  {journal} {Phys. Rev. Lett.}\ }\textbf {\bibinfo {volume} {121}},\
  \bibinfo {pages} {267003} (\bibinfo {year} {2018})}\BibitemShut {NoStop}%
\bibitem [{\citenamefont {Cremin}\ \emph {et~al.}(2019)\citenamefont {Cremin},
  \citenamefont {Zhang}, \citenamefont {Homes}, \citenamefont {Gu},
  \citenamefont {Sun}, \citenamefont {Fogler}, \citenamefont {Millis},
  \citenamefont {Basov},\ and\ \citenamefont
  {Averitt}}]{Cremin2019photoenhanced}%
  \BibitemOpen
  \bibfield  {author} {\bibinfo {author} {\bibfnamefont {K.~A.}\ \bibnamefont
  {Cremin}}, \bibinfo {author} {\bibfnamefont {J.}~\bibnamefont {Zhang}},
  \bibinfo {author} {\bibfnamefont {C.~C.}\ \bibnamefont {Homes}}, \bibinfo
  {author} {\bibfnamefont {G.~D.}\ \bibnamefont {Gu}}, \bibinfo {author}
  {\bibfnamefont {Z.}~\bibnamefont {Sun}}, \bibinfo {author} {\bibfnamefont
  {M.~M.}\ \bibnamefont {Fogler}}, \bibinfo {author} {\bibfnamefont {A.~J.}\
  \bibnamefont {Millis}}, \bibinfo {author} {\bibfnamefont {D.~N.}\
  \bibnamefont {Basov}},\ and\ \bibinfo {author} {\bibfnamefont {R.~D.}\
  \bibnamefont {Averitt}},\ }\bibfield  {title} {\bibinfo {title}
  {Photoenhanced metastable c-axis electrodynamics in stripe-ordered cuprate
  {$\mathrm{La_{1.885}Ba_{0.115}CuO_4}$}},\ }\href
  {https://doi.org/10.1073/pnas.1908368116} {\bibfield  {journal} {\bibinfo
  {journal} {Proceedings of the National Academy of Sciences}\ }\textbf
  {\bibinfo {volume} {116}},\ \bibinfo {pages} {19875} (\bibinfo {year}
  {2019})}\BibitemShut {NoStop}%
\bibitem [{\citenamefont {Weber}\ \emph {et~al.}(2012)\citenamefont {Weber},
  \citenamefont {Yee}, \citenamefont {Haule},\ and\ \citenamefont
  {Kotliar}}]{Weber2012scaling}%
  \BibitemOpen
  \bibfield  {author} {\bibinfo {author} {\bibfnamefont {C.}~\bibnamefont
  {Weber}}, \bibinfo {author} {\bibfnamefont {C.}~\bibnamefont {Yee}}, \bibinfo
  {author} {\bibfnamefont {K.}~\bibnamefont {Haule}},\ and\ \bibinfo {author}
  {\bibfnamefont {G.}~\bibnamefont {Kotliar}},\ }\bibfield  {title} {\bibinfo
  {title} {Scaling of the transition temperature of hole-doped cuprate
  superconductors with the charge-transfer energy},\ }\href
  {https://doi.org/10.1209/0295-5075/100/37001} {\bibfield  {journal} {\bibinfo
   {journal} {{EPL} (Europhysics Letters)}\ }\textbf {\bibinfo {volume}
  {100}},\ \bibinfo {pages} {37001} (\bibinfo {year} {2012})}\BibitemShut
  {NoStop}%
\bibitem [{\citenamefont {Ruan}\ \emph {et~al.}(2016)\citenamefont {Ruan},
  \citenamefont {Hu}, \citenamefont {Zhao}, \citenamefont {Cai}, \citenamefont
  {Peng}, \citenamefont {Ye}, \citenamefont {Yu}, \citenamefont {Li},
  \citenamefont {Hao}, \citenamefont {Jin}, \citenamefont {Zhou}, \citenamefont
  {Weng},\ and\ \citenamefont {Wang}}]{Ruan2016relationship}%
  \BibitemOpen
  \bibfield  {author} {\bibinfo {author} {\bibfnamefont {W.}~\bibnamefont
  {Ruan}}, \bibinfo {author} {\bibfnamefont {C.}~\bibnamefont {Hu}}, \bibinfo
  {author} {\bibfnamefont {J.}~\bibnamefont {Zhao}}, \bibinfo {author}
  {\bibfnamefont {P.}~\bibnamefont {Cai}}, \bibinfo {author} {\bibfnamefont
  {Y.}~\bibnamefont {Peng}}, \bibinfo {author} {\bibfnamefont {C.}~\bibnamefont
  {Ye}}, \bibinfo {author} {\bibfnamefont {R.}~\bibnamefont {Yu}}, \bibinfo
  {author} {\bibfnamefont {X.}~\bibnamefont {Li}}, \bibinfo {author}
  {\bibfnamefont {Z.}~\bibnamefont {Hao}}, \bibinfo {author} {\bibfnamefont
  {C.}~\bibnamefont {Jin}}, \bibinfo {author} {\bibfnamefont {X.}~\bibnamefont
  {Zhou}}, \bibinfo {author} {\bibfnamefont {Z.-Y.}\ \bibnamefont {Weng}},\
  and\ \bibinfo {author} {\bibfnamefont {Y.}~\bibnamefont {Wang}},\ }\bibfield
  {title} {\bibinfo {title} {Relationship between the parent charge transfer
  gap and maximum transition temperature in cuprates},\ }\href
  {https://doi.org/https://doi.org/10.1007/s11434-016-1204-x} {\bibfield
  {journal} {\bibinfo  {journal} {Science Bulletin}\ }\textbf {\bibinfo
  {volume} {61}},\ \bibinfo {pages} {1826} (\bibinfo {year}
  {2016})}\BibitemShut {NoStop}%
\bibitem [{\citenamefont {R\o{}mer}\ \emph {et~al.}(2020)\citenamefont
  {R\o{}mer}, \citenamefont {Maier}, \citenamefont {Kreisel}, \citenamefont
  {Eremin}, \citenamefont {Hirschfeld},\ and\ \citenamefont
  {Andersen}}]{Romer2020pairing}%
  \BibitemOpen
  \bibfield  {author} {\bibinfo {author} {\bibfnamefont {A.~T.}\ \bibnamefont
  {R\o{}mer}}, \bibinfo {author} {\bibfnamefont {T.~A.}\ \bibnamefont {Maier}},
  \bibinfo {author} {\bibfnamefont {A.}~\bibnamefont {Kreisel}}, \bibinfo
  {author} {\bibfnamefont {I.}~\bibnamefont {Eremin}}, \bibinfo {author}
  {\bibfnamefont {P.~J.}\ \bibnamefont {Hirschfeld}},\ and\ \bibinfo {author}
  {\bibfnamefont {B.~M.}\ \bibnamefont {Andersen}},\ }\bibfield  {title}
  {\bibinfo {title} {Pairing in the two-dimensional {Hubbard model} from weak
  to strong coupling},\ }\href
  {https://doi.org/10.1103/PhysRevResearch.2.013108} {\bibfield  {journal}
  {\bibinfo  {journal} {Phys. Rev. Research}\ }\textbf {\bibinfo {volume}
  {2}},\ \bibinfo {pages} {013108} (\bibinfo {year} {2020})}\BibitemShut
  {NoStop}%
\bibitem [{\citenamefont {Le~Tacon}\ \emph {et~al.}(2011)\citenamefont
  {Le~Tacon}, \citenamefont {Ghiringhelli}, \citenamefont {Chaloupka},
  \citenamefont {Sala}, \citenamefont {Hinkov}, \citenamefont {Haverkort},
  \citenamefont {Minola}, \citenamefont {Bakr}, \citenamefont {Zhou},
  \citenamefont {Blanco-Canosa}, \citenamefont {Monney}, \citenamefont {Song},
  \citenamefont {Sun}, \citenamefont {Lin}, \citenamefont {De~Luca},
  \citenamefont {Salluzzo}, \citenamefont {Khaliullin}, \citenamefont
  {Schmitt}, \citenamefont {Braicovich},\ and\ \citenamefont
  {Keimer}}]{LeTacon2011intense}%
  \BibitemOpen
  \bibfield  {author} {\bibinfo {author} {\bibfnamefont {M.}~\bibnamefont
  {Le~Tacon}}, \bibinfo {author} {\bibfnamefont {G.}~\bibnamefont
  {Ghiringhelli}}, \bibinfo {author} {\bibfnamefont {J.}~\bibnamefont
  {Chaloupka}}, \bibinfo {author} {\bibfnamefont {M.~M.}\ \bibnamefont {Sala}},
  \bibinfo {author} {\bibfnamefont {V.}~\bibnamefont {Hinkov}}, \bibinfo
  {author} {\bibfnamefont {M.~W.}\ \bibnamefont {Haverkort}}, \bibinfo {author}
  {\bibfnamefont {M.}~\bibnamefont {Minola}}, \bibinfo {author} {\bibfnamefont
  {M.}~\bibnamefont {Bakr}}, \bibinfo {author} {\bibfnamefont {K.~J.}\
  \bibnamefont {Zhou}}, \bibinfo {author} {\bibfnamefont {S.}~\bibnamefont
  {Blanco-Canosa}}, \bibinfo {author} {\bibfnamefont {C.}~\bibnamefont
  {Monney}}, \bibinfo {author} {\bibfnamefont {Y.~T.}\ \bibnamefont {Song}},
  \bibinfo {author} {\bibfnamefont {G.~L.}\ \bibnamefont {Sun}}, \bibinfo
  {author} {\bibfnamefont {C.~T.}\ \bibnamefont {Lin}}, \bibinfo {author}
  {\bibfnamefont {G.~M.}\ \bibnamefont {De~Luca}}, \bibinfo {author}
  {\bibfnamefont {M.}~\bibnamefont {Salluzzo}}, \bibinfo {author}
  {\bibfnamefont {G.}~\bibnamefont {Khaliullin}}, \bibinfo {author}
  {\bibfnamefont {T.}~\bibnamefont {Schmitt}}, \bibinfo {author} {\bibfnamefont
  {L.}~\bibnamefont {Braicovich}},\ and\ \bibinfo {author} {\bibfnamefont
  {B.}~\bibnamefont {Keimer}},\ }\bibfield  {title} {\bibinfo {title} {{Intense
  paramagnon excitations in a large family of high-temperature
  superconductors}},\ }\href {https://doi.org/10.1038/nphys2041} {\bibfield
  {journal} {\bibinfo  {journal} {Nature Physics}\ }\textbf {\bibinfo {volume}
  {7}},\ \bibinfo {pages} {725} (\bibinfo {year} {2011})}\BibitemShut {NoStop}%
\bibitem [{\citenamefont {Mitrano}\ and\ \citenamefont
  {Wang}(2020)}]{Mitrano2020probing}%
  \BibitemOpen
  \bibfield  {author} {\bibinfo {author} {\bibfnamefont {M.}~\bibnamefont
  {Mitrano}}\ and\ \bibinfo {author} {\bibfnamefont {Y.}~\bibnamefont {Wang}},\
  }\bibfield  {title} {\bibinfo {title} {Probing light-driven quantum materials
  with ultrafast resonant inelastic x-ray scattering},\ }\href
  {https://doi.org/10.1038/s42005-020-00447-6} {\bibfield  {journal} {\bibinfo
  {journal} {Communications Physics}\ }\textbf {\bibinfo {volume} {3}},\
  \bibinfo {pages} {184} (\bibinfo {year} {2020})}\BibitemShut {NoStop}%
\bibitem [{\citenamefont {Wang}\ \emph {et~al.}(2021)\citenamefont {Wang},
  \citenamefont {Chen}, \citenamefont {Devereaux}, \citenamefont {Moritz},\
  and\ \citenamefont {Mitrano}}]{wang2021xray}%
  \BibitemOpen
  \bibfield  {author} {\bibinfo {author} {\bibfnamefont {Y.}~\bibnamefont
  {Wang}}, \bibinfo {author} {\bibfnamefont {Y.}~\bibnamefont {Chen}}, \bibinfo
  {author} {\bibfnamefont {T.~P.}\ \bibnamefont {Devereaux}}, \bibinfo {author}
  {\bibfnamefont {B.}~\bibnamefont {Moritz}},\ and\ \bibinfo {author}
  {\bibfnamefont {M.}~\bibnamefont {Mitrano}},\ }\bibfield  {title} {\bibinfo
  {title} {X-ray scattering from light-driven spin fluctuations in a doped mott
  insulator},\ }\bibfield  {journal} {\bibinfo  {journal} {Communications
  Physics}\ }\href {https://doi.org/10.1038/s42005-021-00715-z}
  {10.1038/s42005-021-00715-z} (\bibinfo {year} {2021})\BibitemShut {NoStop}%
\bibitem [{\citenamefont {Yang}\ \emph {et~al.}(2010)\citenamefont {Yang},
  \citenamefont {L\"auchli}, \citenamefont {Mila},\ and\ \citenamefont
  {Schmidt}}]{Yang2010effective}%
  \BibitemOpen
  \bibfield  {author} {\bibinfo {author} {\bibfnamefont {H.-Y.}\ \bibnamefont
  {Yang}}, \bibinfo {author} {\bibfnamefont {A.~M.}\ \bibnamefont {L\"auchli}},
  \bibinfo {author} {\bibfnamefont {F.}~\bibnamefont {Mila}},\ and\ \bibinfo
  {author} {\bibfnamefont {K.~P.}\ \bibnamefont {Schmidt}},\ }\bibfield
  {title} {\bibinfo {title} {{Effective Spin Model for the Spin-Liquid Phase of
  the Hubbard Model on the Triangular Lattice}},\ }\href
  {https://doi.org/10.1103/PhysRevLett.105.267204} {\bibfield  {journal}
  {\bibinfo  {journal} {Phys. Rev. Lett.}\ }\textbf {\bibinfo {volume} {105}},\
  \bibinfo {pages} {267204} (\bibinfo {year} {2010})}\BibitemShut {NoStop}%
\bibitem [{\citenamefont {Sahebsara}\ and\ \citenamefont
  {S\'en\'echal}(2008)}]{Sahebsara2008hubbard}%
  \BibitemOpen
  \bibfield  {author} {\bibinfo {author} {\bibfnamefont {P.}~\bibnamefont
  {Sahebsara}}\ and\ \bibinfo {author} {\bibfnamefont {D.}~\bibnamefont
  {S\'en\'echal}},\ }\bibfield  {title} {\bibinfo {title} {{Hubbard Model on
  the Triangular Lattice: Spiral Order and Spin Liquid}},\ }\href
  {https://doi.org/10.1103/PhysRevLett.100.136402} {\bibfield  {journal}
  {\bibinfo  {journal} {Phys. Rev. Lett.}\ }\textbf {\bibinfo {volume} {100}},\
  \bibinfo {pages} {136402} (\bibinfo {year} {2008})}\BibitemShut {NoStop}%
\bibitem [{\citenamefont {Shirakawa}\ \emph {et~al.}(2017)\citenamefont
  {Shirakawa}, \citenamefont {Tohyama}, \citenamefont {Kokalj}, \citenamefont
  {Sota},\ and\ \citenamefont {Yunoki}}]{Shirakawa2017ground}%
  \BibitemOpen
  \bibfield  {author} {\bibinfo {author} {\bibfnamefont {T.}~\bibnamefont
  {Shirakawa}}, \bibinfo {author} {\bibfnamefont {T.}~\bibnamefont {Tohyama}},
  \bibinfo {author} {\bibfnamefont {J.}~\bibnamefont {Kokalj}}, \bibinfo
  {author} {\bibfnamefont {S.}~\bibnamefont {Sota}},\ and\ \bibinfo {author}
  {\bibfnamefont {S.}~\bibnamefont {Yunoki}},\ }\bibfield  {title} {\bibinfo
  {title} {Ground-state phase diagram of the triangular lattice {Hubbard model}
  by the density-matrix renormalization group method},\ }\href
  {https://doi.org/10.1103/PhysRevB.96.205130} {\bibfield  {journal} {\bibinfo
  {journal} {Phys. Rev. B}\ }\textbf {\bibinfo {volume} {96}},\ \bibinfo
  {pages} {205130} (\bibinfo {year} {2017})}\BibitemShut {NoStop}%
\bibitem [{\citenamefont {Szasz}\ \emph {et~al.}(2020)\citenamefont {Szasz},
  \citenamefont {Motruk}, \citenamefont {Zaletel},\ and\ \citenamefont
  {Moore}}]{Szasz2020chiral}%
  \BibitemOpen
  \bibfield  {author} {\bibinfo {author} {\bibfnamefont {A.}~\bibnamefont
  {Szasz}}, \bibinfo {author} {\bibfnamefont {J.}~\bibnamefont {Motruk}},
  \bibinfo {author} {\bibfnamefont {M.~P.}\ \bibnamefont {Zaletel}},\ and\
  \bibinfo {author} {\bibfnamefont {J.~E.}\ \bibnamefont {Moore}},\ }\bibfield
  {title} {\bibinfo {title} {{Chiral Spin Liquid Phase of the Triangular
  Lattice Hubbard Model: A Density Matrix Renormalization Group Study}},\
  }\href {https://doi.org/10.1103/PhysRevX.10.021042} {\bibfield  {journal}
  {\bibinfo  {journal} {Phys. Rev. X}\ }\textbf {\bibinfo {volume} {10}},\
  \bibinfo {pages} {021042} (\bibinfo {year} {2020})}\BibitemShut {NoStop}%
\bibitem [{\citenamefont {Liao}\ \emph {et~al.}(2017)\citenamefont {Liao},
  \citenamefont {Xie}, \citenamefont {Chen}, \citenamefont {Liu}, \citenamefont
  {Xie}, \citenamefont {Huang}, \citenamefont {Normand},\ and\ \citenamefont
  {Xiang}}]{Liao2017gapless}%
  \BibitemOpen
  \bibfield  {author} {\bibinfo {author} {\bibfnamefont {H.~J.}\ \bibnamefont
  {Liao}}, \bibinfo {author} {\bibfnamefont {Z.~Y.}\ \bibnamefont {Xie}},
  \bibinfo {author} {\bibfnamefont {J.}~\bibnamefont {Chen}}, \bibinfo {author}
  {\bibfnamefont {Z.~Y.}\ \bibnamefont {Liu}}, \bibinfo {author} {\bibfnamefont
  {H.~D.}\ \bibnamefont {Xie}}, \bibinfo {author} {\bibfnamefont {R.~Z.}\
  \bibnamefont {Huang}}, \bibinfo {author} {\bibfnamefont {B.}~\bibnamefont
  {Normand}},\ and\ \bibinfo {author} {\bibfnamefont {T.}~\bibnamefont
  {Xiang}},\ }\bibfield  {title} {\bibinfo {title} {Gapless spin-liquid ground
  state in the $s=1/2$ kagome antiferromagnet},\ }\href
  {https://doi.org/10.1103/PhysRevLett.118.137202} {\bibfield  {journal}
  {\bibinfo  {journal} {Phys. Rev. Lett.}\ }\textbf {\bibinfo {volume} {118}},\
  \bibinfo {pages} {137202} (\bibinfo {year} {2017})}\BibitemShut {NoStop}%
\bibitem [{\citenamefont {Yan}\ \emph {et~al.}(2011)\citenamefont {Yan},
  \citenamefont {Huse},\ and\ \citenamefont {White}}]{Simeng2011spin}%
  \BibitemOpen
  \bibfield  {author} {\bibinfo {author} {\bibfnamefont {S.}~\bibnamefont
  {Yan}}, \bibinfo {author} {\bibfnamefont {D.~A.}\ \bibnamefont {Huse}},\ and\
  \bibinfo {author} {\bibfnamefont {S.~R.}\ \bibnamefont {White}},\ }\bibfield
  {title} {\bibinfo {title} {Spin-liquid ground state of the <i>s</i> = 1/2
  kagome heisenberg antiferromagnet},\ }\href
  {https://doi.org/10.1126/science.1201080} {\bibfield  {journal} {\bibinfo
  {journal} {Science}\ }\textbf {\bibinfo {volume} {332}},\ \bibinfo {pages}
  {1173} (\bibinfo {year} {2011})},\ \Eprint
  {https://arxiv.org/abs/https://www.science.org/doi/pdf/10.1126/science.1201080}
  {https://www.science.org/doi/pdf/10.1126/science.1201080} \BibitemShut
  {NoStop}%
\bibitem [{\citenamefont {Jiang}\ \emph {et~al.}(2012)\citenamefont {Jiang},
  \citenamefont {Wang},\ and\ \citenamefont {Balents}}]{Jiang2012identifying}%
  \BibitemOpen
  \bibfield  {author} {\bibinfo {author} {\bibfnamefont {H.-C.}\ \bibnamefont
  {Jiang}}, \bibinfo {author} {\bibfnamefont {Z.}~\bibnamefont {Wang}},\ and\
  \bibinfo {author} {\bibfnamefont {L.}~\bibnamefont {Balents}},\ }\bibfield
  {title} {\bibinfo {title} {Identifying topological order by entanglement
  entropy},\ }\href {https://doi.org/10.1038/nphys2465} {\bibfield  {journal}
  {\bibinfo  {journal} {Nature Physics}\ }\textbf {\bibinfo {volume} {8}},\
  \bibinfo {pages} {902} (\bibinfo {year} {2012})}\BibitemShut {NoStop}%
\bibitem [{\citenamefont {Kaneko}\ \emph {et~al.}(2019)\citenamefont {Kaneko},
  \citenamefont {Shirakawa}, \citenamefont {Sorella},\ and\ \citenamefont
  {Yunoki}}]{Kaneko2019photoinduced}%
  \BibitemOpen
  \bibfield  {author} {\bibinfo {author} {\bibfnamefont {T.}~\bibnamefont
  {Kaneko}}, \bibinfo {author} {\bibfnamefont {T.}~\bibnamefont {Shirakawa}},
  \bibinfo {author} {\bibfnamefont {S.}~\bibnamefont {Sorella}},\ and\ \bibinfo
  {author} {\bibfnamefont {S.}~\bibnamefont {Yunoki}},\ }\bibfield  {title}
  {\bibinfo {title} {{Photoinduced $\ensuremath{\eta}$ Pairing in the Hubbard
  Model}},\ }\href {https://doi.org/10.1103/PhysRevLett.122.077002} {\bibfield
  {journal} {\bibinfo  {journal} {Phys. Rev. Lett.}\ }\textbf {\bibinfo
  {volume} {122}},\ \bibinfo {pages} {077002} (\bibinfo {year}
  {2019})}\BibitemShut {NoStop}%
\bibitem [{\citenamefont {Peronaci}\ \emph {et~al.}(2020)\citenamefont
  {Peronaci}, \citenamefont {Parcollet},\ and\ \citenamefont
  {Schir\'o}}]{peronaci2020enhancement}%
  \BibitemOpen
  \bibfield  {author} {\bibinfo {author} {\bibfnamefont {F.}~\bibnamefont
  {Peronaci}}, \bibinfo {author} {\bibfnamefont {O.}~\bibnamefont
  {Parcollet}},\ and\ \bibinfo {author} {\bibfnamefont {M.}~\bibnamefont
  {Schir\'o}},\ }\bibfield  {title} {\bibinfo {title} {{Enhancement of local
  pairing correlations in periodically driven Mott insulators}},\ }\href
  {https://doi.org/10.1103/PhysRevB.101.161101} {\bibfield  {journal} {\bibinfo
   {journal} {Phys. Rev. B}\ }\textbf {\bibinfo {volume} {101}},\ \bibinfo
  {pages} {161101} (\bibinfo {year} {2020})}\BibitemShut {NoStop}%
\bibitem [{\citenamefont {Li}\ \emph {et~al.}(2020)\citenamefont {Li},
  \citenamefont {Golez}, \citenamefont {Werner},\ and\ \citenamefont
  {Eckstein}}]{Li2020eta}%
  \BibitemOpen
  \bibfield  {author} {\bibinfo {author} {\bibfnamefont {J.}~\bibnamefont
  {Li}}, \bibinfo {author} {\bibfnamefont {D.}~\bibnamefont {Golez}}, \bibinfo
  {author} {\bibfnamefont {P.}~\bibnamefont {Werner}},\ and\ \bibinfo {author}
  {\bibfnamefont {M.}~\bibnamefont {Eckstein}},\ }\bibfield  {title} {\bibinfo
  {title} {{$\ensuremath{\eta}$-paired superconducting hidden phase in
  photodoped Mott insulators}},\ }\href
  {https://doi.org/10.1103/PhysRevB.102.165136} {\bibfield  {journal} {\bibinfo
   {journal} {Phys. Rev. B}\ }\textbf {\bibinfo {volume} {102}},\ \bibinfo
  {pages} {165136} (\bibinfo {year} {2020})}\BibitemShut {NoStop}%
\bibitem [{\citenamefont {Tindall}\ \emph {et~al.}(2020)\citenamefont
  {Tindall}, \citenamefont {Schlawin}, \citenamefont {Buzzi}, \citenamefont
  {Nicoletti}, \citenamefont {Coulthard}, \citenamefont {Gao}, \citenamefont
  {Cavalleri}, \citenamefont {Sentef},\ and\ \citenamefont
  {Jaksch}}]{Tindall2020dynamical}%
  \BibitemOpen
  \bibfield  {author} {\bibinfo {author} {\bibfnamefont {J.}~\bibnamefont
  {Tindall}}, \bibinfo {author} {\bibfnamefont {F.}~\bibnamefont {Schlawin}},
  \bibinfo {author} {\bibfnamefont {M.}~\bibnamefont {Buzzi}}, \bibinfo
  {author} {\bibfnamefont {D.}~\bibnamefont {Nicoletti}}, \bibinfo {author}
  {\bibfnamefont {J.~R.}\ \bibnamefont {Coulthard}}, \bibinfo {author}
  {\bibfnamefont {H.}~\bibnamefont {Gao}}, \bibinfo {author} {\bibfnamefont
  {A.}~\bibnamefont {Cavalleri}}, \bibinfo {author} {\bibfnamefont {M.~A.}\
  \bibnamefont {Sentef}},\ and\ \bibinfo {author} {\bibfnamefont
  {D.}~\bibnamefont {Jaksch}},\ }\bibfield  {title} {\bibinfo {title}
  {Dynamical order and superconductivity in a frustrated many-body system},\
  }\href {https://doi.org/10.1103/PhysRevLett.125.137001} {\bibfield  {journal}
  {\bibinfo  {journal} {Phys. Rev. Lett.}\ }\textbf {\bibinfo {volume} {125}},\
  \bibinfo {pages} {137001} (\bibinfo {year} {2020})}\BibitemShut {NoStop}%
\bibitem [{\citenamefont {Zhang}(1990)}]{Zhang1990pseudospin}%
  \BibitemOpen
  \bibfield  {author} {\bibinfo {author} {\bibfnamefont {S.}~\bibnamefont
  {Zhang}},\ }\bibfield  {title} {\bibinfo {title} {Pseudospin symmetry and new
  collective modes of the {Hubbard model}},\ }\href
  {https://doi.org/10.1103/PhysRevLett.65.120} {\bibfield  {journal} {\bibinfo
  {journal} {Phys. Rev. Lett.}\ }\textbf {\bibinfo {volume} {65}},\ \bibinfo
  {pages} {120} (\bibinfo {year} {1990})}\BibitemShut {NoStop}%
\bibitem [{\citenamefont {Mansart}\ \emph {et~al.}(2010)\citenamefont
  {Mansart}, \citenamefont {Boschetto}, \citenamefont {Savoia}, \citenamefont
  {Rullier-Albenque}, \citenamefont {Bouquet}, \citenamefont {Papalazarou},
  \citenamefont {Forget}, \citenamefont {Colson}, \citenamefont {Rousse},\ and\
  \citenamefont {Marsi}}]{Mansart2010}%
  \BibitemOpen
  \bibfield  {author} {\bibinfo {author} {\bibfnamefont {B.}~\bibnamefont
  {Mansart}}, \bibinfo {author} {\bibfnamefont {D.}~\bibnamefont {Boschetto}},
  \bibinfo {author} {\bibfnamefont {A.}~\bibnamefont {Savoia}}, \bibinfo
  {author} {\bibfnamefont {F.}~\bibnamefont {Rullier-Albenque}}, \bibinfo
  {author} {\bibfnamefont {F.}~\bibnamefont {Bouquet}}, \bibinfo {author}
  {\bibfnamefont {E.}~\bibnamefont {Papalazarou}}, \bibinfo {author}
  {\bibfnamefont {A.}~\bibnamefont {Forget}}, \bibinfo {author} {\bibfnamefont
  {D.}~\bibnamefont {Colson}}, \bibinfo {author} {\bibfnamefont
  {A.}~\bibnamefont {Rousse}},\ and\ \bibinfo {author} {\bibfnamefont
  {M.}~\bibnamefont {Marsi}},\ }\bibfield  {title} {\bibinfo {title} {Ultrafast
  transient response and electron-phonon coupling in the iron-pnictide
  superconductor
  {${\text{Ba}{({\text{Fe}}_{1\ensuremath{-}x}{\text{Co}}_{x})}_{2}{\text{As}}_{2}}$}},\
  }\href {https://doi.org/10.1103/PhysRevB.82.024513} {\bibfield  {journal}
  {\bibinfo  {journal} {Phys. Rev. B}\ }\textbf {\bibinfo {volume} {82}},\
  \bibinfo {pages} {024513} (\bibinfo {year} {2010})}\BibitemShut {NoStop}%
\end{thebibliography}%
\end{document}